\documentclass[preprint2]{aastex}

\usepackage{natbib}
\slugcomment{ApJ accepted}

\shorttitle{Obscuration in extremely luminous quasars}
\shortauthors{Polletta et al.}

\def\rp{$r^\prime$}
\def\gp{$g^\prime$}
\def\rp{$r^\prime$}
\def\ip{$i^\prime$}
\def\msun{M$_{\odot}$}
\def\lsun{L$_{\odot}$}
\def\deg{$^{\circ}$}

\def\chandra {{\it Chandra}}
\def\xmm {XMM-{\it Newton}}
\def\spitzer {{\it Spitzer}}
\def\nh {${\rm N_\mathrm{H}}$}
\def\av {${\rm A_\mathrm{V}}$}

\def\kms{\ifmmode {\rm\,km\,s^{-1}}\else
    ${\rm\,km\,s^{-1}}$\fi}
\def\kmsMpc{\ifmmode {\rm\,km\,s^{-1}\,Mpc^{-1}}\else
    ${\rm\,km\,s^{-1}\,Mpc^{-1}}$\fi}
\def\ergAcm2{\ifmmode {\rm\,ergs\,cm^{-2}\,{\rm \AA}^{-1}}\else
    ${\rm\,ergs\,cm^{-2}\,\AA^{-1}}$\fi}
\def\cm2{\ifmmode {\rm\,cm^{-2}}\else
    ${\rm\,cm^{-2}}$\fi}
\def\ergcm2s{\ifmmode {\rm\,ergs\,cm^{-2}\,s^{-1}}\else
    ${\rm\,ergs\,cm^{-2}\,s^{-1}}$\fi}
\def\cgsdeg2{\ifmmode {\rm\,ergs\,cm^{-2}\,s^{-1}\,deg^{-2}}\else
    ${\rm\,ergs\,cm^{-2}\,s^{-1}\,deg^{-2}}$\fi}
\def\sqdeg{\ifmmode {\rm\,deg^{2}}\else
    ${\rm\,deg^{2}}$\fi}
\def\ergsHz{\ifmmode {\rm\,ergs\,s^{-1}\,Hz^{-1}}\else
    ${\rm\,ergs\,s^{-1}\,Hz^{-1}}$\fi}
\def\ergs{\ifmmode {\rm\,ergs\,s^{-1}}\else
    ${\rm\,ergs\,s^{-1}}$\fi}
\def\ergsA{\ifmmode {\rm\,ergs\,s^{-1}\,\AA^{-1}}\else
    ${\rm\,ergs\,s^{-1}\,\AA^{-1}}$\fi}
\def\WHz{\ifmmode {\rm\,W\,Hz^{-1}}\else
    ${\rm\,W\,Hz^{-1}}$\fi}
\def\WHzsr{\ifmmode {\rm\,W\,Hz^{-1}\,sr^{-1}}\else
    ${\rm\,W\,Hz^{-1}\,sr^{-1}}$\fi}
\def\ergscm2Hz{\ifmmode {\rm\,ergs\,cm^{-2}\,s^{-1}\,Hz^{-1}}\else
    ${\rm\,ergs\,cm^{-2}\,s^{-1}\,Hz^{-1}}$\fi}

\def\um{$\mu$m}

\def\lya{Ly$\alpha$}
\def\nv{\ion{N}{5} $\lambda$1240}

\def\civ{\ion{C}{4} $\lambda$1549}

\def\oii{[\ion{O}{2}] $\lambda$3727}

\begin{document}


\title{Obscuration in extremely luminous quasars}


\author{M. Polletta\altaffilmark{1,2}, D. Weedman\altaffilmark{3},
 S. H\"onig\altaffilmark{4}, C.~J. Lonsdale\altaffilmark{1,5}, H.~E.
 Smith\altaffilmark{1,6}, J. Houck\altaffilmark{3}}

\altaffiltext{1}{Center for Astrophysics \& Space Sciences, University of California, San
      Diego, La Jolla, CA  92093--0424, USA}
\altaffiltext{2}{Institut d'Astrophysique de Paris, 98bis bld Arago, 75014 Paris, France}
\email{polletta@iap.fr}
\altaffiltext{3}{Department of Astronomy, Cornell University, Ithaca, NY
      14853, USA}
\altaffiltext{4}{Max-Planck-Institut f\"u Radioastronomie, 53121 Bonn, Germany}
\altaffiltext{5}{Infrared Processing \& Analysis Center, California Institute
      of Technology, 100-22, Pasadena, CA 91125, USA}
\altaffiltext{6}{Deceased 2007 August 16.}

\begin{abstract}
The spectral energy distributions and infrared spectra of a remarkable
sample of obscured AGNs selected in the mid-infrared are modeled with recent
clumpy torus models~\citep{hoenig06} to investigate the nature of the
sources, the properties of the obscuring matter, and dependencies on
luminosity.  The sample contains 21 obscured AGNs at $z$=1.3--3 that were
discovered in the largest \spitzer\ surveys (SWIRE, NDWFS, \& FLS) by means of
their extremely red infrared to optical colors. All selected sources show
the 9.7$\mu$m silicate feature in absorption and have extreme mid-infrared
luminosities ($\nu L_{\nu}$(6$\mu$m) = $L(6\mu m)\simeq$ 10$^{46}$\ergs).

The infrared SEDs and spectra of 12 sources are well reproduced with a
simple torus model, while the remaining 9 sources require foreground
extinction from a cold dust component to reproduce both the depth of the
silicate feature and the near-infrared emission from hot dust. The best-fit
torus models show a broad range of inclinations, with no preference for the
edge-on torus expected in obscured AGNs.

Based on the unobscured QSO mid-infrared MIR luminosity
function~\citep{brown06}, and on a color-selected sample of obscured and
unobscured infrared sources, we estimate the surface densities of obscured
and unobscured QSOs at L(6$\mu$m)$>$10$^{12}$\lsun\, and $z$=1.3-3.0 to be
about 17--22 deg$^{-2}$, and 11.7 deg$^{-2}$, respectively.  Overall we find
that $\sim$35--41\% of luminous QSOs are unobscured, 37--40\% are obscured
by the torus, and 23--25\% are obscured by a cold absorber detached from the
torus.  These fractions constrain the torus half opening angle (measured
from the torus axis) to be $\sim$67\deg. This value is significantly larger
than found for FIR selected samples of AGN at lower luminosity
($\sim$46\deg), supporting the receding torus scenario.

A far-infrared component is observed in 8 objects. The estimated
far-infrared luminosities associated with this component all exceed
3.3$\times$10$^{12}$\lsun, implying SFRs of 600--3000 \msun\ yr$^{-1}$. For
the whole sample, the average contribution from a starburst component to the
bolometric luminosity, as estimated from the PAH 7.7$\mu$m luminosity in the
composite IR spectra, is $\leq$20\% of the total bolometric luminosity.

Five sources are detected in the X-rays, and large column densities
(\nh$\geq$10$^{23}$\cm2) are estimated for all of them.
 \end{abstract}


\section{Introduction}\label{intro}

\subsection{Obscured AGNs at high luminosities}

Multiple X-ray studies show that the fraction of active galactic nuclei
(AGNs) whose emission is heavily absorbed (\nh$\geq10^{22}$\cm2) decreases
with increasing luminosity~\citep[from $>$80 at L$_X$=10$^{42}$\ergs, to
38\% at 10$^{45}$\ergs;][]{akylas06}. However, because of the difficulty in
detecting and identifying absorbed AGNs, it is still unclear whether their
paucity at high luminosities is an observational selection effect or
real~\citep{lafranca05,treister06,akylas06,tozzi06}. In order to overcome
these uncertainties, searches for absorbed QSOs have focused on observations
at wavelengths less affected by absorption, i.e., infrared (IR) and radio,
e.g., FIRST, 2MASS and various \spitzer\
surveys~\citep[e.g.][]{polletta06,martinez06a,lacy07,wilkes02,urrutia05}.

These searches have unveiled a large population of QSOs obscured at optical
wavelengths. Assuming that all optically obscured QSOs are also absorbed in 
X-rays, the fraction of absorbed QSOs would be $\geq$50\% of all
QSOs~\citep{martinez05}. This fraction is still significantly lower than the
fraction measured for AGNs at lower luminosities, 80\%~\citep{osterbrock93,akylas06}.

Current AGN evolutionary models and observations indicate a link between the
absorption properties in AGNs and their luminosities~\citep[see e.g. Figure
4 in ][]{hopkins05b}. More specifically, AGNs in a growing phase are
moderately luminous and heavily absorbed, and as the AGN luminosity
increases the intense radiation field destroys the surrounding matter and
the AGN shines unabsorbed~\citep[e.g.][]{dimatteo05}. A similar trend is
predicted by the receding torus
models~\citep[e.g.][]{lawrence91a,simpson05}. According to these models, the
opening angle of the torus (measured from the torus axis to the equatorial
plane) is larger in more luminous objects. Thus, obscuration should be less
common in more luminous AGNs.

\subsection{The obscuring matter in AGNs}

Obscuration in AGNs is caused by a mixture of neutral and ionized gas, as
well as dust. Absorption by gas is usually observed in the X-rays through
the suppression of soft X-ray emission with an energy cut-off that increases
with the gas column density, \nh. The absorbing gas is mainly located in the
circumnuclear region as suggested by measured
variability~\citep{risaliti05c,risaliti07}, but gas in our Galaxy and in the
host galaxy can also contribute to the overall absorption. Obscuration by
dust is usually observed at optical and IR wavelengths where the continuum
and broad emission lines are reddened and, in some cases, completely
suppressed. In the near- and mid-IR (NIR and MIR), dust obscuration
interplays with thermal re-emission from the putative parsec-scaled dust
torus. Depending on the actual geometry, optical depth and chemistry, dust
absorption features at about 10 and 18$\mu$m due to silicates ($Si$) can be
present in the IR.

According to AGN unification models~\citep[e.g. ][]{antonucci93} and to
torus
models~\citep[e.g.,][]{krolik88,pier92,efstathiou95,granato97,nenkova02,dullemond05,fritz06,hoenig06}
optical depths are expected to be at least roughly correlated from the X-ray
through the MIR, although this correlation might be very weak for clumpy
tori. In obscured AGNs, the line of sight (LOS) intercepts the putative
torus and thus absorbs radiation from the nucleus (broad emission lines,
X-rays and thermal radiation from hot dust), while in unobscured AGNs the
radiation from the nucleus is directly visible and thus unabsorbed.  An
edge-on view of the torus is characterized by red optical through IR colors
and $Si$ features in absorption (at 10 and 18$\mu$m), while a face-on torus
has bluer colors and $Si$ features absent or in emission. Sources with
narrow emission lines in their optical spectra, red colors and strong $Si$
absorption features are usually associated with a highly inclined torus and
thus are expected to be also absorbed in the X-rays. However, there is
emerging evidence for a surprising mismatch between the absorption measured
in the X-rays and that measured in the optical through IR for a significant
fraction of sources, especially at high luminosities.

AGNs selected in the mid-IR because of extremely red optical through IR
colors sometimes show broad optical emission lines, even having absorption
features in their IR spectra in some cases~\citep{brand07}. The comparison
between X-ray absorption and optical obscuration in various X-ray selected
AGN samples shows that about 20--30\% of obscured AGNs are not absorbed in
the X-rays and viceversa~\citep[e.g.][]{perola04,tozzi06,tajer07,gliozzi07}. 
The MIR spectra of a sample of type 2 QSOs with heavy X-ray absorption did
not reveal the $Si$ feature in absorption at 10$\mu$m~\citep{sturm06} as
expected. This mismatch is even more pronounced at high luminosities.
Indeed, absorption signatures in the IR, e.g. the $Si$ absorption feature at
10$\mu$m, are more prominent at the high luminosities of ultra luminous IR
galaxies (ULIRGS)~\citep{sanders96,hao07}, while X-ray absorption is
progressively less common in AGNs with increasing
luminosity~\citep{ueda03,hasinger05}. These results suggest that a non
negligible fraction of obscured AGNs might not be obscured by a torus, but
by dust in either the narrow line region~\citep{sturm05} or in the host
galaxy and that large X-ray column densities are not always associated with
geometrically and optically thick dust distributions~\citep[see
also][]{rigby06,martinez06a}. Alternatively, this mismatch can be explained
by a low dust-to-gas ratio (\av/\nh) or by a different path for the IR and
the X-ray LOSs~\citep{maiolino01a,shi06}.

Large optical depths for $Si$ absorption features ($>$1.7) usually imply a
compact source deeply embedded in a smooth distribution of material which is
both geometrically and optically thick, rather than absorption by a 
foreground screen of dust close to the heating
source~\citep{levenson07,imanishi07}.  Also a detached foreground cold
absorber which is far enough from the heating source to be cold can produce
large $Si$ optical depths. Deep ($\tau_{Si}>$1.7) $Si$ absorption features
are usually considered as indicators of a buried compact source such as an
AGN. However, it is not clear whether the optical depth is a consequence of
a random alignment of $Si$ clouds or a specific dust distribution and
orientation. The observed MIR spectra of obscured AGNs do not favor the
latter scenario because the $Si$ absorption features are not always present
in the spectra of optically obscured and X-ray absorbed AGNs, as would be
expected if the $Si$ absorbers were associated with the same material which
suppresses the broad emission lines and the soft X-ray emission.

In order to investigate the properties of the obscuring matter in extremely
luminous AGNs and asses how often the absorption signatures at optical,
X-ray and infrared wavelengths do not correlate, a comprehensive analysis
of the absorption properties at all wavelengths of obscured AGNs at high
luminosity is necessary, as well as detailed studies of their absorbing
matter. Observations at various wavelengths from X-ray through optical and
IR constrain the absorption along multiple LOSs and thus the
geometry and distribution of the absorbing matter.

In this work, we investigate the observed properties of the obscuring matter
in a sample of obscured and extremely luminous AGNs, and we model their SEDs
with clumpy torus models~\citep{hoenig06} to explore the dust geometries
associated with large obscuration. This study is based on a sample of
extremely luminous and obscured AGNs for which multi-wavelength SEDs and IR
spectra are available. The selected sample includes sources from the 
three widest \spitzer\ extragalactic surveys which were observed with the
\spitzer\ InfraRed Spectrograph~\citep[IRS;
][]{houck04}. In \S~\ref{sample}, we describe the sample selection, and
in \S~\ref{obs_data} the observations and the data used in this work.
The IRS spectra and detected spectral features are presented in
\S~\ref{irs_spectral_prop}. The modeling of the SEDs using clumpy
torus models~\citep{hoenig06} is presented in \S~\ref{models}. The
composite spectra of two sub-samples defined on the basis of the model
results are analyzed in \S~\ref{avg_spe}. The properties (IR colors,
surface density, luminosity, optical depths, and redshift distribution) of
the selected sample are compared with those of other samples of AGNs and
ULIRGs from the literature in \S~\ref{lit_comparison}. The X-ray
properties and a comparison between the absorption seen in the X-ray and the
optical depth in the IR are described in \S~\ref{xray}. Our results and
their implications are discussed in \S~\ref{discussion} and summarized
in \S~\ref{summary}.

Throughout the paper, the term ``absorbed'' refers to X-ray sources
with effective column densities \nh$\geq$10$^{22}$\cm2, and ``obscured'' to
sources with red optical-IR colors, e.g. F(3.6$\mu$m)/F(\rp)$>$20 or
F(24$\mu$m)/F(\rp)$>$100, implying extreme dust extinction. The IR SEDs are
defined as ``red'' if they are as red or redder than a power-law model,
F$_{\nu}\propto \nu^{-\alpha}$, with slope $\alpha$=2. The terms type 1 and
type 2 refer to AGNs with broad and narrow optical emission lines,
respectively, in their optical spectra. We adopt a flat cosmology with H$_0$
= 71\kmsMpc, $\Omega_{M}$=0.27 and $\Omega_{\Lambda}$=0.73~\citep{spergel03}.

\section{Sample description}\label{sample}

There are three large area \spitzer\ surveys with Infrared Array
Camera~\citep[IRAC;][]{fazio04}, Multiband Imaging
Photometer~\citep[MIPS;][]{rieke04}, and optical photometric coverage, the
50\,deg$^2$ \spitzer\ Wide-Area Infrared Extragalactic
Survey~\citep[SWIRE\footnote{http://swire.ipac.caltech.edu/swire/swire.html};][]{lonsdale03,lonsdale04};
the 9 deg$^2$ NOAO Deep Wide-Field Survey (NDWFS) of the Bootes
field~\citep{jannuzi99,murray05}; and the 3.7 deg$^2$ Extragalactic First Look
Survey~\citep[E-FLS\footnote{http://ssc.spitzer.caltech.edu/fls/};][]{fadda06}.

The initial objective for \spitzer\ IRS spectroscopy of sources chosen from
these surveys was to understand populations of new sources that would not
have been known prior to \spitzer.  Consequently, spectroscopy has
emphasized sources which are optically very faint, typically mag($R$)$>$24,
and IR bright, i.e. F(24$\mu$m)$>$1\,mJy. Various selection criteria based
on IRAC colors also entered choices of spectroscopic targets, but the
unifying theme of existing spectroscopic samples is their high ratio of
mid-infrared to optical flux, IR/opt =
$\nu$F$_{\nu}$(24\,\um)/$\nu$F$_{\nu}$($R$).  Most sources chosen for
spectroscopy have IR/opt $\geq$ 10.  For 58 sources observed in
Bootes~\citep{houck05,weedman06b}, 70 sources observed in the
FLS~\citep{yan07,weedman06c}, and 20 sources in the SWIRE Lockman Hole
field~\citep{weedman06a}, the majority are at high redshift ($z\sim$2) and
many have infrared spectra dominated by the strong silicate ($Si$) absorption
feature at rest frame 9.7\,$\mu$m.  The presence of this absorption combined
with the red colors and high luminosities of these sources led these
previous observers to the conclusion that IR bright sources with large
IR/optical flux ratios are dominated by obscured
AGN (or QSOs because of their very high luminosities).  A minority of
sources in these studies show strong PAH emission features characteristic of
starbursts, but these sources generally do not show strong $Si$
absorption, so different populations of AGN and starbursts are observed.

If the interpretation of the obscured sources as AGN is correct, this
implies a significant population of luminous, obscured QSOs. It is
essential, therefore, to verify the AGN classification and to understand the
obscured sources in context of models for AGN.

Our goal in this work is to determine quantitative AGN models for the most
luminous of these obscured sources, with two primary objectives. The first
is to investigate the properties of the AGN emission and the nature of
the obscuration in AGNs at high luminosities. The second is to use the SEDs
to define the color criteria which describe such obscured AGNs in the IRAC
and MIPS color space in order to compare surface densities between obscured
and unobscured QSOs at high luminosities and high redshifts. To achieve
these goals, we have selected for detailed modeling the most luminous of the
obscured sources with AGN-dominated MIR SEDs from the archival \spitzer\
spectroscopy in the survey fields mentioned. 

Our selection criteria are the following: high redshift, i.e. z$>$1; large
MIR luminosities, i.e. 6$\mu$m luminosities,
$\nu$L$_{\nu}$(6$\mu$m) = L(6$\mu$m) $\geq$10$^{12}$\lsun; the presence of
$Si$ in absorption in the IRS spectra; and red MIR SEDs consistent with
being AGN-dominated~\citep[see e.g.][]{polletta06,alonso06}. There are 21
sources that satisfy all these criteria; 13 are from the
literature~\citep{houck05,weedman06a,yan07}, and 8 are from our own IRS
observing programs. As a consequence of these selection criteria, all
sources have 24$\mu$m fluxes greater than 1mJy and have relatively large
IR/optical flux ratios, i.e. F(24$\mu$m)/F(\rp)$\geq$400. To our knowledge,
this sample contains all sources with available IRS spectra in the
literature or the \spitzer\ public archive that satisfy the selection
criteria described. However, this sample is not meant to be complete or
unbiased since it is mainly based on the availability of an IRS spectra,
therefore it is not essential to be exhaustive.

The selected sample includes the most luminous obscured QSOs currently
known. The large MIR luminosities of the selected targets enable us to
measure the emission from the hottest dust in the torus with negligible
contribution from the host galaxy, providing a laboratory to test the
predictions of torus models in obscured AGNs. Up to now, these kinds of SEDs
could be sampled only using high spatial resolution data in nearby
AGNs~\citep[e.g. in NGC 1068;][]{hoenig06}. By combining samples from the
widest \spitzer\ surveys, we were able to find a significant number of these
rare objects. 

The list of selected sources, coordinates, optical and IR fluxes are
reported in Table~\ref{basic_data}. For simplicity, we use throughout the
paper simplified names for the selected sources. Official IAU names are
reported in Table~\ref{basic_data}. For the SWIRE sources, the first two
letters in their names designate the field where they were discovered (LH
for Lockman Hole, N1 for ELAIS-N1, and N2 for ELAIS-N2). The first letter of
the NDWFS source names is B for Bootes. The names of the FLS sources
correspond to those in~\citet{yan07}.

\section{Observations and Data Analysis}\label{obs_data}

Multi-band photometric data in the optical from the ground and in the IR
from \spitzer\ are available for all sources. The available data and
references are listed in Table~\ref{basic_data}. All sources are detected
with MIPS at 24\,\um\ with fluxes ranging from 1.0 to 10.6 mJy, and median
flux of 2.6 mJy. All sources have been observed with IRAC in 4 bands, from
3.6 to 8.0\,\um\ with varying limits. The FLS sources have few detections
with IRAC, probably because of their less sensitive observations compared to
the NDWFS and SWIRE surveys. Eight sources have been observed in 5 optical
bands, from U- to z-band, 3 in 4 bands, 5 in 3 bands, and 5 in 1 band.
Optical magnitudes for Bootes sources are in NDWFS Data Release 3 (DR3).
Optical and IR fluxes for the FLS sources were obtained
from~\citet{yan07,sajina07}, and those for the SWIRE sources were taken from
the latest internal catalogs that will be released in the Data Release
6~\citep[for details on the data reduction see][]{surace05}.

The IRAC fluxes of the sources in the NDWFS survey were measured from the
post-BCD images in the \spitzer\ archive. Aperture fluxes were measured within
a 2\farcs 9 radius aperture at their position and background subtracted.
Aperture corrections as derived from the SWIRE survey were applied. The
aperture corrections are 1.15, 1.15, 1.25, and 1.43, for the four IRAC
bands, respectively.

All sources have been observed in the optical \rp\ or R band and 16/21 have
been detected. The F(24$\mu$m)/F(R) flux ratio ranges from 575 to 25000
with a median value of 3500.

IRS data were in part taken from the archive and in part from our own
projects~\citep{houck05,weedman06a}. The spectroscopic observations were made
with the IRS Short Low module in order 1 (SL1) and with the Long Low module
in orders 1 and 2 (LL1 and LL2), described in \citet{houck04}\footnote{The
IRS was a collaborative venture between Cornell University and Ball
Aerospace Corporation funded by NASA through the Jet Propulsion Laboratory
and the Ames Research Center} providing low resolution spectral coverage
from $\sim$8\,$\mu$m to $\sim$35\,$\mu$m. The main parameters of the IRS
observations are listed in Table~\ref{irs_log}.

Since the objects have similar properties in terms of SED and fluxes, we
applied the same reduction procedure to all spectra for uniformity. The
reduction method is described in detail in~\citet{weedman06a} and briefly
summarized here. Six SWIRE sources (see details in Table~\ref{irs_log}) were
processed with version 11.0 of the SSC pipeline; the remaining sources were
processed with version 13.0. Extraction of source spectra was done with the
SMART analysis package~\citep{higdon04}. Since the selected sources are
faint and the spectra are dominated by background signal, we restricted the
number of pixels used to define the source spectrum to a width of
only 4 pixels (scaling with wavelength) in order to improve the spectral
signal-to-noise (S/N).  The flux correction necessary to change the fluxes
obtained with the narrow extraction to the fluxes that would be measured
with the standard extraction provided for the basic calibrated data was
measured by extracting an unresolved source of high S/N with both
techniques, and this correction (typically $\sim$ 10\% but varying with
wavelength) was applied to all sources.

A search in the literature provided additional data for 6 sources in
the sample, an ISO 15$\mu$m flux measurement for N2\_08~\citep{gonzalez05},
and MAMBO 1.2mm flux upper limits for all 5 E-FLS sources~\citep{lutz05}.

The SEDs and IRS spectra of all sources in the sample are shown in
Figure~\ref{sed_models} and ~\ref{irs_spectra}. The displayed spectra have
been boxcar-smoothed to nine times the approximate resolution of the
different IRS modules (0.6\,\um\ for SL1, 0.9\,\um\ for LL2, and 1.2\,\um\
for LL1). The MIPS 24\,$\mu$m fluxes typically agree within 10\,\% of the
24\,\um\ IRS flux.

 \begin{figure*}[ht!]
  \epsscale{2.0}
   \plotone{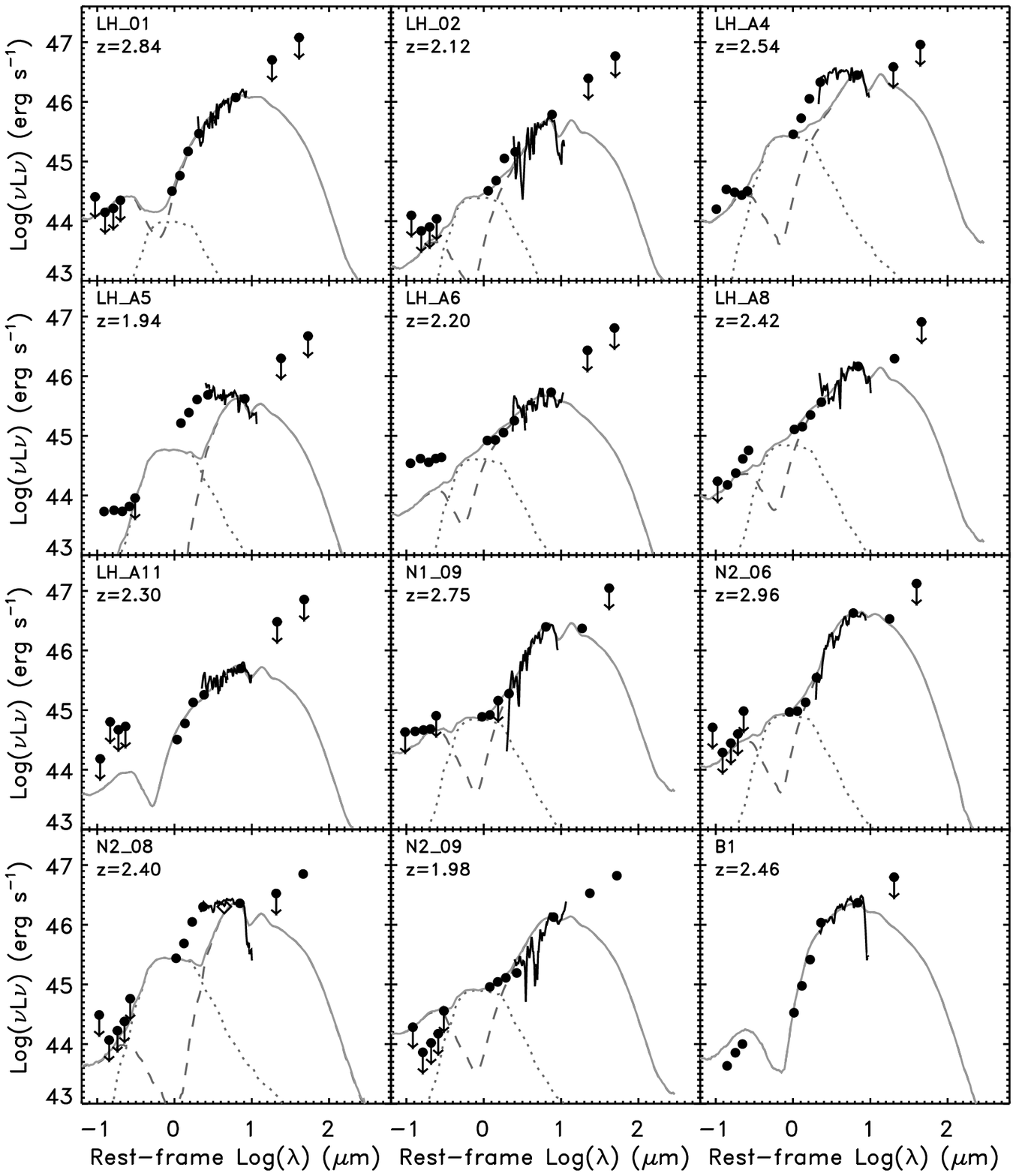}
      \caption{Rest-frame spectral energy distributions in $\nu L_{\nu}$
       $vs$ $\lambda$ (black symbols and curve) and best-fit model (solid
       grey curve) obtained from a torus model (dashed grey curve) and an
       elliptical template (dotted grey curve) to fit the residuals in the
       near-IR. Open diamonds represent ISO 15$\mu$m (in N2\_08) and MAMBO
       1.2\,mm (in all E-FLS sources) data~\citep{gonzalez05,lutz05}. 
       Downward arrows represent 5$\sigma$ upper limits for the optical and
       infrared data points, and 3$\sigma$ upper limits for millimeter data. 
       Source names and redshifts are annotated.}
         \label{sed_models}
   \end{figure*}
     \addtocounter{figure}{-1}
  \begin{figure*}
      \plotone{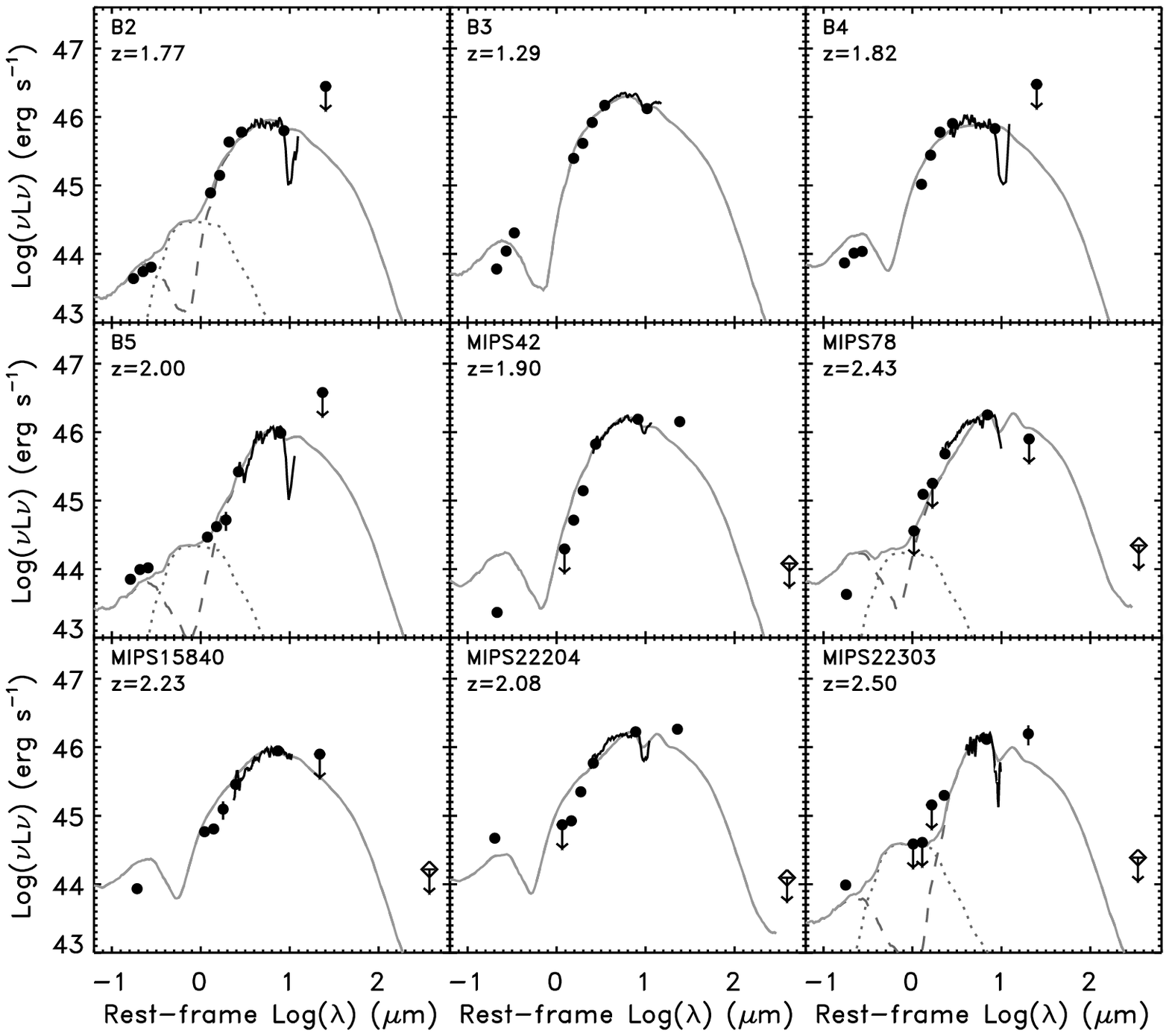}
      \caption{\it Continued}
   \end{figure*}

 \begin{figure*}[ht!]
  \epsscale{2.0}
   \plotone{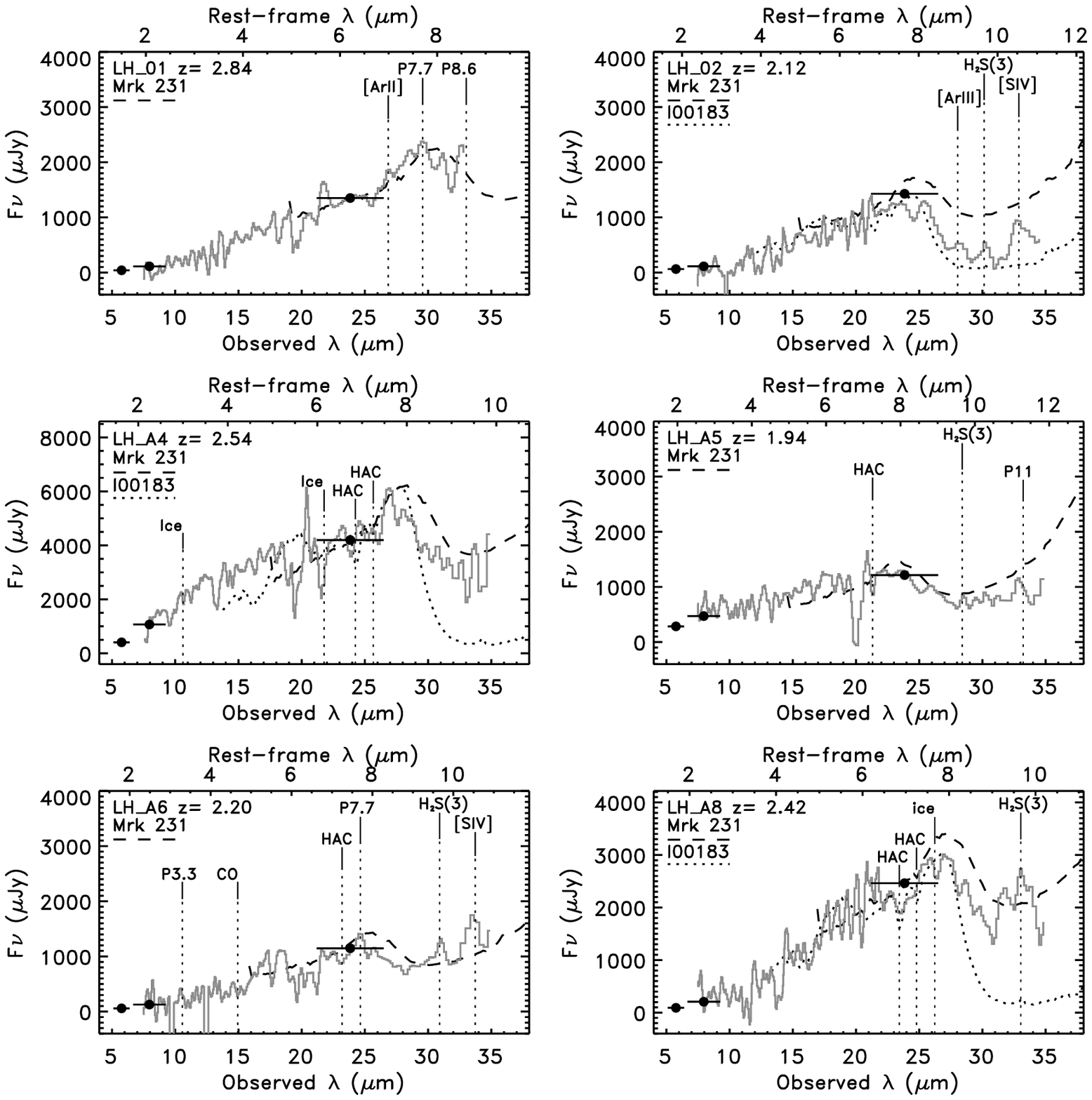}
      \caption{Observed IRS spectra (grey solid curve) and broad-band IR
       photometric data (black full circles) of all selected sources.  The
       expected location of some spectral features are annotated at the
       corresponding observed wavelength (P refers to PAH emission
       features). The IRS spectra of the Seyfert 1 Mrk\,231 (dotted curve)
       and of the heavily obscured AGN and ULIRG\,IRAS 000183$-$7111 (dashed
       curve) are overplotted for comparison. The name and redshift of each
       source are annotated and rest-frame wavelengths are reported on the
       upper horizontal axis.}
         \label{irs_spectra}
  \end{figure*}
      \addtocounter{figure}{-1}
  \begin{figure*}
      \plotone{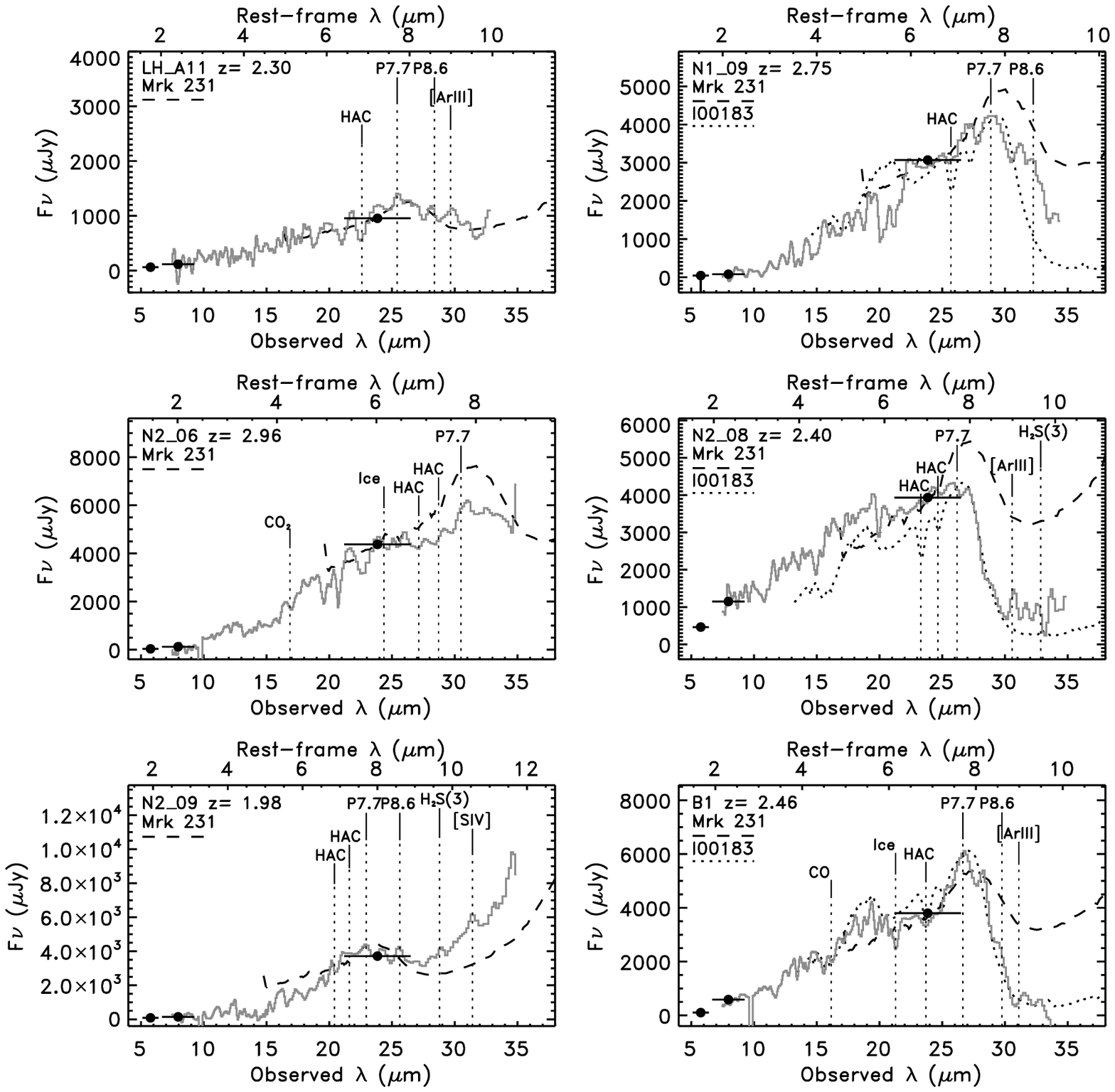}
      \caption{\it Continued}
   \end{figure*}
      \addtocounter{figure}{-1}
  \begin{figure*}
      \plotone{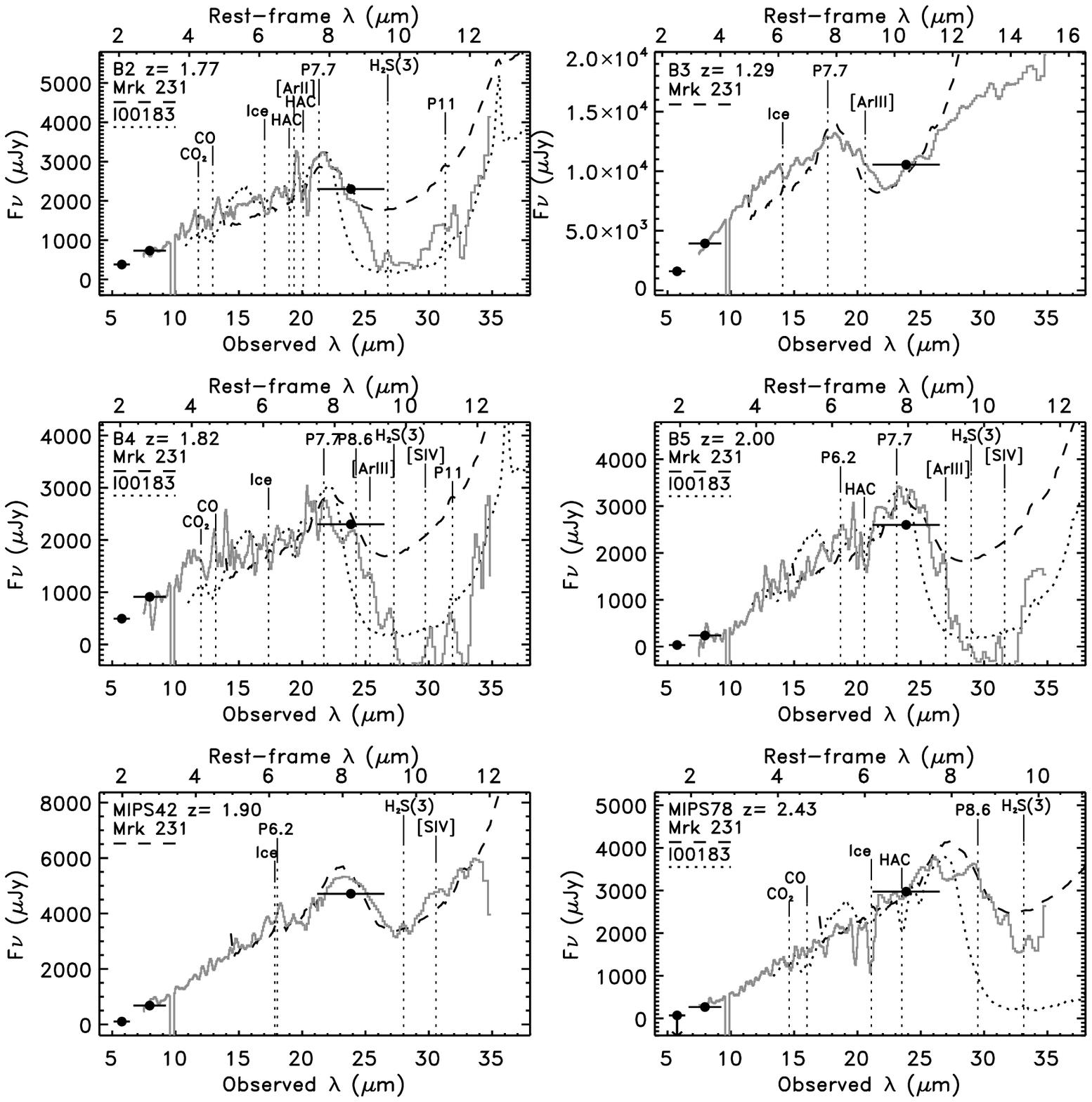}
      \caption{\it Continued}
   \end{figure*}
      \addtocounter{figure}{-1}
  \begin{figure}
      \epsscale{1.0}
      \plotone{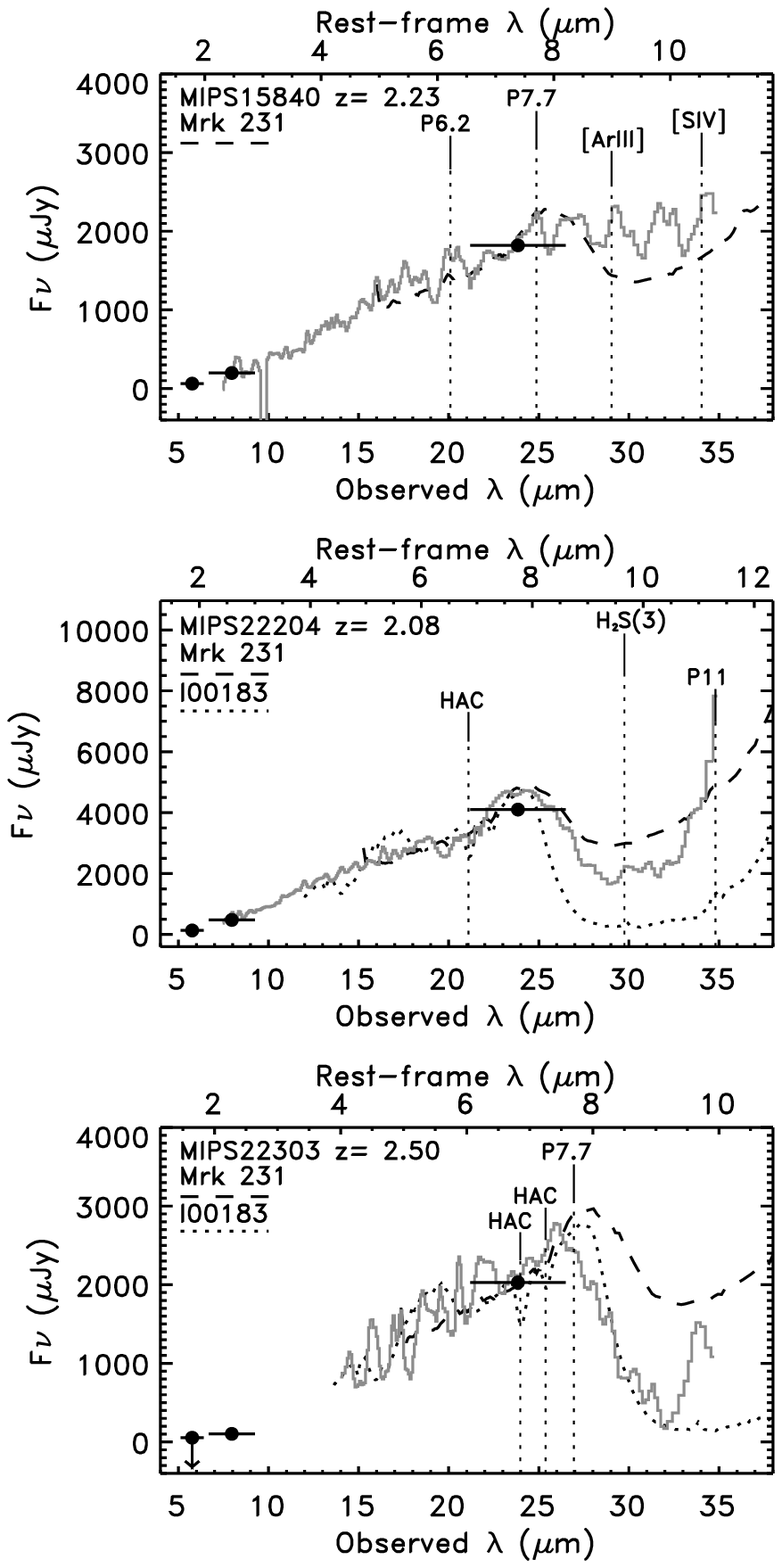}
      \caption{\it Continued}
   \end{figure}

\section{Spectral properties of the IRS spectra}\label{irs_spectral_prop}

The IRS spectra of all 21 sources are shown in Figure~\ref{irs_spectra}. The
spectra of an absorbed Seyfert 1~\citep[Mrk 231; ][]{weedman05}, and a heavily
obscured ULIRG~\citep[IRAS F00183$-$7111; I00183 hereinafter,
][]{tran01,spoon04} are also shown for comparison. Both reference spectra
exhibit the $Si$ absorption feature at 9.7\,\um\, but with different optical
depths, $\tau_{Si}$(Mrk 231) = 0.65~\citep{spoon02} and
$\tau_{Si}$(I00183)$\geq$5~\citep{spoon04}. The spectrum of I00183 also
shows absorption features at 4.26 and 4.67$\mu$m due to CO$_2$ and CO gas,
and at 6.85 and 7.25 \um\ due to hydrogenated amorphous carbons
(HAC)~\citep{spoon04}.

Spectroscopic redshifts are derived from the IRS spectra based on the
location of the $Si$ feature in comparison with the spectra of Mrk 231 and
I00183. Only in 2 cases (LH\_A4, and B3) are optical spectroscopic redshifts
available~\citep{polletta06,desai06}. For most of the sources, IR
spectroscopic redshifts were previously
reported~\citep{weedman06a,yan07,houck05}, although we have revised some of
them. The estimated redshifts are reported in Table~\ref{lum_tab}. Based on
the comparison with previous $z$ determinations and with the spectroscopic
redshifts, we estimate that uncertainties associated with these redshifts
are as much as $\pm$0.2 for sources with the poorest S/N.

The $Si$ feature in absorption is clearly observed in 18 objects (LH\_02,
LH\_A4, LH\_A5, LH\_A6, LH\_A8, LH\_A11, N1\_09, N2\_08, N2\_09, B1, B2, B3,
B4, B5, MIPS42, MIPS78, MIPS22204, MIPS22303), and only marginally
observed in 3 objects (LH\_01, N2\_06, MIPS15840). The apparent optical
depth of the $Si$ feature, $\tau_{Si}$, is measured as $ln(F_{9.7\mu
m}^{int}/F_{9.7\mu m}^{obs}$), where $F_{9.7\mu m}^{obs}$ is the observed
flux at 9.7\,\um, at the minimum of the $Si$ feature, and
$F_{9.7\mu m}^{int}$ is the intrinsic flux at 9.7\,\um\ that would arise from an
extrapolated, unabsorbed continuum. Since in most of the cases, the SEDs and
spectra do not sample the wavelength region beyond the $Si$ feature, we cannot
use the standard method to derive the intrinsic continuum~\citep[see e.g.
][]{spoon04}. The intrinsic flux at 9.7\,\um\ is thus estimated using two
different methods, 1) by extrapolating a 4--8\,\um\ rest-frame power-law fit
to 9.7\,\um, and 2) by normalizing an unobscured AGN template to the
observed 7.4$\mu$m flux, after redshifting it to the redshift of each
source, and reddening it using the Galactic Center extinction
curve~\citep{chiar06} until the observed $Si$ feature is well reproduced.
The $Si$ optical depth, $\tau_{Si}$, is measured only in sources at $z<2.7$,
18 sources. For the remaining we estimate a lower limit since we have only
an upper limit to the flux at the bottom of the $Si$ feature. The measured
values, reported in Table~\ref{lum_tab}, vary from $>$0.2 to 3.4.

The expected location of other spectral features common in starburst
galaxies and AGNs are also annotated in Figure~\ref{irs_spectra}: 1)
absorption features due to molecular gas, CO$_2 \lambda$4.26\um,
and~$\lambda$4.67\um, to water ice at $\lambda$6.15 and 7.67\um, and to HACs
at $\lambda$6.85,and 7.25\um; 2) emission features associated with
polycyclic aromatic hydrocarbons (PAHs) at $\lambda$=3.3, 6.22, 7.7, 8.6,
11.3\,\um, and with low and high ionization emission lines,
[ArII]$\lambda$6.99\um, H$_2S(3)\lambda$9.66\um, and [SIV]$\lambda$10.54\um.
Absorption features like CO and HACs would imply the presence of warm gas
along the LOS and features like CO$_2$, water ice and $Si$ would
imply the presence of shielded cold molecular clouds~\citep{spoon04,spoon07}. 
Although there are sometimes features in the individual spectra that are
consistent with these spectral features, higher S/N spectra would be
necessary to confirm their detection.  No spectrum shows any secure
feature except the $Si$ absorption.

\section{Modeling the Spectral energy distributions with torus models}\label{models}

The optical-IR SEDs and re-binned IRS spectra of all 21 sources are shown in
Figure~\ref{sed_models}. The IR SEDs, from 3.6 to 24$\mu$m, and IRS spectra
are modeled with a grid of torus models from~\citet{hoenig06} and the
residuals in the near-IR (NIR) are fit with a galaxy template to represent
the host galaxy. The model parameters are the cloud density distribution
(from compact to extended), the vertical radial distribution (with various degrees of
flaring or non-flaring), the number of clouds along the LOS
($\sim$optical depth), and the torus inclination (from face-on to edge-on).
The best model is the one that gives the best fit, based on a $\chi^2$-test,
to the five broad-band photometric data points from 3.6 to 24$\mu$m and to a
maximum of six additional data points derived by interpolating the IRS
spectrum at $\lambda$= 3, 5, 7, 8.5, 9.7, and 12\,$\mu$m in the rest-frame.

The best-fit models are shown in Figure~\ref{sed_models} and the best-fit
parameters are listed in Table~\ref{model_params}. In order to fit the
stellar component, we adopt a 3\,Gyr old elliptical template~\citep[from
GRASIL;][]{silva98} normalized at the observed flux of the residuals
(model-subtracted SED) at 3.6\,\um\ (in the Z-band for LH\_A5). The choice of
an elliptical template is justified by the evidence of ellipticals as hosts
in the vast majority of quasars~\citep[e.g.][]{dunlop03}. In order to
check the validity of our assumption on the host-galaxy, we estimate the
associated R-band ($\lambda^{rest}=0.7\mu m$) absolute magnitudes, M$_R$,
and compare them with those measured for the large sample of quasars
in~\citet{dunlop03}. The measured M$_R$ range from $-$25.9 to $-$22.2, and
the median value is $-$23.7. These values are consistent with those found
by~\citet{dunlop03}, M$_R$ = $-$23.53$\pm$0.09, but our sample shows a wider
dispersion that can be attributed to the large uncertainty of our method and
to the different redshift range. A host galaxy of later type cannot be
ruled out, but the low number of detections in the optical, the lack of NIR
(JHK) data, and contamination from the AGN light do not allow us to better
constrain the host type. In most of the cases, the elliptical template
provides an acceptable fit. Only in one source, LH\_A8, the SED of the
stellar component is not well fit with an elliptical template because of an
excess of emission at $\lambda^{rest}<$0.3\,\um\, and a late spiral template
could provide a better fit. Note that a heavily obscured starburst galaxy
would have an optical-NIR SED that is very similar to that of an elliptical
galaxy.

The reduced $\chi^2$ obtained by comparing the model with the data at
$\lambda>1\mu$m in the rest-frame are reported in Table~\ref{model_params}.
We will refer to these as T (for torus) models (see also
Table~\ref{model_params}). In 11 cases, a T model well reproduces the
observed IR SED and spectrum ($\chi^2_{\nu}<$1).  In the remaining 10 cases,
poorer fits are obtained (1$<\chi^2_{\nu}<$10). In 3 cases the model
underestimates the NIR emission (LH\_A4, LH\_A5, and N2\_08), in 4 cases the
model fails to reproduce the depth of the $Si$ feature (B1, B4, MPS78, and
MIPS22303), and in the remaining 3 cases the large $\chi^2_{\nu}$ values are
caused by either a noisy IRS spectrum (LH\_02, and N2\_09), or a poor fit at
short wavelengths (N2\_09, and MIPS22304). There are also 2 sources with
good $\chi^2_{\nu}$ but their fits do not well reproduce the $Si$ feature.
This is due to the fact that the feature is not well sampled by the
representative values for the IRS spectrum. There is indeed only one data
point in the $Si$ feature.

The main cause of failure in the fits with large $\chi^2_{\nu}$ is the
inability of the models to reproduce both a prominent NIR emission and a
deep $Si$ absorption feature. Yet, sources with deep $Si$ features are
expected to have weak NIR emission~\citep[e.g.][]{pier92,levenson06}. A NIR
excess requires that hot dust is seen directly by the observer (typically
happening only in less obscured objects), while the $Si$ absorption feature
requires the presence of a significant amount of optically and geometrically
thick dust~\citep[e.g.][]{pier92,granato97,levenson06,imanishi07}, and this
cooler dust component should absorb the hot dust emission from the vicinity
of the obscured AGN.

\begin{figure*}[ht!]
  \epsscale{2.0}
   \plotone{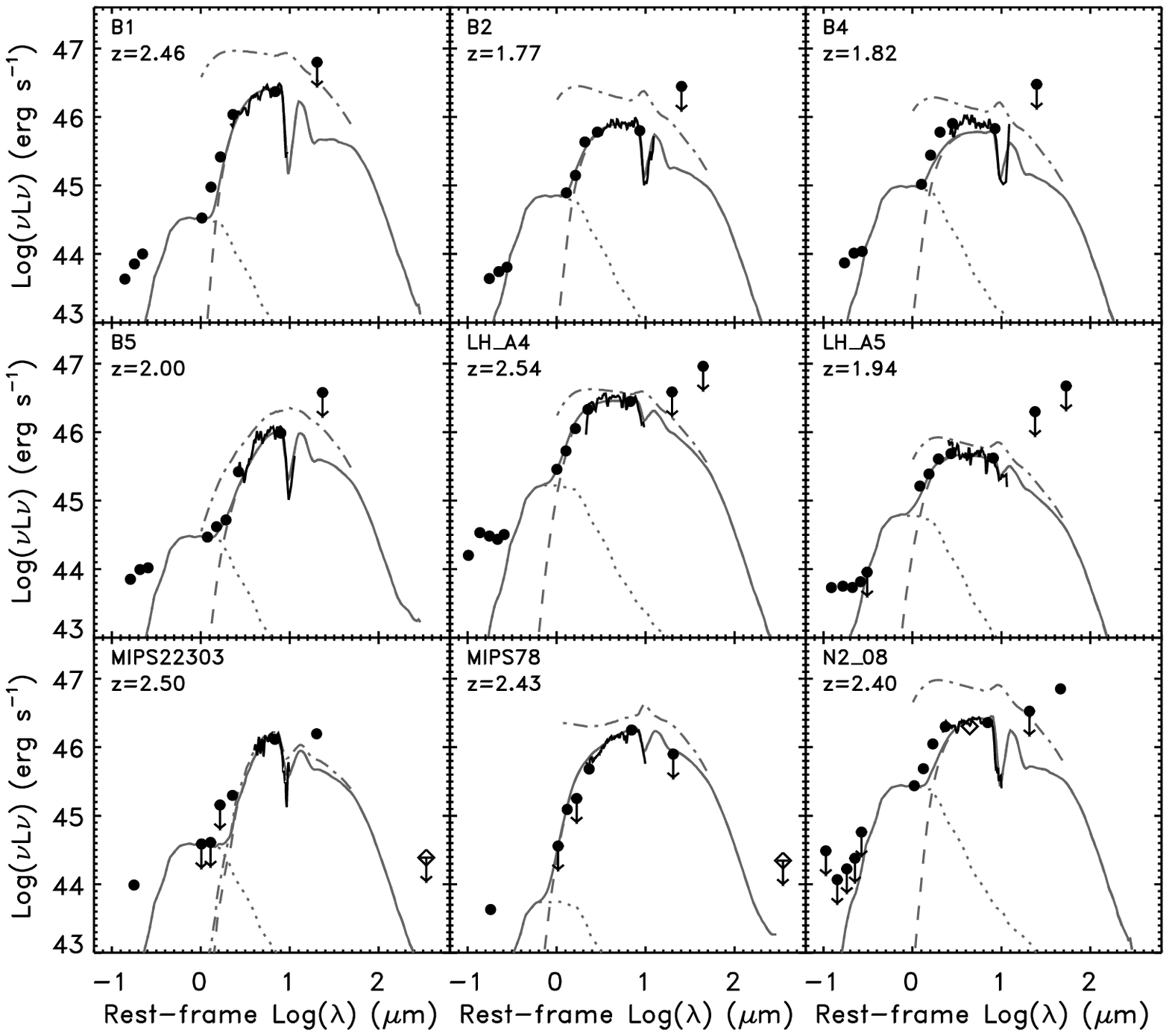}
      \caption{Rest-frame spectral energy distributions in $\nu L_{\nu}$
       $vs$ $\lambda$ (black symbols and curve) and best-fit model
       (solid grey curve) obtained from a torus+cold absorber model (dashed
       grey curve) and an elliptical template (dotted grey curve) to fit the
       residuals in the near-IR. The dot-dashed curve represents the torus
       model, in the 1--50$\mu$m rest-frame wavelength range, before
       including the cold absorber. Symbols as in Figure~\ref{sed_models}. 
       The name and redshift of each source are annotated.}
         \label{sed_ca_models}
   \end{figure*}

We attempted, therefore, to model the objects with poor fits, as well
as the rest of the sample with an alternative model that includes a cold
absorber detached from the torus (T+C models hereinafter). We assume
the Galactic Center extinction curve~\citep{chiar06} for the cold absorber. 
We do not model re-emission from this absorber for simplicity and because
this is expected to occur at FIR wavelengths. The number of degrees of
freedom in the $\chi^2_{\nu}$ estimates is 4 in case of T models and 5 in
case of T+C models. The cold absorber is able to produce deeper $Si$
features, and significantly improves the fits for 9 sources, as shown in
Figure~\ref{sed_ca_models}. In 2 cases, for B2 and B5, the $\chi^2_{\nu}$
have actually increased, but they are still $<$1. In these two sources the
$\chi^2_{\nu}$ is more sensitive to the broad-band photometric data than to
the IRS spectrum because of our limited sampling in the observed deep $Si$
feature. But the two fits are equally good and from a visual inspection of
the fits, we choose the model with the cold absorber component as preferred
model because it better fits the $Si$ feature.

The T+C models were adopted for these 2 sources and for 7 sources for which
the T+C model gives a significantly better fit (LH\_A4, LH\_A5, N2\_08, B1,
B2, B4, B5, MIPS78 and MIPS22303). In two cases (B4, and N2\_08), even the
T+C model does not provide a good fit to the NIR emission, although it is an
improvement with respect to the T model. No improvement was obtained for the
3 sources with poor $\chi^2_{\nu}$ (LH\_02, N2\_09, and MIPS22204),
therefore for these we keep the simple T model. In addition to the T+C
models, we also show in Figure~\ref{sed_ca_models} the torus model before
applying the extinction due to the cold absorber (see dot-dashed curve in
Figure~\ref{sed_ca_models}). The cold absorber can absorb up to 84\% of
torus emission in the near- and mid-IR (1--50\,$\mu$m), as shown in
Figure~\ref{sed_ca_models}.

In modeling the SEDs, we neglected the optical data. However, we notice that
in most of the cases the predicted optical flux from the torus model is
lower than the observed optical emission. The optical light might be
associated with the host galaxy or with scattered light from the AGN.  In 6
sources detected in the optical, the SED shows an upturn or a blue continuum
toward shorter wavelengths, i.e. in the rest-frame far-ultraviolet. This
feature can be reproduced by either a population of young stars or by
emission from the AGN accretion disk. The latter scenario is favored by
the optical spectrum of source LH\_A4 (aka SW104409). Its optical spectrum
is dominated by a blue faint continuum, narrow emission lines, e.g. \lya\
and \civ\, with asymmetric and weak broad components~\citep{polletta06}. The
most likely explanation for the properties of the observed optical spectrum
is that it is dominated by scattered light, and the scattering fraction is
estimated to be $<$1\%~\citep{polletta06}.

In one case (N2\_06), the LOS is at only 30\deg\ from the torus
axis, implying an almost clear view of the nuclear region. Since no
additional absorber is required by the data, we expect to see the nuclear
emission directly. This seems in apparent contradiction with the faint
optical emission of this object. Its best-fit model, however, predicts one optically
thick cloud, associated with the torus, along the LOS (see
Table~\ref{model_params}). The presence of such a cloud in front of the
optical source would be enough to absorb the intrinsic optical emission in
this source. The estimated apparent $Si$ optical depth in the model is
$\tau_{Si}$=0.07--0.34 which corresponds to a visual optical depth of
0.4--2.1 or to a suppression factor of the optical flux of 40\%--0.8\%.
However, $\tau_{Si}$ is poorly constrained by the data and in the model
because this source is at $z$=2.96. The red MIR SED of this object would
imply a much higher optical depth than derived by the $Si$ feature. Because
of its high $z$, the dust distribution in N2\_06 is thus poorly constrained.
A deeper $Si$ might be present implying a higher torus inclination or the
presence of a cold absorber. On the other hand, it is also possible that our
best-fit model is correct and that the source has only a weak $Si$ absorption
feature and a low inclination torus, and one optically thick cloud is
responsible for suppressing the optical and NIR nuclear light. This scenario
might be quite common in obscured QSOs as suggested by a recent study of the
IRS spectra of a sample of type 2 absorbed QSOs~\citep{sturm06}.

\subsection{Model parameters}\label{mod_params}

Here, we analyze the model parameters of the best-fit models. In case of the
9 sources for which the T+C model is preferred, we consider only the
parameters obtained with such a model (see Table~\ref{model_params}). 

The cloud density distribution, which can be considered as an indicator for
a compact or extended torus, is expressed by a power-law, $n_r(r)\propto
r^{-a}$, where $r$ is the torus radius. In the models grid, the index $a$
varies from 1 to 3 in steps of 0.5. Larger values of the index $a$ indicate
more compact distributions. Since values lower than 1 are not supported by
theoretical considerations for an accretion scenario~\citep{beckert04}, we
set a minimum value for the parameter $a$ of unity. In models with $a\geq$1,
the depth of the $Si$ feature increases for larger values of
$a$~\citep[see Fig. 10 in][]{hoenig06}. 

Trends of $Si$ strength with $a$ are different for our models based on an
accretion torus compared to the models in ~\citet{levenson07}.  They do not
model the dust distribution in a torus but instead use different dust
distributions (slab and shell) around a central heating source and they do
not consider the combination of a clumpy medium and an additional absorber
as proposed here. According to~\citet{levenson07}, the strength of the $Si$
absorption feature is a function of the temperature contrast in dust
material, and, thus, a deep absorption feature favors a more extended dust
distribution, i.e. $a$=0--1. However, for the majority of our sources,
a more compact dust distribution is favored; $a$=3 is chosen in 9 cases,
$a$=2 in 10 cases, and smaller values, $a\leq$1.5, are preferred in only 2
sources.

The clouds' vertical distribution (flaring or non-flaring) is approximated
by a power-law, $H(r)\propto r^b$, where $H$ is the torus scale height and
$r$ the torus radius. In the models grid, the index $b$ can assume values 1,
1.5, and 2. A non-flaring distribution ($b$=1) is preferred by 17 sources,
and moderately flaring ($b$=1.5) is preferred only by 4 sources.

The torus inclination is defined by the angle $\theta$ between the torus
axis and the LOS. A face-on torus has $\theta$=0\deg\ and an edge-on torus
has $\theta$=90\deg. In the models grid, the angle $\theta$ varies from 0 to
90\deg\ in steps of 15\deg. Our sources do not show a preferred torus
inclination, indeed $\theta$ ranges from 15 to 90\deg, with most of the
sources (13 sources) at intermediate values ($\theta$=30--45\deg).  All but
two (B5, and MIPS22303) of the sources modeled with the T+C model and one of
those with weak $Si$ absorption feature (N2\_06) favor a torus with little
inclination, $\theta$=0--30\deg. All of the others have inclined tori,
$\theta\geq$45\deg. Therefore, optical obscuration or
$Si$ in absorption do not necessarily imply a LOS intercepting the
torus~\citep[see also][]{rigby06,brand07}.

An additional model parameter is the number of clouds along the LOS,
$N_0^{LOS}$. This parameter is the average number of clouds obtained from
five different cloud arrangements for the same set of model parameters, and
is thus indicative of the extinction to the AGN emission produced by the
torus, $\tau_V^T\simeq N_0^{LOS}$~\citep{natta84}.
The apparent optical depth in the $Si$ feature of the T and T+C models,
$\tau_{\rm Si}^{T}$, and $\tau_{\rm Si}^{T+C}$, are measured following the
same procedure applied to the data (see \S~\ref{irs_spectral_prop}).
The apparent optical depths so derived are reported in
Table~\ref{model_params}. The line-of-sight optical depths associated with
the torus in the T models are $\tau_\mathrm{Si}^{T}\leq$1.04, with a median
value of $\tau_\mathrm{Si}^{T}=$0.13 or 0.46 depending on the method applied
to estimate $\tau_\mathrm{Si}^{T}$. The optical depths associated with the
cold absorber are $\tau_\mathrm{V}^{C}$=4--25, with a median value of 15. 
The apparent optical depths for the sources fit with the T+C model,
$\tau_{Si}^{T+C}$, range from 0.6 to 2.9, with a median value of
1.6, or from 0.4 to 1.4 with a median value of 0.9 depending on the method
used to measure it.

 \begin{figure}[ht!]
  \epsscale{1.0}
   \plotone{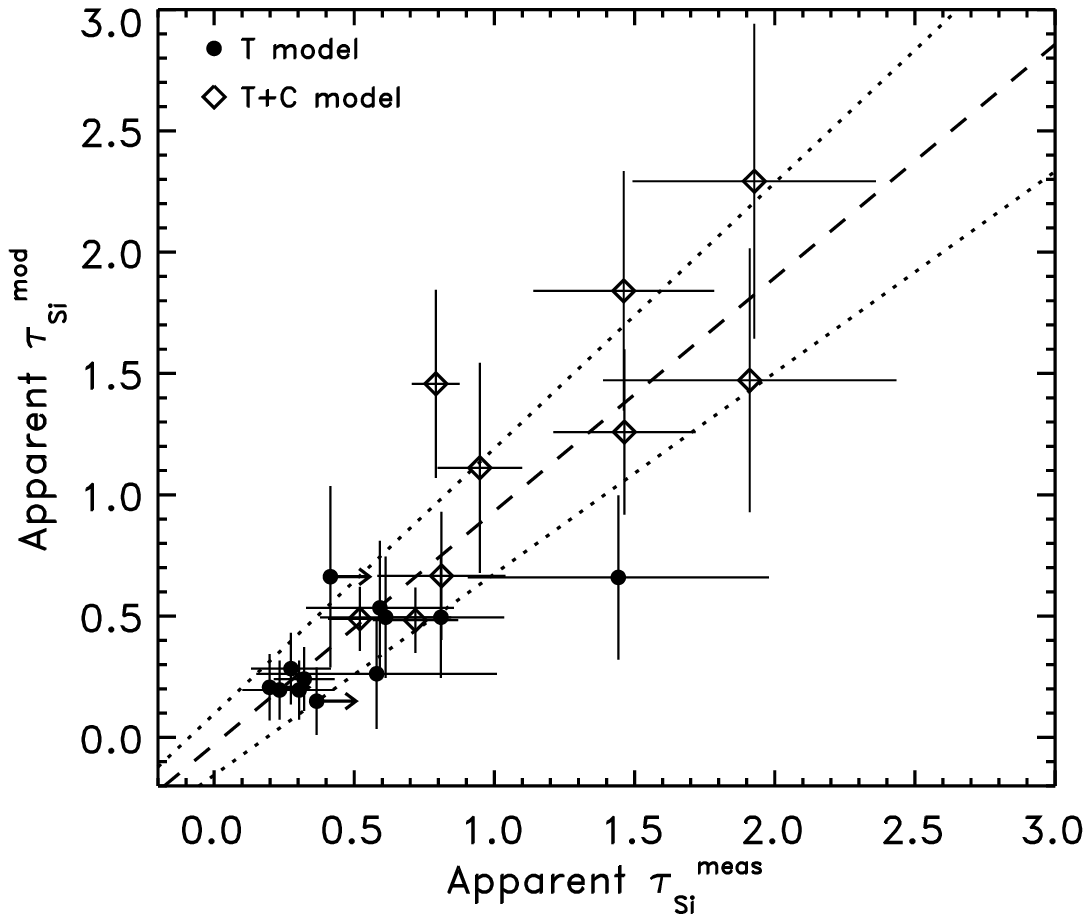}
      \caption{Comparison between the apparent $Si$ optical depth at
        9.7$\mu$m measured from the SED and spectrum, as described in
        \S~\ref{irs_spectra}, and from the torus models, as described
        in \S~\ref{models}. Full circles represent sources fit with a T
        model, diamonds represent sources fit with a T+C model.  The dashed
        line represents the best linear fit between the modeled and the
        measured $\tau_{Si}$, and the dotted lines include the 1$\sigma$
        uncertainties on the best-fit parameters.}
         \label{taus_comparison}
   \end{figure}

In Figure~\ref{taus_comparison}, we compare the modeled optical depths in
the $Si$ feature, $\tau_{Si}^{mod}$, with the observed optical depths,
$\tau_{Si}^{meas}$ (see \S~\ref{irs_spectral_prop}). The modeled
optical depths range from 0.01 to 2.9, considering both the T and T+C
models. The modeled and measured $\tau_{Si}$ satisfy the following
relation, $\tau_{Si}^{mod.}$\,=\,0.96$\pm$0.13$\times
\tau_{Si}^{meas.}-0.03\pm$0.13. Although there is a general agreement
between the two estimates, it is clear that these kind of measurements are
characterized by large uncertainties, especially in objects at high redshift
for which the $Si$ feature is not well sampled, and that it is a difficult
task to model both the NIR--MIR continuum and the $Si$ absorption feature.

In summary, we find that there is no preferred torus inclination associated
with the detection of a $Si$ feature in absorption. However, in the cases
where the absorption feature is well detected and can be modeled by the
torus model, the inclination is always higher than $\theta$=45\deg,
consistent with the LOS intercepting the torus. Sources with a
less inclined torus require an additional absorber to explain the observed
$Si$ feature or show a weak $Si$ feature. A compact non-flaring torus is preferred by the majority of the
sources. The clouds' radial density distribution indeed indicates that the
torus emission region is compact, or that the near- and mid-IR emission is
dominated by dust in the vicinity of the nucleus. The preferred non-flaring
torus in the majority of these luminous sources is consistent with the
predictions of the receding torus models~\citep[e.g.][]{simpson05,hoenig07}.
According to these models, the opening angle of the torus increases at
larger luminosities. Our results are in agreement with the predictions
from~\citet{hoenig07}. These authors claim that flaring should not occur in
high luminosity sources because large clouds at large distances from the AGN
should be driven away by the radiation pressure.

\subsection{Far-IR emission}

According to current AGN evolutionary models, obscured and extremely
luminous AGNs are believed to represent a specific and rare phase in the
evolution of an AGN, when the central super-massive black hole (SMBH) is
fully grown and still surrounded by a large amount of dust and
gas~\citep{sanders88,dimatteo05,hopkins05b}. High luminosity AGNs are
believed to be triggered by large scale galaxy mergers and, therefore, to be
accompanied by intense starburst activity. In order to test whether this
scenario applies to our sample, we search for starburst signatures in our
objects. Obscured starburst galaxies are generally heavily extincted at
optical wavelengths and are weak X-ray sources compared to AGN. Thus, the
only observations that could reveal a starburst are PAH features at MIR
wavelengths and strong continuum from cool dust at far-IR (FIR) and sub-mm
wavelengths. An analysis of the starburst contribution based on the PAH
features is discussed in section~\ref{disc_sb}, here we analyze the FIR
properties.

Most of the power produced by a starburst emerges in the FIR. The typical
FIR emission of powerful starbursts is characterized by luminosities
$\simeq$10$^{11.0-12.5}$\lsun, and peaks at 60--100
$\mu$m~\citep{sanders96}. Such luminosities correspond to star-formation
rates (SFRs) of $\sim$20 to 550 \msun\ yr$^{-1}$~\citep{kennicutt98}.
Higher rates, up to $\sim$5000\msun\,yr$^{-1}$, and luminosities, up to
3$\times$10$^{13}$\lsun\ are measured in some high-$z$ ULIRGs/sub-mm
galaxies (SMGs) and attributed to starbursts~\citep[e.g.][]{chapman04b}.

\begin{figure*}
  \epsscale{2.0}
   \plottwo{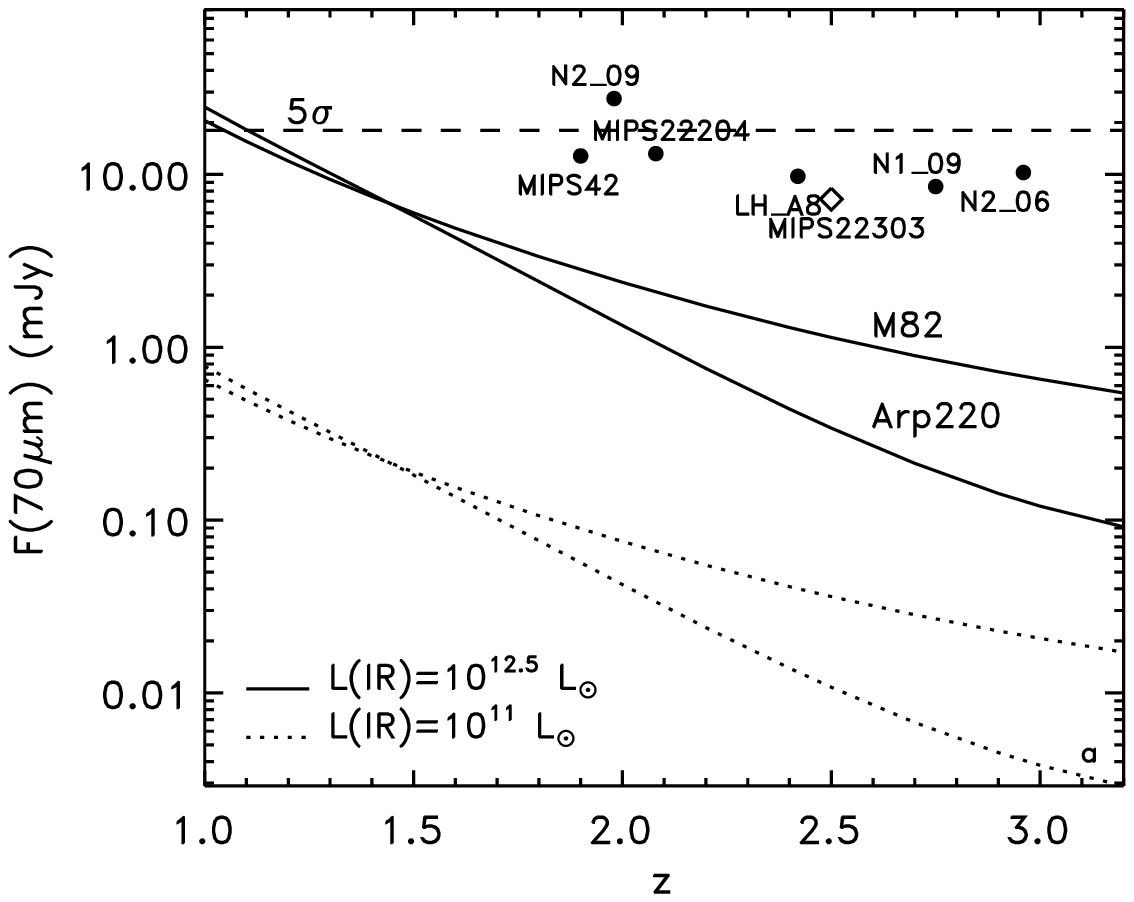}{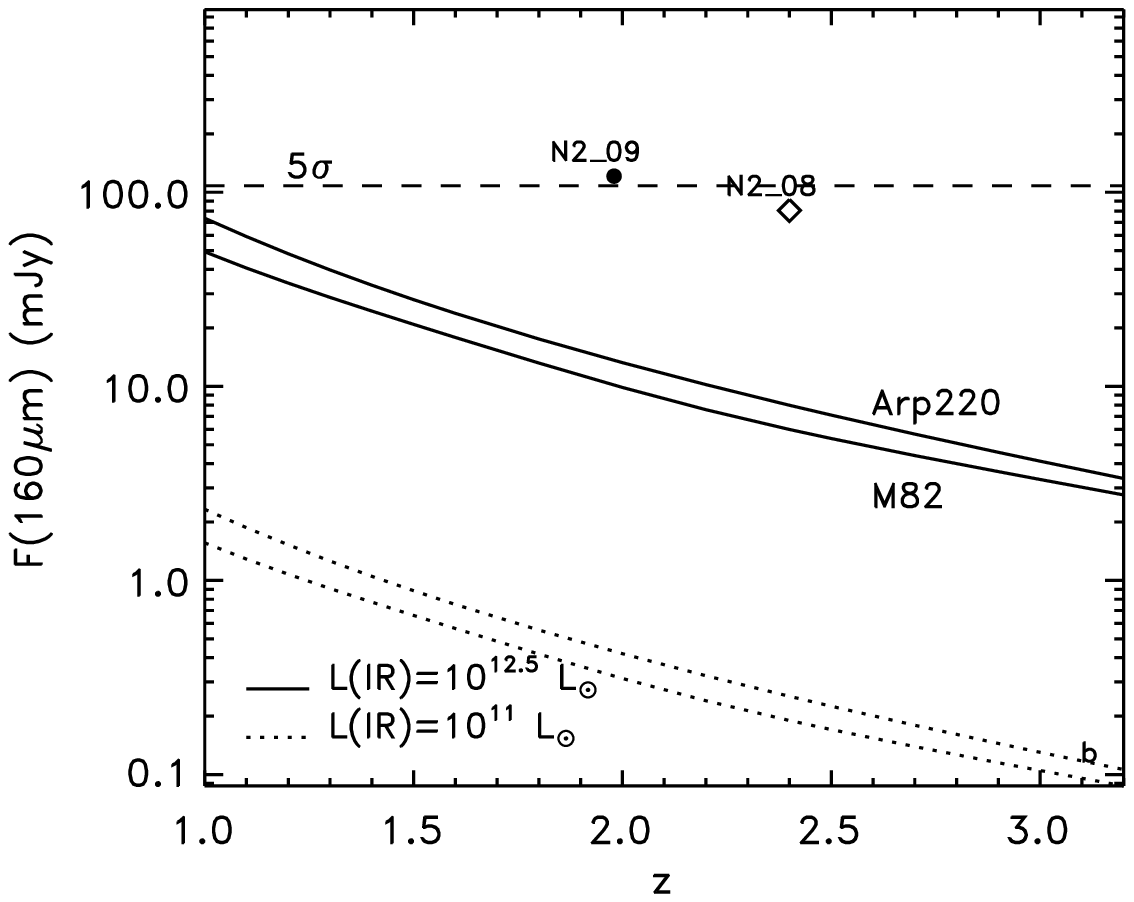}
      \caption{Predicted MIPS 70$\mu$m (panel a) and 160$\mu$m (panel b)
          fluxes for the starburst galaxies M 82 and Arp 220 with IR
          luminosities equal to 10$^{11}$\lsun\ (dotted curves) and
          10$^{12.5}$\lsun\ (solid curves). The horizontal dashed line shows
          the SWIRE 5$\sigma$ limit. Full black circles and diamonds
          represent sources in the selected sample detected at 70$\mu$m or
          160$\mu$m fit with a T and a T+C model, respectively. Source names
          are annotated.}
         \label{mips_z}
   \end{figure*}

In our sample, 5 SWIRE sources are detected at 70 or 160$\mu$m and
3 E-FLS sources are detected at 70$\mu$m. Note that both MIPS FIR
observations are available only for SWIRE sources (13 sources), and MIPS
70$\mu$m pointed observations are available for the E-FLS
sources~\citep{sajina07}. The SWIRE observations in the FIR are sensitive
enough only to reveal FIR luminosities $>$10$^{13}$\lsun\ at the observed
redshifts (the 5$\sigma$ limits are 18 mJy, and 108 mJy at 70, and
160$\mu$m, respectively). In Figure~\ref{mips_z}, we compare the SWIRE
5$\sigma$ detection limit and the detected fluxes at 70 and 160$\mu$m of our
sources with those expected for two starburst galaxies, M\,82 and Arp\,220,
with an IR luminosity of 10$^{11}$\lsun\ and 10$^{12.5}$\lsun\ at 1$<z<$3.2.
The figure clearly shows that we do not expect to detect the FIR emission of
starburst with such luminosities at $z>$1.2.

What is then the origin of the detected FIR emission in our objects ?  We
investigate three possible origins, the torus, AGN-heated dust at large
distances from the nucleus, and an exceptionally powerful starburst. The
spectrum of our torus models peaks at around 10$\mu$m, corresponding to a
temperature of about 300 K, and falls off at longer wavelengths. In two
cases, the 70$\mu$m detections or upper limits agree well with our models
(N1\_09, N2\_06), but in the other six cases, the observed 70 or
160$\mu$m fluxes are significantly higher than the model predictions
(LH\_A8, N2\_08, N2\_09, MIPS42, MIPS22204, and MIPS22303) and require an
additional component to be explained. The SEDs of these sources resemble
that of the starburst/AGN composite source CXO-J1417 discussed
in~\citet{lefloch07}. The FIR luminosity and SFR estimated for CXO-J1417 are
4.5$\times$10$^{12}$\lsun, and $\sim$750\msun\,yr$^{-1}$, respectively, as
in a powerful starburst.

\begin{figure*}[ht!]
  \epsscale{2.0}
   \plotone{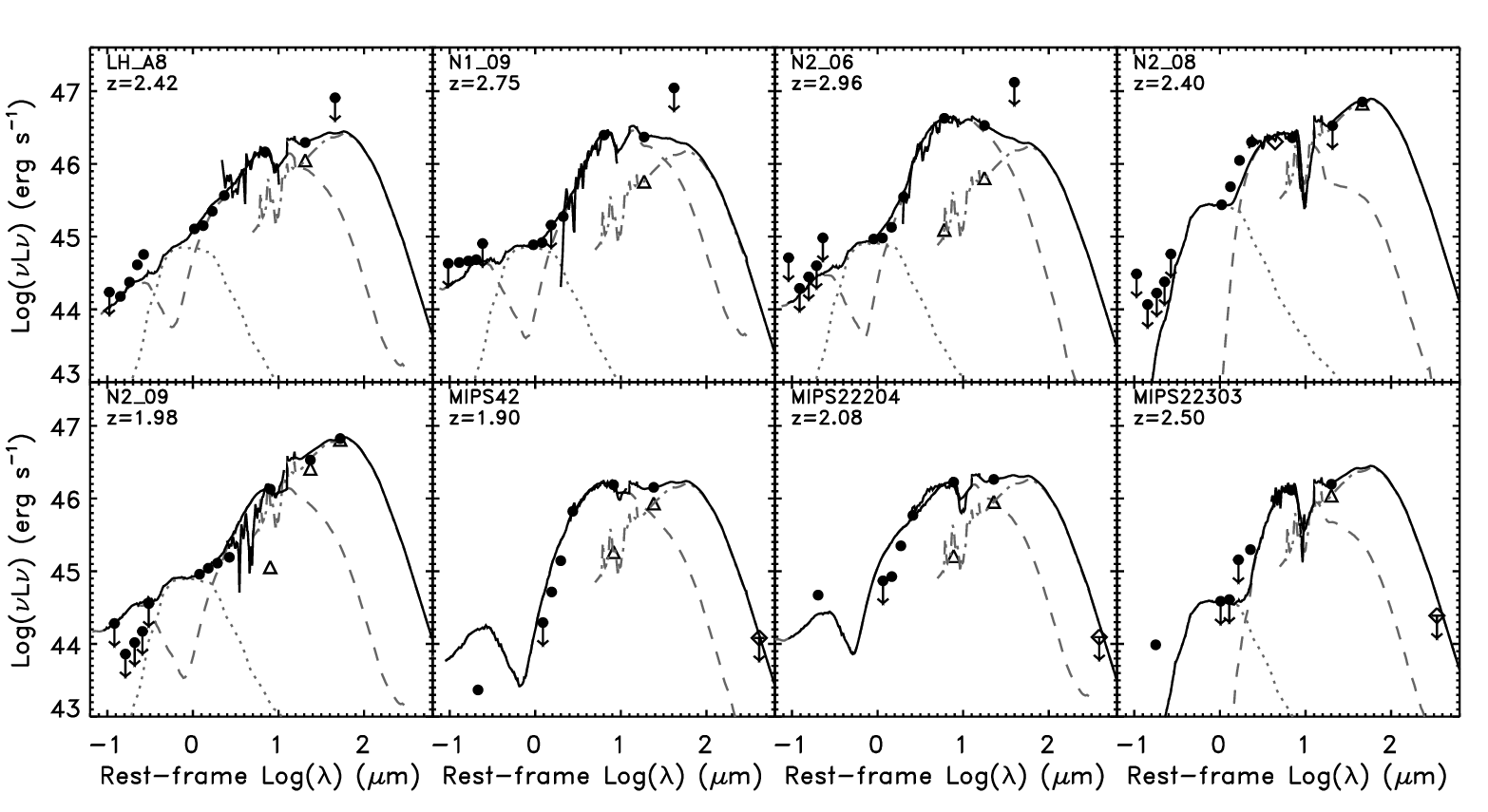}
      \caption{Rest-frame spectral energy distributions in $\nu L_{\nu}$
       $vs$ $\lambda$ (black symbols and curve) of sources detected at 70 or
       160$\mu$m. Also shown are the best-fit model (solid black curve)
       obtained from a torus (+cold absorber) (dashed grey curve), an
       elliptical template to fit the residuals in the near-IR (dotted grey
       curve), and a starburst template (dot-dashed grey curve) to fit the
       residuals in the far-IR (triangles). Symbols as in
       Figure~\ref{sed_models}.  The name and redshift of each source are
       annotated.}
         \label{sed_to_sb}
   \end{figure*}

In order to quantify the amount of luminosity detected in the FIR, we model
the observed fluxes at rest-frame $\lambda>$5$\mu$m with a starburst
template. We first derive the residual fluxes after subtracting the torus
contribution to the observed FIR fluxes and then fit them with starburst
templates. Two different starburst templates, M\,82 and
Arp\,220~\citep{silva98}, are used for the fits, but we only show the
results obtained with the M\,82 template because it provides better fits in
the majority of the cases and similar good fits in the remaining cases. The
starburst fits, combined with the torus and host galaxy fits, are shown in
Figure~\ref{sed_to_sb}. The estimated starburst luminosities and
contribution to the total IR luminosity are reported in Table~\ref{lum_tab}.
In half of the FIR-detected sources the torus contribution to the total IR
luminosity is lower than that of the additional FIR component. The remaining
objects divide equally between those for which the two contributions are
similar and those for which the torus contribution is larger. The estimated
starburst FIR luminosities are all greater than 3.3$\times$10$^{12}$\lsun.
These luminosities imply SFRs $>$ 600\,\msun\ yr$^{-1}$~\citep{kennicutt98},
which are consistent only with those measured in the most powerful
starbursts.

Another possible origin of the FIR component is re-emission of the energy
absorbed by the cold absorber discussed in \S\ref{models}. There are two
sources detected at long wavelengths and modeled with the T+C model, N2\_08
and MIPS22303. The estimated absorbed luminosities in the 1--50\,$\mu$m
wavelength range are 1.8$\times$10$^{47}$\,\ergs, and
5.25$\times$10$^{45}$\,\ergs, for N2\_08 and MIPS22303, respectively. These
correspond to 2.8 and 0.24 times the FIR luminosities derived from the
starburst template ($L^{SB}_{FIR}$). Although these ratios depend on the
cold absorber covering factor and on the efficiency in re-emitting the
absorbed energy, these values indicate that this hypothesis is energetically
viable.

Although we can reproduce the FIR component with a starburst template or
with re-emission from the cold absorber, its origin remains undetermined.
Large FIR luminosities, as measured in our sources, even when cold molecular
gas is detected, do not probe the presence of a powerful starburst, e.g. in
QSOs where \oii\ emission is not detected~\citep{ho05b}. Thus, both
explanations remain plausible and do not exclude others. In order to
understand the origin of the FIR component we would need more FIR
measurements, higher signal-to-noise IR spectra, and near-IR spectra.
Multiple FIR detections would help to constrain the radiation field, high
S/N IR spectra would allow us to search for and, eventually, measure PAH
emission (see \S\ref{disc_sb}), and near-IR spectra could be used to
constrain the SFR from the \oii\ emission line.

\subsection{AGN bolometric luminosity}\label{agn_lums}

Assuming the best-fit torus model and integrating it from 1000\AA\ to
1000$\mu$m, we can derive a lower limit to the AGN bolometric luminosity.
This estimate of AGN bolometric luminosity does not include the AGN optical
light that is not reprocessed by the dust and it is not emitted along our
LOS. Thus, the true AGN bolometric luminosity should be corrected by a
factor that depends on the torus covering factor and anisotropy of the torus
re-emission. The covering factor of a torus can vary from 20 to 50\%,
implying AGN bolometric luminosities that can be 2--5 times larger than our
estimates. The uncertainty associated with the redshift estimates ($\pm$0.2)
corresponds to a 30\% (20\%) uncertainty in the luminosity estimate for
objects at $z\simeq$2 ($z\simeq$3). We do not include the luminosity
associated with the host galaxy. Note that the discrepancy between the
observed optical data and the best-fit models does not affect the results,
since the bulk of the luminosity is emitted around 10$\mu$m in the
rest-frame, where the SED is well reproduced by the models. The estimated
bolometric luminosities, reported in Table~\ref{lum_tab}, range from
10$^{46}$ to 10$^{47}$\ergs.

\begin{figure}[ht!]
  \epsscale{1.0}
   \plotone{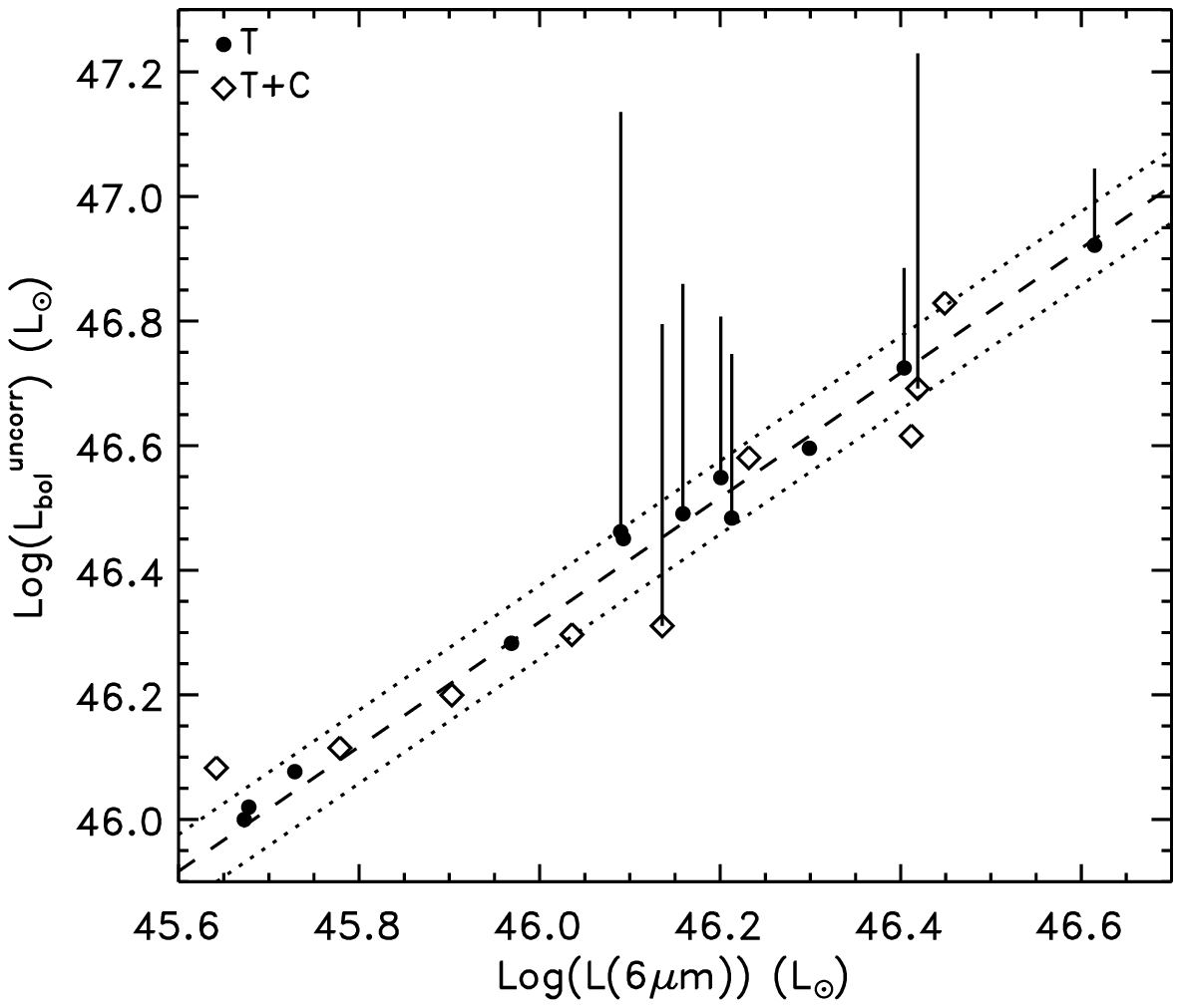}
      \caption{Comparison between the measured 6$\mu$m luminosity and the
        estimated AGN bolometric luminosity derived by integrating the best
        fit torus model between 0.1 and 1000$\mu$m.  The dashed line
        corresponds to a linear fit with slope 1, and the dotted lines
        correspond to the 1$\sigma$ dispersion. The bolometric luminosities 
        derived by including the FIR luminosity in sources with FIR
        detections correspond to the top of the solid vertical lines. The
        additional FIR luminosity is derived by fitting the residuals
        in the FIR with an M82 template (see Figure~\ref{sed_to_sb}).}
         \label{lmir_lbol}
   \end{figure}

The estimated AGN bolometric luminosities are compared with the 6$\mu$m
luminosities in Figure~\ref{lmir_lbol} in order to investigate whether it is
possible to determine AGN bolometric luminosities from the observed 6$\mu$m
luminosities without modeling the SED and without having measurements at
long wavelengths. We find that the two luminosities are highly correlated
and that the AGN bolometric luminosity can be derived from the 6$\mu$m
luminosity adopting the following relationship:
\begin{equation}
Log(L_{bol}^{AGN}) = Log(L(6\mu m))+0.32\pm 0.06.
\label{lbol_eq}
\end{equation}
This relationship gives a bolometric luminosity that is seven times
smaller than what would be derived assuming the bolometric corrections
obtained for a sample of optically-selected, type 1 QSOs by~\citet{elvis94a},
$Log(L_{bol}) = Log(L(6\mu m))+1.16$. 
Such a difference is mainly due to the fact that $L_{bol}$
in~\citet{elvis94a} includes both the AGN optical emission that is not
included in our analysis, and the emission from the host galaxy. Their
$L_{bol}$ also include a significant contribution in the FIR from cool dust
which is not included in our models and thus in our estimates of
$L_{bol}^{AGN}$. Moreover, their QSOs are about 100 times less MIR luminous
than our sample, and, thus, it is also probable that their median template
does not well represent the MIR--FIR SEDs of our objects.

For the 8 objects for which FIR data are available (see previous section),
we also estimate a bolometric luminosity including the FIR component (see
values in Figure~\ref{lmir_lbol}). In those cases the bolometric
luminosities are on average 2.5 times larger than estimated from the torus
model. This implies a significant amount of cool dust which dominates the
bolometric luminosity and which is located further from the nucleus than the
torus.  This dust could be illuminated by radiation from the AGN that is not
intercepted by the torus, which would be, qualitatively, consistent with the
T+C models described above. Overall, by taking into account the torus
covering factor, and a FIR component, the bolometric luminosities of our
objects can be up to 7--17 times larger than estimated using
equation~\ref{lbol_eq}.

\section{Composite IR spectra}\label{avg_spe}

\begin{figure}[htbp]
  \epsscale{1.0}
   \plotone{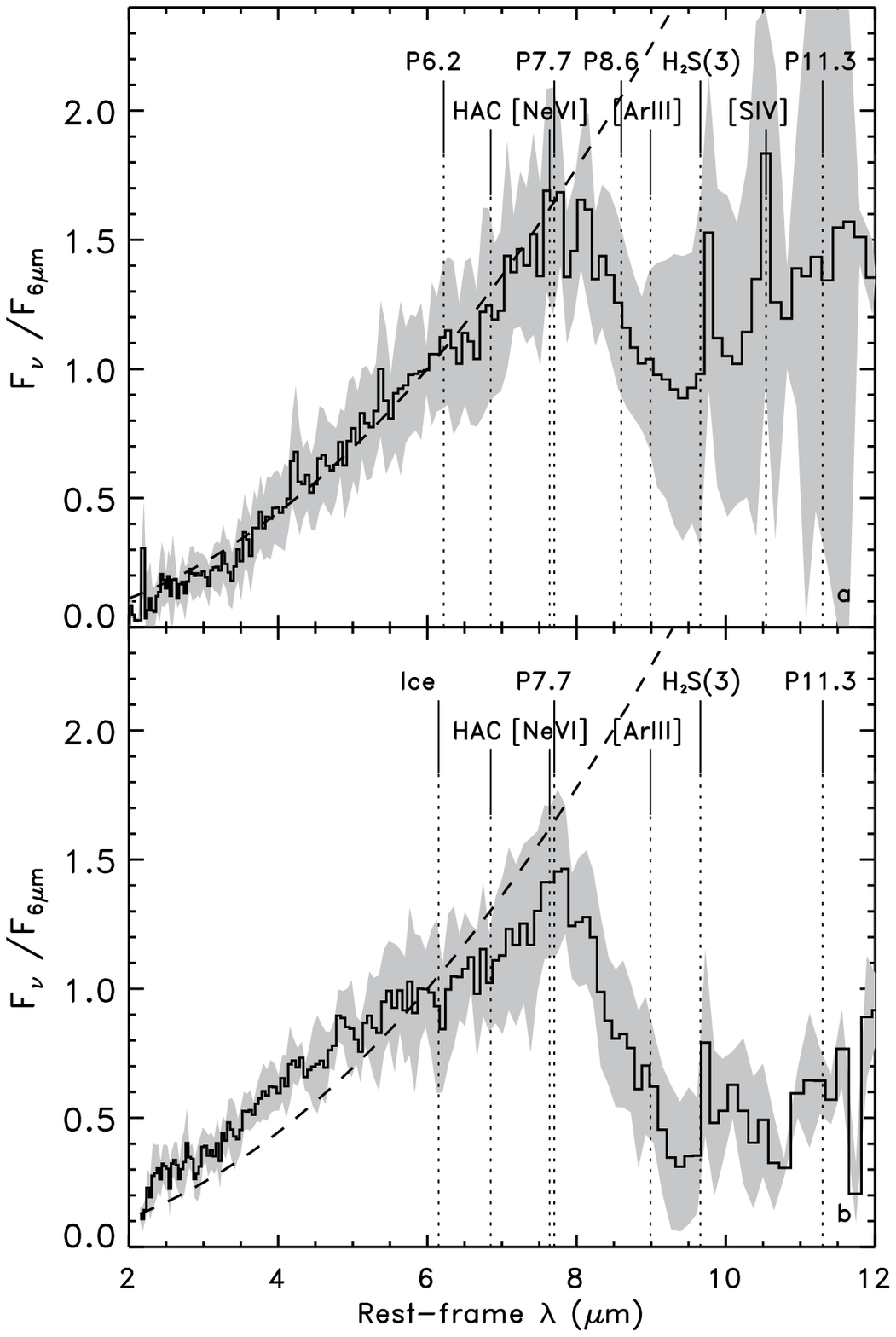}
      \caption{Median normalized IRS spectra of two sub-group of sources: (a)
         12 sources modeled with the T model; and (b) 9 sources modeled with
         T+C model. The shaded area represents the mean absolute deviation.
         The dashed curve represents a power-law of slope
         $\alpha$=2 (F$_{\nu}\propto \nu^{-\alpha}$) normalized at 6$\mu$m.
         Some spectral features that might be present are annotated.}
         \label{irs_avg_spe}
   \end{figure}

In order to search for weak features in the IRS spectra of Figure 2, the
individual spectra, after being boxcar-smoothed to three times the
approximate resolution of each module, were combined to obtain a composite
spectrum with higher S/N. Composite spectra were made for 2 sub-samples: 1)
sources fit with the T model (12 sources); and 2) sources fit with the T+C
model (9 sources). The composite spectra were obtained by taking the median
of all available data in the individual spectra in bins of
$\Delta(Log(\lambda))$=0.005 with
$\lambda$ in $\mu$m, after normalizing them at 6$\mu$m in the rest-frame.
The width of each bin is increased if less than 3 measurements are available
and it is reduced if more than 100 measurements are available. The
uncertainty associated with the median value per bin corresponds to the
uncertainty of the mean. The two composite spectra and associated 1$\sigma$
uncertainties are shown in Figure~\ref{irs_avg_spe}. The composite spectrum
of the T group is characterized by a large dispersion, especially at
$\lambda\simeq$10\,$\mu$m. This large dispersion indicates that we combined
sources with a broad range of depths in the $Si$ feature.

Some spectral features other than $Si$ that might be detected in the two
composite spectra are highlighted, e.g., absorption features due to ice at
6.15$\mu$m and hydrocarbons (HAC) at 6.85$\mu$m, the atomic emission
features [ArII]$\lambda$6.99\um, [NeVI]$\lambda$7.64\um, and
[SIV]$\lambda$10.54$\mu$m, the molecular hydrogen emission line
H$2$(S3)$\lambda$9.66$\mu$m, and the 6.2, 7.7, 8.6, and 11.3$\mu$m PAH
emission features. Of these, the strongest feature that may be detected in
both composite spectra is the 7.7$\mu$m PAH. This PAH is the strongest
feature in starburst galaxies~\citep{sturm00,brandl06}. 

The strongest feature observed in the composite spectra is associated with
molecular gas H$_2$, H$2$(S3)$\lambda$9.66$\mu$m. The detection of such a
feature cannot be confirmed because of the limited signal-to-noise, but
would likely indicate the presence of shocked molecular hydrogen outside the
MIR absorbing material, as can be expected in case of high-velocity galaxy
collisions, starburst super-winds, or AGN-driven outflows. Further
observations would be necessary to test these scenarios.

Absorption due to HAC might be present in both composite spectra.
This might be associated with the material that is also responsible for the
deep $Si$ feature, as in IRAS 00183$-$7111~\citep{spoon04}.

In order to facilitate a comparison between the two spectra, we overplotted
a power-law model with slope of 2 normalized at 6$\mu$m in both panels of
Figure~\ref{irs_avg_spe}. The main differences among the spectra are the
2--6$\mu$m spectral slopes and the depths and shapes of the
$Si$ feature.  In group (1) (T model), the FWHM of the $Si$ feature is
1.8$\mu$m, the apparent optical depth is 0.4, and the 2--6$\mu$m rest-frame
power-law slope is 3.0. In group (2) (T+C model), the FWHM of the $Si$
feature is 3.9$\mu$m, the apparent optical depth is 0.9, and the 2--6$\mu$m
rest-frame power-law slope is 1.7. It is interesting to note that, contrary
to the expectation from torus models that the steepest NIR spectra are
associated with the shallowest $Si$ absorption feature, the spectrum with
the deepest $Si$ feature (group 2) shows instead the flattest NIR slope.
This result supports the scenario proposed in \S~\ref{models} for the
sources fit with the T+C models in which the hot dust close to the AGN is
directly visible, and an additional absorber is required to reproduce the
deep $Si$ feature.

\section{Comparison with other AGN samples}\label{lit_comparison}

\subsection{Fraction of obscured QSOs}\label{obs_frac}

In order to estimate the total surface density of the QSOs in our
IRS sample, we first define empirically the colors that encompass our IRS
sample.  This is done by analyzing the location of our sources in various IR
color-color diagrams. Then, we select a reference flux-limited sample of IR
sources from a large and contiguous area which occupy the same location in
the selected color-color diagrams. To select our reference sample we used
the SWIRE catalog in the Lockman Hole field.

\begin{figure*}[htbp]
  \epsscale{2.2}
   \plottwo{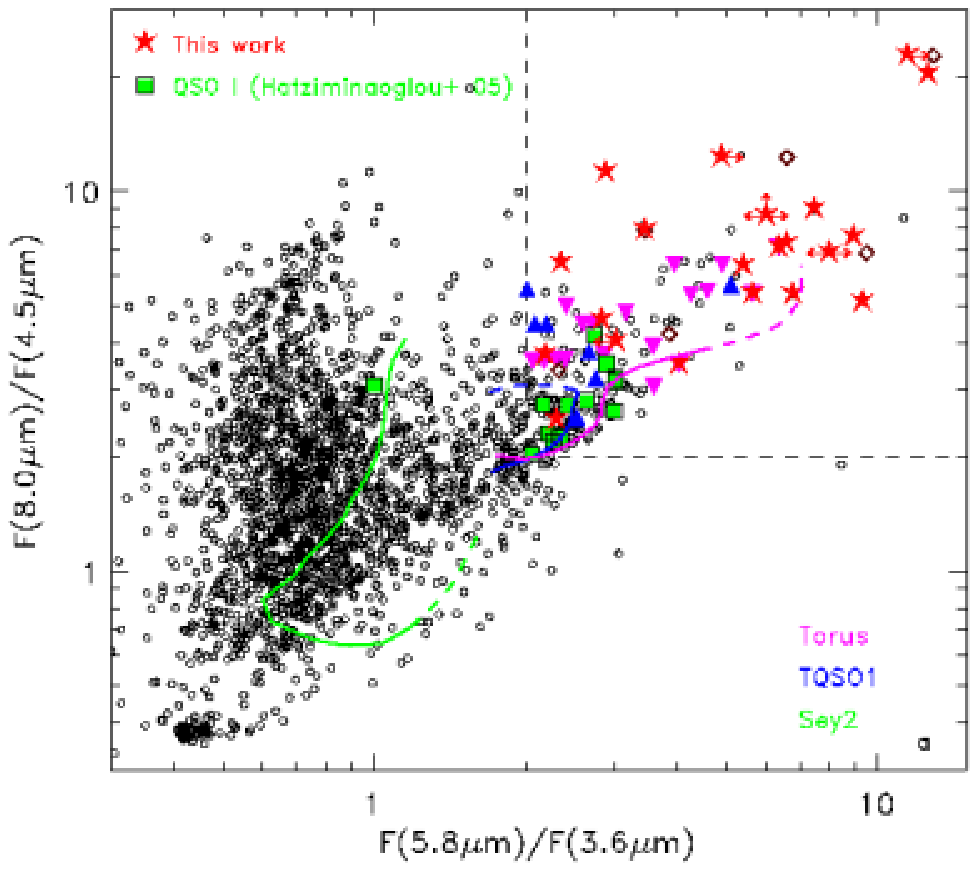}{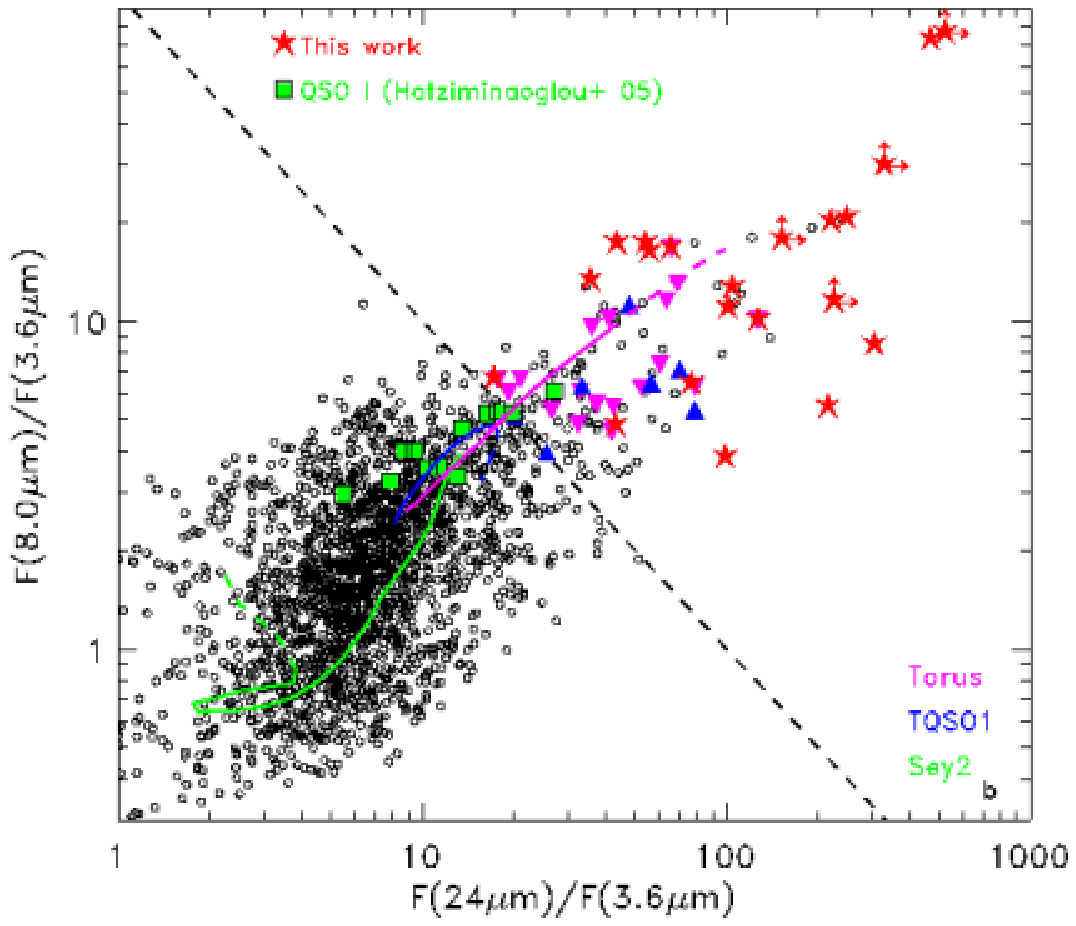}
      \caption{{\it Panel a:} IRAC color-color diagram. {\it Panel b:}
       IRAC-MIPS color-color diagram. The sources discussed in this paper
       are shown as red stars. Black circles represent all of the sources in
       a randomly selected 1\,deg$^2$ field with F(24$\mu$m)$>$209$\mu$Jy
       and F(8$\mu$m)$>$47$\mu$Jy. The green full squares represent the
       SDSS/SWIRE type 1 QSOs with $1<z<3$ from~\citet{hatziminaoglou05a}. 
       The curves show the color of templates of various types as annotated
       (solid: 0.1$<z<$2, dashed: 2$<z<$3). The regions to the right of the
       dashed lines represent the color criteria applied to select sources
       with colors consistent with those in our sample: unobscured QSO
       candidates (cyan triangles) and obscured QSO candidates (magenta
       downward triangles) (see \S~\ref{obs_frac}).}
\label{agn_diagrams}
   \end{figure*}

It has been established that the majority of sources with red IRAC colors
from \spitzer\ surveys have the colors expected from SEDS of AGN at various
redshifts \citep{lacy04,stern05,hatziminaoglou05a,barmby06}. AGNs have such
colors because of their hot dust emission. We thus analyzed several
color-color diagrams combining the 4 IRAC bands and the MIPS 24\,$\mu$m
band. From this analysis, we conclude that we can select sources with
similar colors as our IRS sample by using only two color-color diagrams. 
One diagram combines all 4 IRAC bands and it is shown in the left panel of
Figure~\ref{agn_diagrams}. The other diagrams is made by combining 3 bands,
IRAC[3.6], IRAC[8.0], and MIPS[24], and it is shown in the right panel of
Figure~\ref{agn_diagrams}. In both panels, we show our IRS sample and 1540
sources in a 1\,deg$^2$ randomly selected field from the SWIRE survey. The
1540 sources are all detected in 4 IRAC bands and at 24$\mu$m to the SWIRE
limits in the Lockman Hole field (4.2, 7.5, 46.1, 47.3, and 209$\mu$Jy at
3.6, 4.5, 5.8, 8.0 and 24$\mu$m, respectively).

The highly luminous and obscured QSOs in our IRS sample occupy the areas in the
color-color diagrams with the reddest colors. The only template that shows
similar colors corresponds to a highly obscured AGN~\citep{polletta06}.
Based on the areas in the diagrams occupied by our IRS sample, we define the
following multi-color criteria to locate similar sources:
F(5.8)/F(3.6)$>$2, F(8.0)/F(4.5)$>$2, and
Log(F(8.0)/F(3.6))+Log(F(24)/F(3.6)$>$2.0. These color criteria are
also shown in Figure~\ref{agn_diagrams} by dashed lines.

Having defined this color space for the highly luminous, obscured QSOs in
our spectroscopic sample, we select from a random 3.5 deg$^2$ field within
the SWIRE Lockman Hole field all sources detected in the 4 IRAC bands and
with 24$\mu$m fluxes greater than 1 mJy with colors consistent with the
color criteria mentioned above and shown in Figure~\ref{agn_diagrams}.

Using all of these criteria, we find a total of 82 sources in 3.5 deg$^{2}$
in this random sample of SWIRE sources. To evaluate in more detail if all of
these sources are similar to our obscured QSOs, we first visually
inspected all of the available images to remove any source with spurious
colors. We found 1 confused source, 1 saturated star and 2 nearby galaxies
with bad aperture photometric data. Thus the sample was reduced from 82 to
78 sources (22 deg$^{-2}$). Next, we examined the overall SEDs of the 78
sources, including the optical data.  Based on this analysis, we
estimate that all selected sources have SEDs consistent with being AGN
dominated and with those in our sample.  We have previously found that
sources classified as starbursts by IRS spectra can be distinguished from
AGN in the SWIRE samples based on their SEDs~\citep{weedman06a}, so we can
have confidence that the SED analysis does correctly locate AGNs. 

To further test that our criteria select only AGN-dominated sources at
$z$=1.3--3.0, we applied them to the large sample of 48 sources
in~\citet{yan07}, for which a MIR classification, \spitzer\ IRAC and MIPS
24$\mu$m data, and spectroscopic redshifts are available. Our criteria
select 7 sources from this sample, of which 4 are included in our sample.
The remaining 3 (MIPS8034, MIPS15958, and MIPS22277) are all AGN-dominated
in the MIR with EW$_{7.7\mu m}<$0.22, two fall into the redshift and
luminosity range of our sample, but one (MIPS8034) is at lower redshift
($z$=0.95) and lower luminosity (L(5.8$\mu$m) = 1.6$\times$10$^{11}$ \lsun).
Note that the other two sources are not formally included in our IRS sample
because MIPS15958 has a MIR luminosity of 10$^{11.95}$\lsun, that is just below
our selection threshold, and the IRS spectrum of MIPS22277 shows some PAH
emission. This test indicates that our sample is highly reliable in
selecting AGN, but it might include some (1/7 or 14\%) sources at lower
redshift and MIR luminosities.

We can perform an additional test on the degree of contamination from low
redshift and low luminosity sources using the available optical
spectroscopic data in the selected sample. Among the 78 selected sources, 12
sources have optical spectroscopic data. All 12 sources with spectroscopic
data have redshifts from 1.3 to 3.0, with a median value of 2.5. This is the
same range of redshifts found for our IRS sample, indicating that our
selection succeeds in selecting sources in a similar redshift range. If 1
out of 7 sources in our sample, or 14\%, were at lower $z$, as suggested by
the analysis of the \citet{yan07} sample, we would have expected 1-2 low-$z$
sources in our spectroscopic sample. Since no low-$z$, low-$L$ interlopers
were found in the spectroscopic sample, we can set a limit on the
contamination fraction of $<$8\%, which corresponds to 1 out of 12 sources,
the minimum number we could have found in our spectroscopic sample. A
redshift greater than 1.3 and the bright 24$\mu$m flux assure that also the
MIR luminosity is high and consistent with those measured in our sample.

\begin{figure}[htbp]
  \epsscale{1.0}
   \plotone{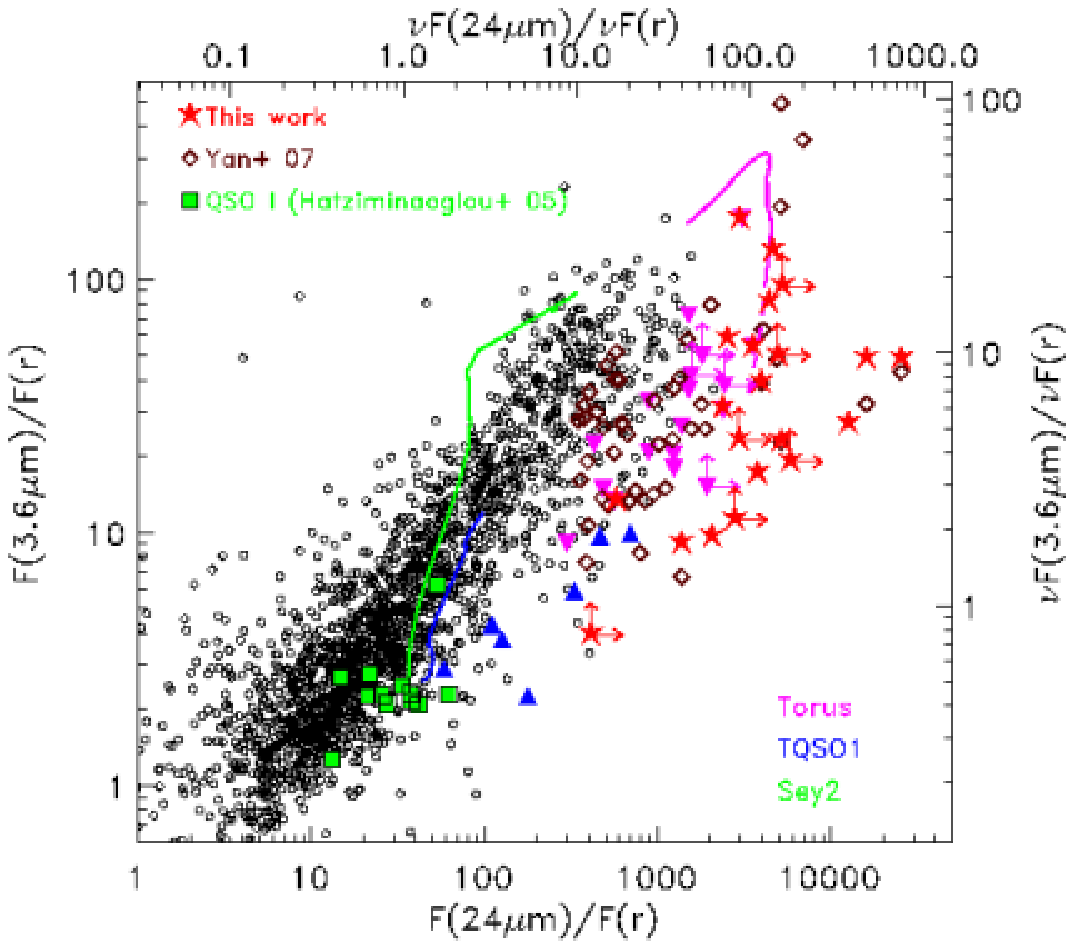}
      \caption{IR/optical flux ratios of the MIR-selected QSO sample (red
        stars) and of the SWIRE sources with similar IR colors (triangles)
        (selected as shown in Fig.~\ref{agn_diagrams}). The
        F(3.6$\mu$m)/F(\rp) flux ratio shows one of the criteria applied to
        separate obscured (magenta downward triangles) and unobscured (blue
        upward triangles) QSO candidates.  Symbols as in
        Figure~\ref{agn_diagrams}.}
         \label{iropt_colors}
   \end{figure}

Since our selection of the 78 sources is only based on IR colors, it is
possible that some of these sources are not as obscured as our IRS sample.
We use the F(3.6$\mu$m)/F(optical) flux ratio, where for the optical we used
the three bands \gp, \rp, and \ip, to separate obscured and unobscured
candidates. Based on previous studies~\citep{polletta06}, we consider
obscured candidates those with F(3.6$\mu$m)/F(\gp, \rp, and \ip) greater
than 15, 13, and 10, respectively, and unobscured candidates those with at
least one flux ratio lower than those thresholds. This criterion yields 61
obscured AGN and 17 unobscured AGN candidates. The optical-IR flux ratios of
the QSO candidates are compared with those of our IRS sample and with other
samples from the literature in Figure~\ref{iropt_colors}.

In order to verify our simple classification, we examined the optical
spectral properties of the sources with available optical spectra and
spectroscopic classification. A spectral classification is available for six
sources, and they are all AGN according to the observed emission lines; 2
show narrow emission lines and 4 show broad emission lines. Our
classification agrees with the spectroscopic classification, in the sense
that those with narrow lines are classified as obscured and those with broad
lines as unobscured, in 5 out of 6 cases. The only exception is a source
with \lya, \nv, and \civ\ emission lines with FWHM$\sim$1700 km s$^{-1}$,
intermediate between a narrow and a broad emission line AGN. Although this
comparison confirms our classification, we want to stress that our method
based on IR/optical flux ratios is based only on a handful of sources whose
availability of optical spectra is biased in favor of AGNs. Thus, any
selection based on the criteria mentioned above should be applied with such
a caveat in mind.

Based on this result alone, we would conclude that the surface density of
obscured QSOs to the limits of the SWIRE survey is 17 deg$^{-2}$ (61 in 3.5
deg$^2$). However, there are a few sources with low F(3.6$\mu$m)/F(optical)
flux ratios, but with extremely red IR SEDs where the 3.6$\mu$m flux is much
lower than the fluxes at longer wavelengths. Thus, it is likely that some
unobscured QSO candidates might actually be obscured or reddened type 1
AGNs. In such a case, we might be underestimating the surface density of
obscured QSOs.

Indeed our color selection criteria rule out most of the type 1 QSOs such as
those in~\citet{hatziminaoglou05a}. The IR colors for the SDSS/SWIRE type 1
QSOs in the 1.3--3.0 redshift range are also shown for comparison in
Figure~\ref{agn_diagrams}. While there is a large overlap between the
SDSS/SWIRE type 1 QSO sample and our obscured sample in IRAC colors (left
panel in Figure~\ref{agn_diagrams}), the SDSS/SWIRE type 1 QSOs are all
clustered towards the bluest side of the color distribution. Moreover, there
is almost a complete separation in the F(24)/F(3.6) color between the
two samples, with all the SDSS/SWIRE type 1 QSOs having
F(24)/F(3.6)$<$30, and all but one of the obscured QSOs having
F(24)/F(3.6)$>$30. This difference is likely due to the large NIR extinction
affecting our sample which is absent in the SDSS/SWIRE type 1 QSO sample and
highlights the ability of this flux ratio to distinguish obscured from
unobscured QSOs. Thus, our color selection criteria disfavor the classical
type 1 QSOs. If we consider all selected sources as obscured AGN candidates,
their surface density increases to 22 deg$^{-2}$ (78 in 3.5 deg$^2$).

In order to derive the fraction of obscured QSOs among the 24$\mu$m bright
AGN population at $z$=1.3--3.0 and with MIR luminosities $>$10$^{12}$\lsun,
we need to estimate the surface density of unobscured, type 1 QSOs with the
same redshift range and MIR luminosities as our IRS sample. We estimate such
a surface density from the 8$\mu$m luminosity function derived
by~\citet{brown06} for all AGN with F(24$\mu$m)$>$1mJy. Assuming the same
QSO template as in~\citet{brown06}, the derived 8$\mu$m absolute magnitude
corresponding to L(6$\mu$m)=10$^{12}$\lsun, is M$_{8\mu m}=-$28.39. The
estimated surface density of all type 1 AGNs with M$_{8\mu m}<-$28.39 in the
redshift range 1.3$<z<$3.0 is 11.7 deg$^{-2}$. A value only slightly
smaller, 9.4 deg$^{-2}$, is derived assuming the optical luminosity function
for type 1 AGNs measured by~\citet{richards06} and M$_B<-$22.3, which
corresponds to M$_{8\mu m}<-$28.39 assuming the same template. Using the
result derived from the MIR-selected QSO sample, the ratio of obscured to
unobscured MIR luminous QSOs in the 1.3--3.0 redshift range is, therefore,
17--22/11.7, or $\sim$1.5--1.9. This value is consistent with the fraction
of obscured AGNs measured in MIR-radio selected samples~\citep{martinez05}.

\subsection{Comparison with other IRS MIR-selected samples}\label{comparison_selection}

As described in \S~\ref{sample}, the main selection criteria applied to
carry out IRS observations of \spitzer\ sources were bright 24$\mu$m fluxes
and faint optical magnitudes~\citep{houck05}. In addition, other selection
criteria were applied in some cases to favor the selection of starburst or
AGN systems. In order to favor starburst-dominated galaxies, \citet{yan06}
also require large F(24$\mu$m)/F(8$\mu$m) flux ratios. \citet{weedman06a}
require an X-ray detection to select AGN and a bump in the IRAC SED to
select starbursts.
 \begin{figure}[htbp]
  \epsscale{1.0}
   \plotone{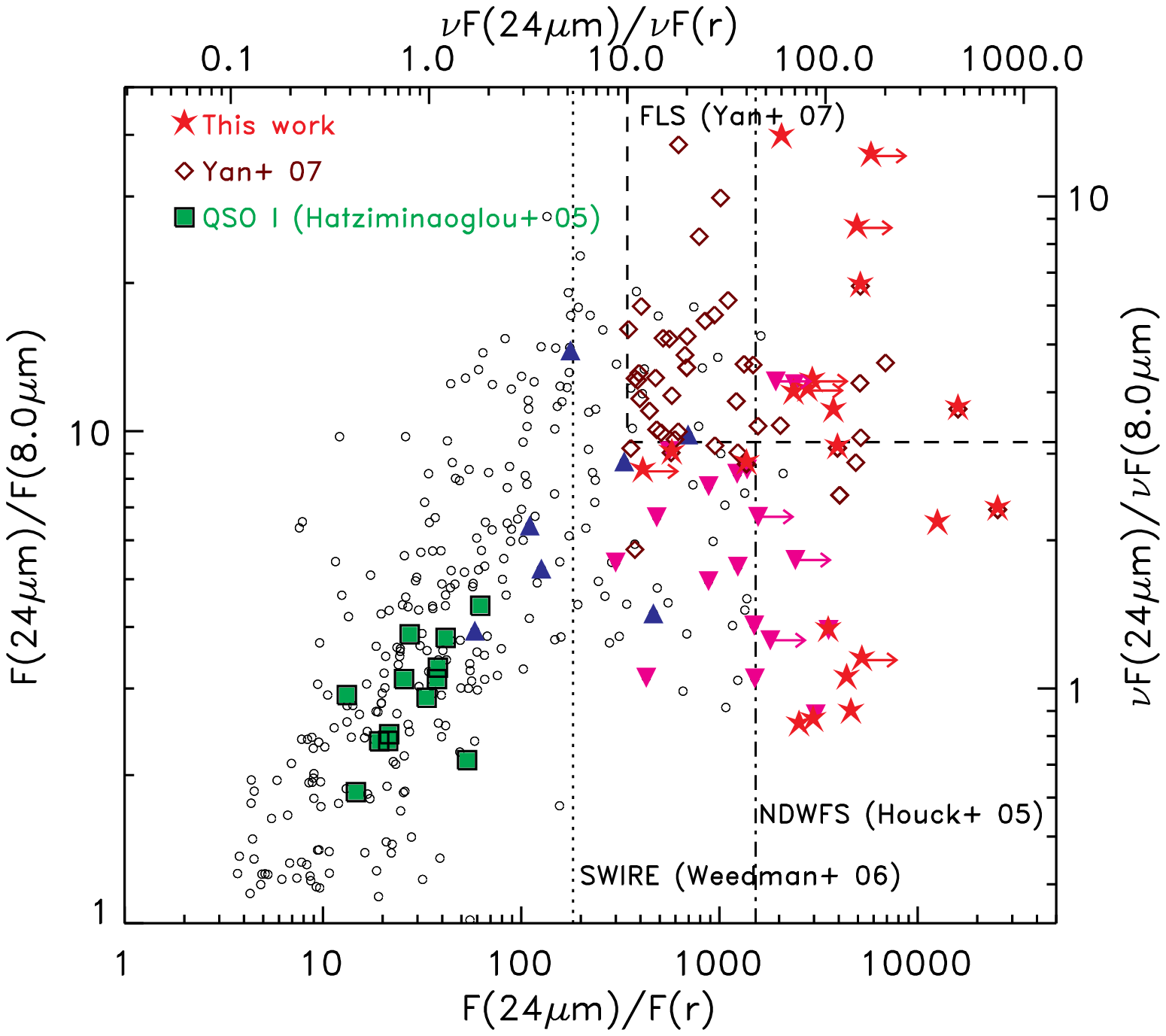}
      \caption{Color-color diagram showing the selection of the MIR-selected
        samples from which some of the sources in this work are drawn. The
        dotted line shows F(24$\mu$m)$>$0.9mJy and mag(\rp)$>$22
        from~\citet{weedman06a} (note that they also require an X-ray
        detection). The dot-dashed line shows F(24$\mu$m)$>$0.75mJy and
        mag(R)$>$24.5 from~\citet{houck05}. The region enclosed by the
        dashed lines shows Log($\nu F(24\mu m)/\nu F(8\mu m)$)$\geq$0.5
        and Log($\nu F(24 \mu m)/\nu F$(\rp))$\geq$1.0 from~\citet{yan07}. For
        comparison we show all the SWIRE sources with F(24$\mu$m)$>$1mJy in
        1 deg$^2$ (open black circles), the SDSS/SWIRE type 1 QSO
        from~\citet{hatziminaoglou05a} (green squares), the MIR-selected
        sources in~\citet{yan07} (brown diamonds), and the SWIRE sources
        detected at 8 and 24$\mu$m selected as shown in
        Figure~\ref{agn_diagrams} (downward magenta triangles and blue
        triangles). Symbols as in Figure~\ref{agn_diagrams}.}
         \label{miropt_selection}
   \end{figure}

In Figure~\ref{miropt_selection}, we compare the F(24$\mu$m)/F(8$\mu$m) and
F(24$\mu$m)/F(\rp) flux ratios of the sources in our IRS sample with the
selection criteria applied to select the sample from which a large fraction
of our sources was drawn~\citep{houck05,yan06,weedman06a}. We also show for
comparison the colors of the sources in~\citet{yan06}, of those selected by
the IR criteria described in the previous section, of the SDSS/SWIRE type 1
AGN from~\citet{hatziminaoglou05a}, and of a random sample of SWIRE sources
with F(24$\mu$m)$>$1mJy in 1 deg$^2$.

It is interesting to note the separation in F(24$\mu$m)/F(8$\mu$m) flux
ratios between the selected sample and the starburst and composite sample
from~\citet{yan06}. Lower F(24$\mu$m)/F(8$\mu$m) flux ratios favor
AGN-dominated sources, but high ratios do not exclusively select starburst
galaxies. Indeed half of our sources show large F(24$\mu$m)/F(8$\mu$m) and a
large fraction of sources in~\citet{yan06} contain an important AGN
component~\citep{sajina07}.

Our sample shows the largest F(24$\mu$m)/F(\rp) ratios of all samples. The
sources selected in the previous section to derive the surface density of
our IRS sample show F(24$\mu$m)/F(\rp) flux ratios in between the SDSS/SWIRE type 1
QSOs and the IRS sources. This indicates that the sources presented here
might actually be more extreme than the sample selected above. 

\subsection{Luminosity, $\tau_{Si}$ and redshift}

\begin{figure*}[ht!]
  \epsscale{2.2}
   \plottwo{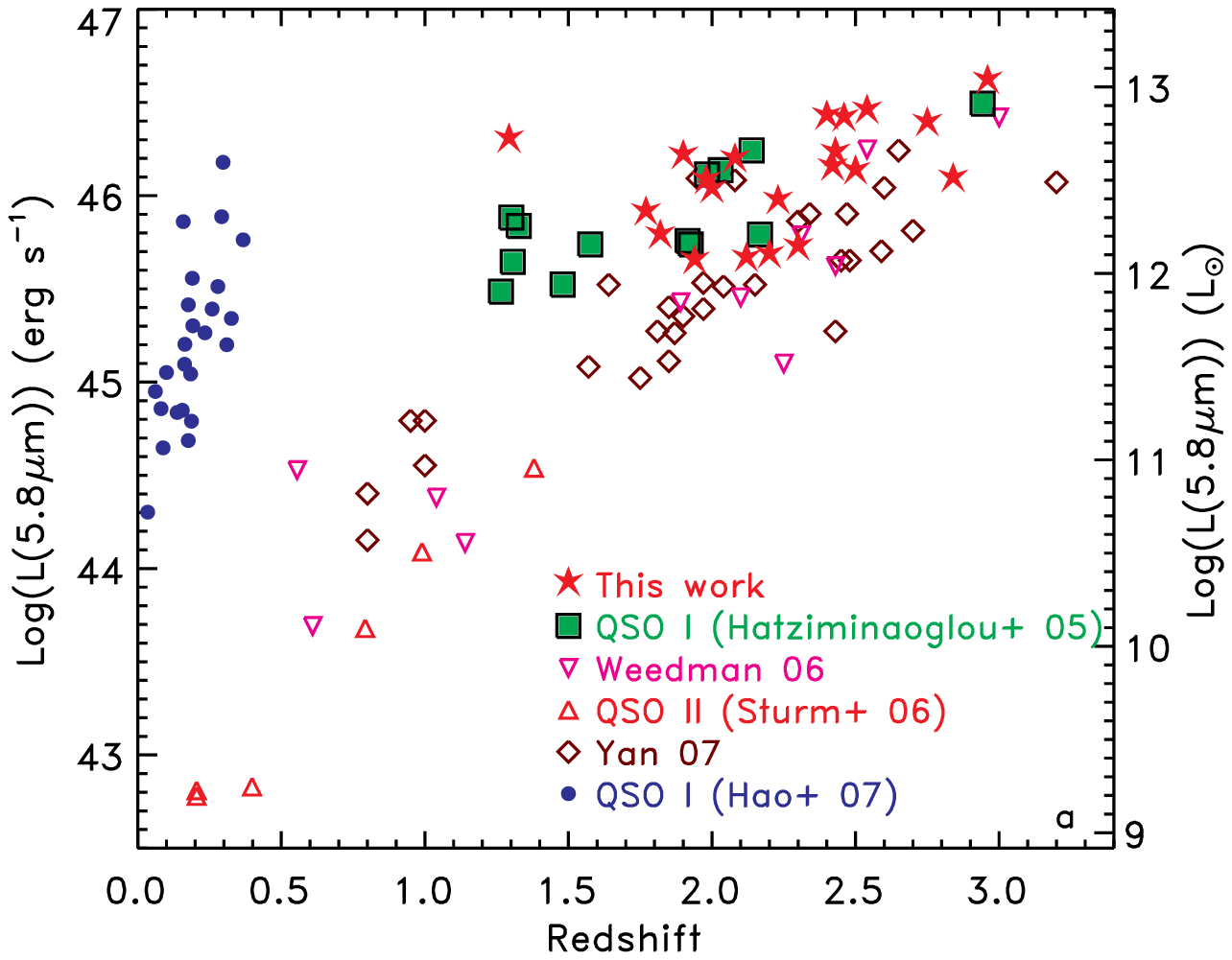}{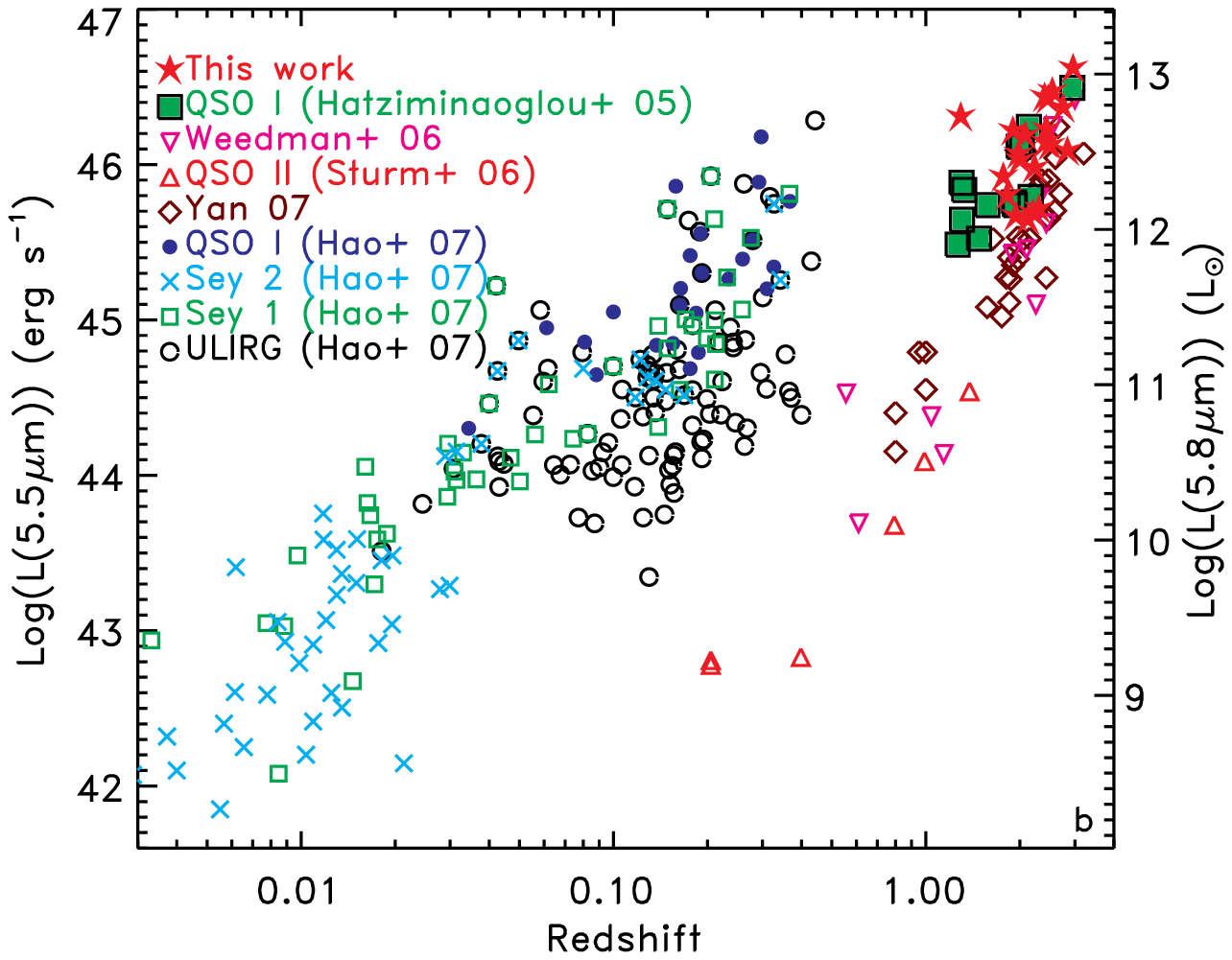}
      \caption{{\it Panel a:} Mid-IR (6$\mu$m rest-frame) luminosities
        $vs$ redshift for the obscured AGNs in this work (red full
        stars), and various AGN samples, X-ray selected type 2 QSOs
        ~\citep[red triangles;][]{sturm06}, X-ray selected optically
        faint AGNs~\citep[reversed magenta triangles;][]{weedman06a},
        IR-selected AGNs and composite sources~\citep[brown
        diamonds;][]{yan07}, SDSS/SWIRE selected type 1 QSOs~\citep[green
        squares;][]{hatziminaoglou05a}, and type 1 QSOs~\citep[full blue
        circles;][]{hao07}. {\it Panel b:} The same data shown in the
        left panel are shown here with, in addition, the values for the Seyfert
        1 (green squares), Seyfert 2 (cyan crosses) and ULIRGs (open black
        circles) from~\citet{hao07}.}
         \label{lmir_z}
   \end{figure*}

The MIR (6$\mu$m) luminosities and redshifts of the AGN sample presented
here are compared with those of other AGN samples from the literature in
Figure~\ref{lmir_z}. The literature samples include X-ray selected type 2
QSOs~\citep{sturm06}, optically and IR selected (SDSS/SWIRE) type 1 QSOs
with 1 $<$ $z$ $<$ 3~\citep{hatziminaoglou05a}, X-ray selected, optically
faint ($r>$22) and IR bright (F(24\,$\mu$m)$>$1\,mJy)
sources~\citep{weedman06a}, MIR selected AGNs and composite
sources~\citep{yan06}, and various types of local AGNs, type 1 QSOs, Seyfert
1, Seyfert 2, and ULIRGs~\citep{hao07}. Note that some of the sources in our
sample are in common with those in~\citet{weedman06a} (5 sources), and with
those in~\citet{yan07} (5 sources).

Our sources are among the most luminous MIR sources in this compilation. The
only sources with similar MIR luminosities are the SDSS/SWIRE type 1 QSOs
from~\citet{hatziminaoglou05a} and the most luminous type 1 QSOs
from~\citet{hao07}. The composite sources in~\citet{yan07} and the type 2
QSOs from~\citet{sturm06} are on average 3 and 250 times fainter,
respectively.

\begin{figure}[htbp]
  \epsscale{1.0}
   \plotone{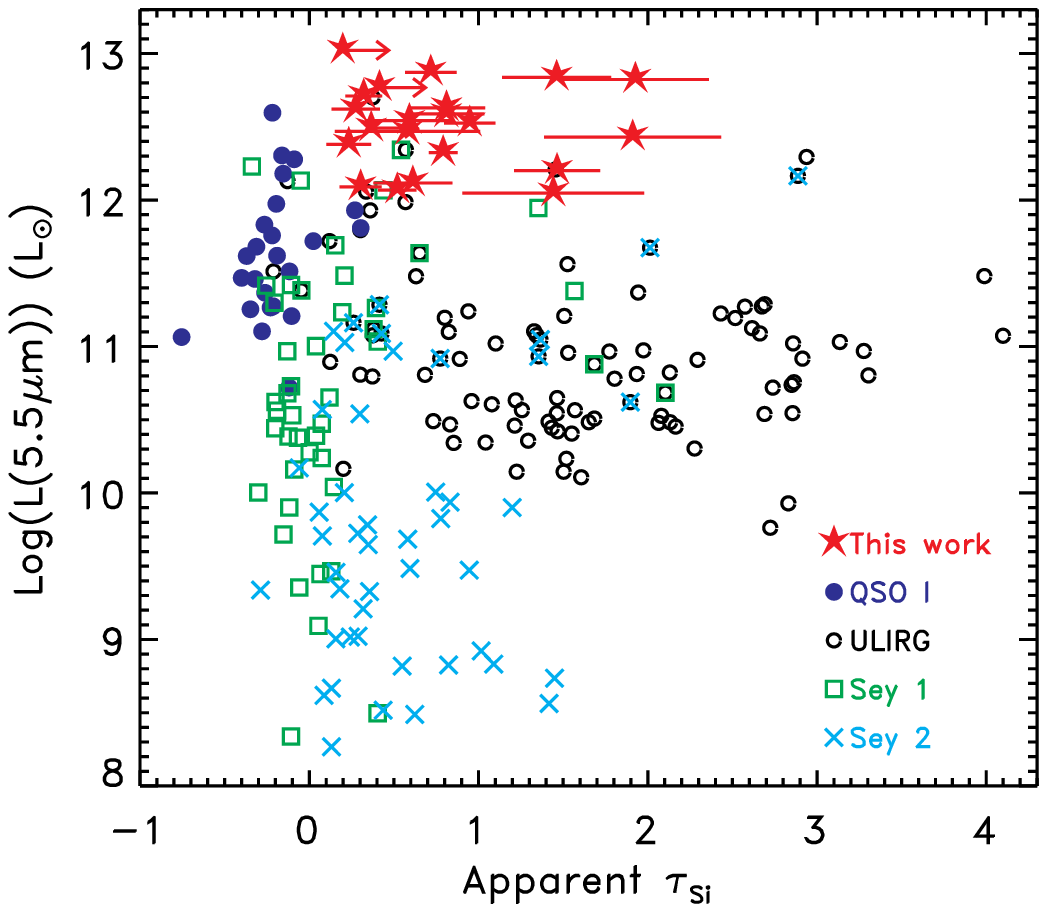}
      \caption{ Mid-IR (5.55$\mu$m rest-frame) luminosities $vs$ the $Si$
        apparent optical depth at 9.7$\mu$m of the obscured AGNs in this work (red
        full stars), and of various AGN samples from~\citet{hao07}: type 1
        QSOs (full blue circles), ULIRGs (empty black circles), Seyfert 1
        (green open squares), and Seyfert 2 (cyan crosses). Rightward
        pointing arrows represent lower limits to $\tau_{Si}$.}
         \label{lmir_si}
   \end{figure}

The MIR (5.5$\mu$m) luminosities and the measured apparent $Si$ optical
depth, $\tau_{Si}$, of our AGN sample are compared with those of other AGN
samples from some of the samples in the above compilation in
Figure~\ref{lmir_si}. The literature samples include all AGNs
in~\citet{hao07}: type 1 QSOs, ULIRGs, Seyfert 1, and Seyfert 2 galaxies.
Our sample covers a new parameter space in the MIR luminosity-absorption
distribution of all known AGNs, as illustrated in Figure~\ref{lmir_si}. The
selected obscured QSOs show the highest MIR luminosities ($L(5.5\mu
m)>10^{12}$\lsun), consistent only with the most luminous type 1 QSOs and
Seyfert 1 galaxies, and also show large $Si$ optical depths, up to
$\tau_{Si}$=3.2, consistent with those observed in ULIRGs and in some
Seyfert 2 galaxies.

\section{X-ray properties}\label{xray}

X-ray data from \chandra\ are available for 11 sources, but only 5 are
detected~\citep[LH\_02, LH\_A4, LH\_A5, LH\_A6, and
LH\_A8;]{weedman06a,polletta06}. \chandra\ 5 ksec observations are available
for the 5 sources in the Bootes field~\citep{murray05}, but none of the
selected sources is detected. We report an upper limit to the broad-band
(0.3-8 keV) X-ray flux for these sources corresponding to their 90\%
completeness limit (10$^{-14}$
\ergcm2s)~\citep{murray05}. LH\_A11 is part of the 70 ksec SWIRE/\chandra\
survey~\citep{polletta06}, but was not detected, thus we report an upper
limit to the broad band flux of 10$^{-15}$\ergcm2s.
\chandra\ observations for B2 are scheduled (CXO Proposal 08700396) and \xmm\
observations are pending for N2\_06, B1, N2\_08, and B3 (\xmm\ Proposal
050326). In Table~\ref{xray_data}, we list observed X-ray fluxes, estimated
effective column densities and luminosities for the detected 5 sources, and
upper limits to the fluxes for the remaining 6 sources. Here, we compare the
absorption measured in the X-rays with the extinction measured in the IR and
use the MIR to X-ray luminosity ratio to investigate the dust covering
fraction in the selected sample.

\subsection{X-ray absorption and IR extinction}

\begin{figure}[htbp]
  \epsscale{1.0}
   \plotone{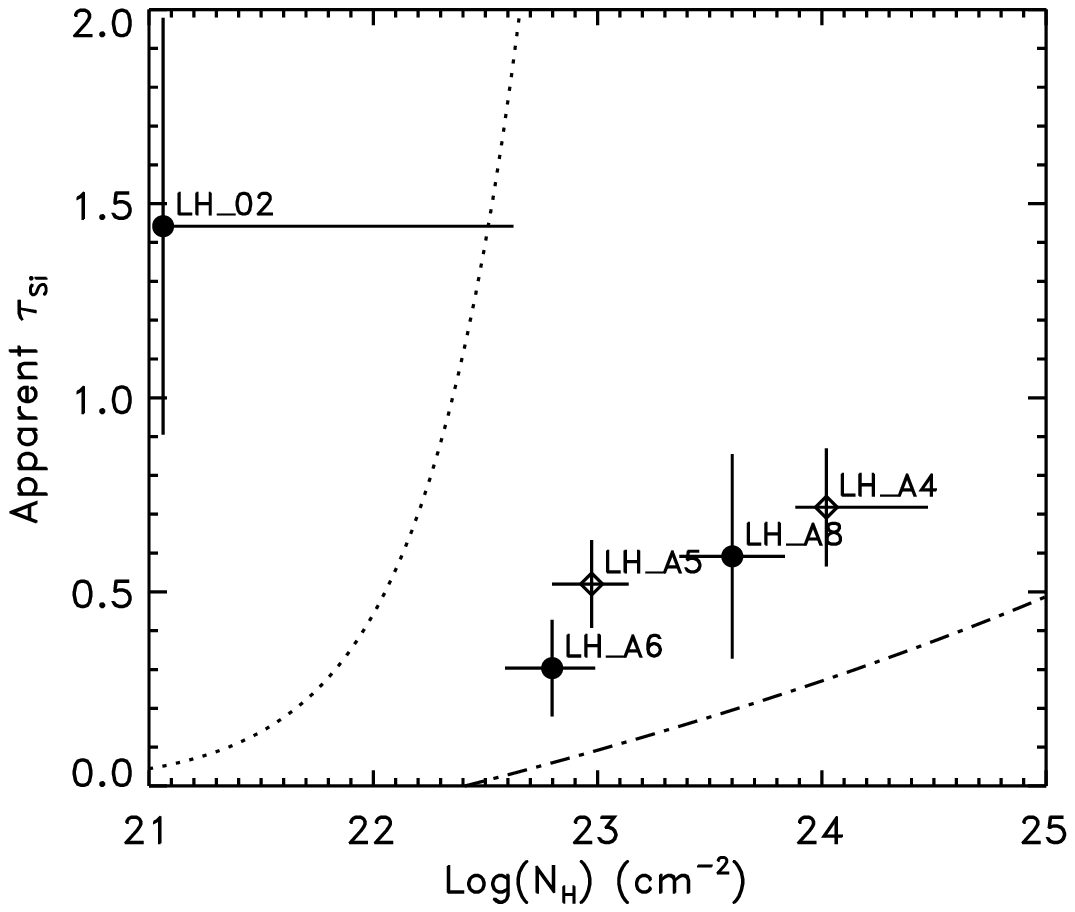}
      \caption{Comparison between the apparent $Si$ optical depth and the
        column density estimated from the X-ray hardness ratio. The dotted
        line corresponds to the expected agreement in case both absorptions
        occur in the same medium and assuming a Galactic gas-to-dust ratio
        of 100. The dashed curve corresponds to the linear fit found for a
        large compilation of AGNs by~\citet{shi06}.}
         \label{tau_nh}
   \end{figure}

The effective gas column densities responsible for absorption in the X-rays
are derived from the X-ray hardness ratios using the method described
in~\citet{polletta06}. The estimated values are reported in
Table~\ref{xray_data}. The absorption measured in the X-rays and the dust
extinction measured in the IR, $\tau_{Si}$, are compared in
Figure~\ref{tau_nh}. We also show a curve (dotted curve) corresponding to
equivalent \nh\ and $\tau_{Si}$ assuming a gas-to-dust ratio of 100, instead
of the Galactic value, a dust density of 2.5 g/cm$^3$ density and 0.1$\mu$m
radius per dust grain, or \nh$\simeq$2.26$\times$10$^{22}\times\tau_{Si}$
\cm2.

In four sources, the \nh\ is significantly larger than that derived in the
IR. The only exception, where the \nh\ is lower, is LH\_A2 which is
characterized by a soft X-ray spectrum. However, there is evidence, based on
the X-ray/IR luminosity ratio, that in LH\_A2 the observed X-ray spectrum
corresponds to a small fraction of the intrinsic X-ray emission due to
scattering, while the intrinsic emission is absorbed by a Compton-thick
column density (\nh$>$10$^{24}$\cm2). This source is also characterized by
the highest IR extinction.

The large \nh\ values compared to the extinction measured in the IR by
$\tau_{Si}$ are easily explained. The apparent $Si$ optical
depth underestimates the true extinction in the IR because its
measurement does not take into account re-emission of radiation at the same
wavelengths where absorption takes place. Moreover we expect the correlation
between the two measurements of absorption to have a wide dispersion because
the absorption measured in the X-rays can be highly
variable~\citep[see e.g.][]{risaliti07}, and because the two columns, in the X-ray
and in the IR, are measured along different paths as the emitting sources
have different spatial locations and distributions. It is also quite
plausible that, in addition to the absorption produced by the medium
that is responsible for the extinction seen in the IR, the X-ray source
suffers also absorption by gas clouds, possibly ionized, close to the
nucleus that may be dust-free, and thus not contribute to any emission or
absorption in the IR.

A similar comparison was presented by~\citet{shi06} for a large compilation
of AGNs of different type. Instead of the apparent $Si$ optical depth,
$\tau_{Si}$, they define the $Si$ strength as an indicator of IR extinction.
The $Si$ strength is equivalent to $e^{-\tau_{Si}}-$1, thus for our sources
it ranges from $-$0.3 to $-$0.8. We report their best linear fit obtained
for their entire AGN samples in Figure~\ref{tau_nh} after converting the
$Si$ strength to the apparent $\tau_{Si}$. By comparing our sources with
their relationship  and their Fig. 3, we find that our sources occupy the
lower boundary of the $Si$ strength-\nh\ relation they observe, and overlap
with the most IR obscured sources in their sample which are all Seyfert 2
galaxies. Thus, the \nh\ and $\tau_{Si}$ of our sources are consistent with
those found in other AGNs, but they overlap with those observed in the most
IR obscured ones. This is a consequence of our sample selection.

\subsection{Dust covering fraction}

The dust covering fraction can be estimated by comparing the thermal
reprocessed emission, e.g. L(6$\mu$m), and the luminosity of the heating
source. The best proxy for the intensity of the heating source is given by
the ultraviolet (UV) luminosity, since UV photons are more efficiently
thermalized than those at other wavelengths (both X-ray and IR). Since we do
not know the intrinsic (before absorption) UV luminosity because it is
mostly absorbed and only few sources are detected in the optical, we adopt
the absorption-corrected X-ray luminosity as a proxy for the intensity of
the heating source. Since the ratio between the UV and X-ray luminosity
varies with luminosity~\citep[e.g.][]{steffen06}, in the sense that more
luminous AGN have lower X-ray/UV luminosity ratios, any trend of this ratio
with the X-ray luminosity must be carefully interpreted. With this caveat in
mind, we adopt the ratio between L(6$\mu$m) and L$_{0.3-8 keV}^{corr}$ as an
approximate indicator of dust covering fraction.

 \begin{figure}[htbp]
  \epsscale{1.0}
   \plotone{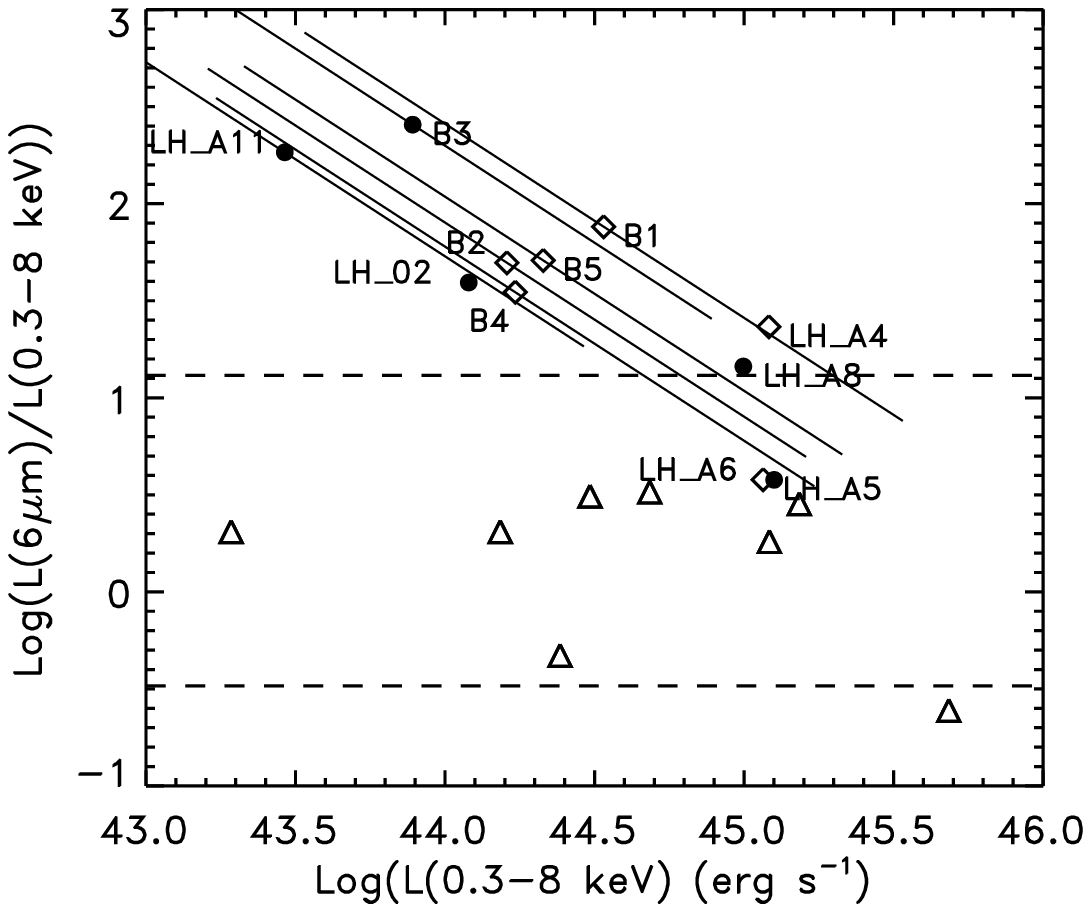}
      \caption{Mid-IR (6$\mu$m rest-frame) over X-ray luminosity as a
         function of X-ray luminosity for the sources presented in this work
         (black full circles and open diamonds) and the type 2 QSOs
         in~\citet{sturm06} (black triangles). Full circles are sources
         fitted with the T model and open diamonds are sources fitted with
         the T+C model. The X-ray luminosities have been corrected for
         absorption. Large error bars on the luminosity ratio (a factor of
         10 in the X-ray luminosity) are reported for sources for which only
         an upper limit to the X-ray flux is available. The dashed lines
         enclose the region where the majority of AGNs lie~\citep[adapted
         from][]{polletta07}.}
         \label{lmirlx_lx}
   \end{figure}

In Figure~\ref{lmirlx_lx}, we show the MIR over X-ray luminosity ratios of
our sample. In case of non X-ray detections, we show the X-ray luminosity
corresponding to the upper limit to the flux. Since the flux is lower than
the assumed upper limit, the intrinsic luminosity could be much lower than
estimated, but since we do not correct for absorption, it can also be higher
than estimated. The luminosity ratios for these sources are thus reported
with an uncertainty of a factor of 10.

We also report for comparison the luminosity ratios typical of AGNs (region
enclosed by the dashed lines in Figure~\ref{lmirlx_lx}) and of the X-ray
selected type 2 QSOs from~\citet{sturm06} (hereinafter S06 sample). The
range of ratios for AGNs was obtained from a compilation of ratios for
various types of AGN~\citep[adapted from][]{polletta07}. These authors find
that the observed range of MIR-to-X-ray luminosity ratios in AGNs is
characterized by a wide dispersion, from 0.3 to 13, with no dependency on
AGN type and luminosity, confirming previous results~\citep[see
e.g.][]{lutz04,sturm05,sturm06,horst06}. The sources in the S06 sample are
heavily absorbed in the X-rays and their MIR emission is mostly due to the
AGN, but they do not show the $Si$ feature in absorption. The ratios for
the S06 sample are consistent with those typical of AGNs.

Interestingly, most (9 out of 11) of the sources in our sample show higher
ratios than the literature samples and the S06 sample. The observed offset
is not due to a different range of X-ray luminosities, as it is observed
across the entire luminosity range of our sample. Yet, it is difficult to
establish whether there is a clear difference among our sample and the
others. For example, the X-ray luminosity of LH\_02 is likely
underestimated. Indeed its soft X-ray spectrum and low X-ray luminosity
suggest that its observed X-ray emission is scattered light and the
intrinsic luminosity is completely absorbed. Moreover, the uncertainties
associated with the absorption corrected X-ray luminosities can be quite
large since these are typically faint sources and the column densities are
derived from the hardness ratios rather than from X-ray spectral modeling.
Deeper X-ray data for these sources or X-ray observations of similar depth
as those already available for the rest of the sample would be necessary to
confirm such a result. If confirmed, this offset in L$_{MIR}$/L$_{X}$ might
be simply due to a selection bias in our sample. A bias could have been
introduced by the powerful AGN emission at MIR wavelengths which might be an
indicator for a specific dust geometry, e.g. a more compact cloud
distribution in the vicinity of the nucleus. Also the presence of $Si$
absorption might have introduced a bias in favor of dustier sources and thus
with larger dust covering factors. However, we do not expect a correlation
between deep $Si$ features in absorption and covering factors, i.e. large
L(MIR)/L(X-ray) luminosity ratios, if the medium producing $Si$ absorption
is clumpy as suggested by~\citet{nenkova02,dullemond05,hoenig06,shi06}.
Indeed, the LOS might not intercept any cloud, even if the covering fraction
is large, and thus not exhibit the $Si$ feature in absorption, whereas a
source with a low covering factor might be seen through an optically thick
cloud and show a deep $Si$ feature.

\section{Discussion}\label{discussion}

\subsection{Starburst signatures in obscured AGNs}\label{disc_sb}

For assigning the ultimate luminosity source of an obscured galaxy, it is
necessary to know what fraction arises from an AGN and what fraction from a
starburst. Our modeling discussed above indicates that the observed NIR and
MIR emissions are mainly produced by dust heated by an AGN.  Because of
suggestions that the obscuring material may be associated with starbursts
within the clouds that surround the AGN, we search for signatures of a
starburst component in the spectra of these obscured AGNs. Because of poor
S/N within the individual spectra (see Figure~\ref{irs_spectra}), there are
no confident detections of the features which are diagnostics of a
starburst, such as PAHs and low ionization emission lines. Some of these
features might be present in some spectra, but higher S/N spectra are
required to confirm them. Here, we analyze the composite spectra in
Figure~\ref{irs_avg_spe} to search for weak PAH features that are not
visible in individual objects.

In none of the composite spectra, however, there are any confident
detections of PAH features.  Limits can be put on the strength of these
features relative to the underlying continuum.  The spectrum in
Figure~\ref{irs_avg_spe} with the highest limit for a PAH feature is the
composite for sources fit with the T model
(panel $a$ in Figure~\ref{irs_avg_spe}), in which a feature near
7.7$\mu$m might be a PAH feature whose peak F(7.7$\mu$m) exceeds by
$\leq$ 10\% the flux density of the underlying continuum at 7.7$\mu$m,
F(7.7$\mu$m)$^{cont}$. If this feature is real, the ratio
F(7.7$\mu$m)/F(7.7$\mu$m)$^{cont}$, where F(7.7$\mu$m)
represents the flux of the total (continuum+line) at 7.7$\mu$m, would be
1.1. For objects classified as pure starbursts without any AGN contribution,
this ratio always exceeds 5 and is typically
$\sim$10~\citep{genzel98,brandl06,weedman06a,houck07}. Lower values are
observed in some SMGs, but always greater than 2~\citep{valiante07}. If a
starburst is present in this spectrum, therefore, it is very diluted by the
underlying continuum of the AGN component, as seen in other AGNs where PAH
emission appears to be weak or absent in the presence of a strong AGN
continuum in the MIR~\citep{lutz98,clavel00,valiante07}.

For use with spectra of poor S/N for which the total PAH fluxes are
difficult to determine,~\citet{houck07} derive an empirical relationship
between the luminosity of the 7.7$\mu$m feature and the bolometric
luminosity of a starburst, $L_{bol}^{SB}$, using the starburst spectra and
luminosities in \citet{brandl06}.  This relation is 
\begin{equation}
log L_{bol}^{SB} = log[\nu L_{\nu}^{SB} (7.7\mu m)] + 0.78,
\label{lbol_lpah}
\end{equation}
for $\nu$L$_{\nu}^{SB}$ (7.7$\mu$m) measured from the F(7.7$\mu$m)$^{line}$
at the peak of the 7.7$\mu$m feature. If the feature has lower equivalent
width than in a pure starburst, as in our composite spectrum, a correction
is necessary to subtract the underlying continuum, F(7.7$\mu$m)$^{cont}$,
from F(7.7$\mu$m) in order to determine F(7.7$\mu$m)$^{line}$, and thus $\nu
L_{\nu}^{SB}$ (7.7$\mu$m) that is due to the starburst.

The composite spectrum in panel $a$ of Figure~\ref{irs_avg_spe} also
indicates that $\nu$F(7.7$\mu$m)$^{cont}$ = $\nu$F(6$\mu$m). Therefore, the
starburst 7.7$\mu$m luminosity can be derived as $\nu L_{\nu}^{SB}$
(7.7$\mu$m) $\leq$ 0.1 $\times L_{cont}(7.7 \mu m)$ = 0.1$\times L(6 \mu m)$. 
Figure~\ref{lmir_lbol}, and Table~\ref{lum_tab} illustrate that the median
$log(L(6\mu m)/erg s^{-1})$ = 46.10, yielding a median $log (\nu
L_{\nu}^{SB} (7.7\mu m))$ = 45.10 for the most luminous PAH feature that
would satisfy the observed limit in the composite spectrum.  This limiting
starburst would then have $log L_{bol}^{SB}/erg s^{-1}$ = 45.88 (equal to
$\sim$2$\times$10$^{12}$\lsun\ or to a SFR$\simeq$350\,\msun\,yr$^{-1}$) arising from
the starburst, using equation~\ref{lbol_lpah}.  This starburst luminosity,
compared to the median AGN bolometric luminosity, log $L_{bol}^{AGN}/erg
s^{-1}$ = 46.47, (see Figure~\ref{lmir_lbol} and Table~\ref{lum_tab}),
indicates, therefore, that any starburst which is present in the limit of
the observed composite spectrum has, on average, a bolometric luminosity $<$ 26\% of the
bolometric luminosity arising from the AGN, and thus $<$20\% of the total
(AGN+Starburst) bolometric luminosity.

Eight sources (out of 13 SWIRE sources and 5 E-FLS sources) are detected in
the FIR (70 and/or 160$\mu$m). For these sources, the FIR luminosities (see
Table~\ref{lum_tab}) are consistent with those observed in the most luminous
starburst galaxies ($\sim$10$^{13}$\lsun), but exceed the limiting starburst
luminosities implied by the absence of PAH features (=
10$^{45.88}$\ergs=2$\times$10$^{12}$\lsun). In two cases (N1\_09, N2\_06),
these powerful FIR luminosities are consistent with those predicted by the
torus models.  However, in the other 6 cases there is evidence for an
additional FIR component whose origin is still unclear. The FIR luminosity
could arise from a starburst, or from dust far from the torus but still
heated by the AGN.

\subsection{Obscuration at high-luminosities}

A comparison between the MIR properties of our sample with other samples
from the literature shows that the selected sources can be as luminous as
the most luminous MIR selected type 1 AGNs in the same redshift range.
The estimated ratio between obscured and unobscured AGN at these
luminosities ($L_{6\mu m}\simeq$10$^{45.5-46.5}$\ergs) is about 1.5--1.9,
and the surface density is 17--22 obscured QSOs deg$^{-2}$ to the limits of
the SWIRE survey.

However, as discussed in the previous sections, the ratio between obscured
and unobscured QSOs derived here does not correspond to the ratio between
QSOs seen through the obscuring torus and those seen through the torus
opening cone. Indeed, from our study, it emerges that the obscuring matter
along the LOS is not always associated with the torus. Only about half (12
sources) of obscured QSOs are seen through the torus. Overall and including
the unobscured AGN population, we conclude that 35--41\% of MIR selected red
QSOs are unobscured, 37--40\% are obscured by the torus, and the remaining
23--25\% are obscured, but not by the torus. These values constrain the
covering factor of the torus at these extreme luminosities to be
$\sim$37--40\%, and the half torus opening angle, measured from the torus
axis, to be $\sim$67\deg. This is significantly larger than the value of
46\deg\ estimated for FIR selected Seyfert galaxies~\citet{schmitt01}. Our
results thus support the receding torus scenario, but are based on the
interpretation that a large fraction of obscured QSOs are obscured by dust
detached from the torus located along or outside the torus opening cone.

The presence of foreground cold dust detached from the torus that might
absorb the nuclear emission was first suggested by~\citet{keel80} and,
subsequently supported by other studies at optical, IR, radio, and X-ray
wavelengths~\citep{lawrence82,rigby06,ogle06b,martinez06a,brand07,sajina07b,urrutia07}.
In our Galaxy, a gas-rich spiral, $\tau_{Si}$ is estimated to be $\sim$ 2.5
in an equatorial direction~\citep{rieke89}. It is possible, therefore, to
produce the observed obscuration with dust in a normal spiral galaxy if this
host galaxy is viewed edge-on, or in a very disturbed system as a merging
galaxy.

Whether the obscuring dust is in a gas-rich spiral host or in a starburst
component, it clearly suggests that these extremely luminous and obscured
AGNs are not hosted by dust-free elliptical galaxies as commonly seen in
luminous AGNs in the nearby universe~\citep[e.g.][]{dunlop03}. The sources
in this study are thus powerful AGNs and they might be hosted by dusty
galaxies.

The cold absorber will also absorb the nuclear X-ray emission if located
along its path. Thus, we can expect that all obscured QSOs, including those
obscured only by the cold absorber and those obscured by the torus, will
have large column densities in the X-ray (\nh$>$10$^{22}$\cm2). Are our
predictions consistent with the findings from X-ray studies?

The lack of X-ray data for the majority of the sources in our sample does
not allow us to verify that these sources are also absorbed in the X-rays.
Only 5 sources have available X-ray data, and they all show large column
densities (\nh$\geq$10$^{23}$\cm2), strongly suggesting that the same might
be true for all these sources. At the X-ray luminosities of our sources,
$\sim$10$^{45}$\ergs, the observed ratio between X-ray absorbed and
unabsorbed AGN in X-ray selected AGN samples is, unfortunately, still
largely unconstrained, $\sim$0.6--1.5:1~\citep[e.g.][]{akylas06,tajer07},
mainly because of the difficulty of detecting absorbed sources in the
X-rays. These values are lower than our predicted 1.5--1.9:1 ratio. On the
other hand, when the selection effects against absorbed sources are
minimized and the spectroscopic completeness is maximized, the absorbed to
unabsorbed fraction can be as high as 2.5:1~\citep{wang07}. Since this ratio
is higher than our estimates, it is more in agreement with our predictions.

Instead of adding the cold absorber, an alternative way to reproduce the
observed SEDs is to include an additional dusty component to reproduce the
observed NIR emission. The spectrum of this NIR component must decline
steeply above $\sim$8$\mu$m, and thus have a very narrow range of
temperatures $>$ 700 K. This NIR component could be produced by direct
emission from warm dust around the AGN and not be associated with the torus,
e.g. in the narrow line region~\citep{efstathiou95,ruiz00,axon01}, but
models do not usually reproduce the high temperatures observed in our
sample~\citep{groves06}. A more viable origin for this hot dust component
might be scattered light in the NIR. This last hypothesis requires the
presence of optically thin dust in the NLR which scatters the nuclear NIR
emission into the LOS~\citep{efstathiou95,pier93}. This scenario, initially
suggested theoretically~\citep{pier93,efstathiou95}, is also
supported by speckle interferometric NIR observations of the Seyfert 2
galaxy NGC 1068~\citep{weigelt04}. Reconstructed $K$-band images show that
the nuclear emission is extended parallel to the outflow and jet direction,
instead of perpendicular as expected for the torus. Observations in
polarized light would be necessary to test this scenario for our objects.
Other possible explanations are a more complex geometry, or a different
extinction curve.

\section{Summary and Conclusions}\label{summary}

In this work, we modeled the IR SEDs and IRS spectra of a sample of 21 MIR
selected red AGNs at $z$=1.3--3.0 with extreme MIR luminosities ($\nu
L_{\nu}$(6$\mu$m) = $L(6\mu m)\simeq$ 10$^{46}$\ergs) with the goal of
investigating the properties of their obscuring matter and dependence on
luminosity. The sample was drawn from the three largest \spitzer\ surveys
(SWIRE, NDWFS, \& FLS) by means of their extremely red infrared colors and
presence of a $Si$ absorption feature in their IRS spectra. Eighteen sources
clearly show the $Si$ absorption feature with various optical depths
(Figure~\ref{irs_spectra}). In the remaining three sources, the $Si$
absorption is only tentatively observed.

The obscured sources are as MIR luminous as the most luminous MIR
selected type 1 AGNs currently known at similar redshifts, but are
characterized by $Si$ optical depths typical of ULIRGs and Seyfert 2
galaxies, that are usually less luminous MIR sources. The presented sample
thus covers a new parameter space in the L(MIR)--$\tau_{Si}$ region.

The observed SEDs and spectra are fit with the clumpy torus models developed
by~\citet{hoenig06}. From the modeling, it emerges that the obscuring matter
along the LOS is not always associated with the torus. Only about half (12
sources) of obscured QSOs are seen through the torus and nine sources
require, in addition to a torus, a cold absorber to reproduce the depth of
the $Si$ feature. The additional component is characterized by large optical
depths ($\tau_V^{CA}$=4--25). Seven of the sources fit with the additional
cold absorber are characterized by tori at low inclination angles,
$\theta=$0--30\deg. The low inclination angles are required by the observed
NIR emission which implies a view of the hottest dust components from the
inner parts of the torus. Even though all sources show the $Si$ feature in
absorption, we conclude from the modelling that there is no preferred torus
inclination in our sample such that these obscured sources have been
selected because they are observed through the torus.

The best-fit models indicate that a compact non-flaring torus is preferred
by the majority of the sources. The torus emission region is compact (more
concentrated towards the nucleus) with the NIR emission being mainly due to
dust in the vicinity of the nucleus. The preferred non-flaring torus in the
majority of these luminous sources is also consistent with the predictions
of the receding torus models~\citep[e.g.][]{simpson05,hoenig07}.

Based on an estimate of the surface density of unobscured QSOs derived from
MIR luminosity functions~\citep{brown06}, and of obscured QSOs derived from
a color-selected sample of IR sources, the estimated surface densities are
about 17--22 deg$^{-2}$ for the obscured QSOs, and 11.7 deg$^{-2}$
for the unobscured ones at $z$=1.3--3.0. The estimated ratio between
obscured and unobscured AGNs at these luminosities ($L(6\mu
m)\simeq$10$^{45.5-46.5}$\ergs) and redshift range is, thus, $\sim$1.5--1.9.
Overall we find that $\sim$35-41\% of MIR selected red AGNs are unobscured,
23-25\% are obscured but not by the torus, and the remaining 37-40\% by the
torus. The ratio of obscured to unobscured QSQs constrains the torus half
opening angle, measured from the torus axis, to be $\sim$67\deg. This value
is significantly larger than found for FIR selected samples of AGN at lower
luminosity~\citep[$\sim$46\deg][]{schmitt01}, supporting the receding torus
scenario.

Five sources are detected in the X-rays, and large column densities
(\nh$\geq$10$^{23}$\cm2) are estimated for all of them. For these objects,
the dust covering fraction estimated from the ratio between the MIR
and X-ray luminosity are on average higher than what is usually measured in
AGNs. However, the large uncertainties and limited sub-sample of detected
X-ray sources, do not allow us to confirm such a result.  Whether these
sources are characterized by an excess of MIR emission with respect to the
X-ray emission and whether this is due to a peculiar dust distribution with
respect to the typical AGN population introduced by our selection is thus
not clear.

From the SEDs of the models, we estimate that the AGN bolometric
luminosities, $L_{bol}^{AGN}$, range from 10$^{46}$ to 10$^{47}$\ergs, and
the AGN bolometric luminosity can be derived from the 6$\mu$m luminosity
using the relation $Log(L_{bol}^{AGN})$=$Log(L(6\mu m))+0.32\pm 0.06$. The
bolometric luminosity of the whole system can be much higher, however, if
the luminosity escaping the torus is taken into account and if an additional
FIR component is included. Such a FIR component is not included in the AGN
torus models, but it is observed in 8 cases. In half of the 8 FIR-detected
sources the torus contribution to the total IR luminosity is lower than that
of the additional FIR component.  For two of the remaining objects, the two
contributions are similar and in the other two, the torus contribution is
larger.

The estimated FIR luminosities for these 8 sources are all greater than
3.3$\times$10$^{12}$\lsun, which would imply SFRs of 600--3000 \msun\
yr$^{-1}$~\citep{kennicutt98} if these FIR luminosities arise from
starbursts.  However, none of the sources show the PAH features
characteristic of starbursts; a limit to the starburst component is
estimated from the limit on PAH 7.7$\mu$m luminosity in the composite IR
spectra.  From this limit, we estimate that, on average, the contribution to
the bolometric luminosity from a starburst characterized by PAH emission is
$\leq$20\% of the total bolometric luminosity.  This leaves unanswered the
question of whether the extreme FIR luminosities of these sources arise from
heavily obscured starbursts or arise from AGN luminosity heating cool dust 
that is distant from the torus.

\acknowledgments

We gratefully acknowledge the anonymous referee for a careful reading of the
manuscript and for useful comments and suggestions that improved the paper.
MP kindly thanks R. Maiolino, N. Levenson, D. J. Axon, and E. Treister for
stimulating discussions, B. Siana for providing the optical spectral
classification, L. Hao for providing the data used in Figure~\ref{lmir_si}
in machine readable format, J. Surace for help with the IRAC measurements,
M. Lacy for providing information on the latest FLS catalog, S. Croom for
his advice on estimating the surface density of unobscured type 1 QSOs, and
to A. Afonso-Luis for providing information on the SWIRE MIPS 70 and
160$\mu$m observations.  MP acknowledges financial support from the
Marie-Curie Fellowship grant MEIF-CT-2007-042111. This work is based on
observations made with the {\it Spitzer Space Telescope}, which is operated
by the Jet Propulsion Laboratory, California Institute of Technology under
NASA contract 1407. Support for this work, part of the {\it Spitzer Space
Telescope} Legacy Science Program, was provided by NASA through an award
issued by the Jet Propulsion Laboratory, California Institute of Technology
under NASA contract 1407. Support for this work by the IRS GTO team at
Cornell University was provided by NASA through Contract
Number 1257184 issued by JPL/Caltech. This work is dedicated to the memory
of Harding Eugene (Gene) Smith.

Facilities: \facility{Spitzer(IRS,IRAC,MIPS)}, \facility{CXO(ACIS)}.


\clearpage

\topmargin=2cm
\footskip=0in
\begin{deluxetable}{l@{}ll@{}c@{}c@{}c@{}c@{}c@{} c@{}c@{}c@{}c@{} c@{}c@{}c@{} c}
\tabletypesize{\scriptsize}
\rotate
\tablecaption{Optical and infrared data of the obscured QSOs sample\label{basic_data}}
\tablewidth{0pt}
\tablehead{
\colhead{Survey}  & 
\colhead{Source}  & 
\colhead{IAU} &
\colhead{F$_\mathrm{U}$\tablenotemark{a}} &   
\colhead{F$_{g^{\prime} or B}$\tablenotemark{a}} & 
\colhead{F$_{r^{\prime} or R}$\tablenotemark{a}} & 
\colhead{F$_{i^{\prime} or I}$\tablenotemark{a}} & 
\colhead{F$_z$\tablenotemark{a}} &   
\colhead{F$_{3.6\mu m}$} & 
\colhead{F$_{4.5\mu m}$} & 
\colhead{F$_{5.8\mu m}$} & 
\colhead{F$_{8.0\mu m}$} & 
\colhead{F$_{24\mu m}$} &
\colhead{F$_{70\mu m}$} &
\colhead{F$_{160\mu m}$} &
\colhead{Ref.}\\
\colhead{Name}      & 
\colhead{Name}      & 
\colhead{Name}      & 
\colhead{($\mu$Jy)} &
\colhead{($\mu$Jy)} &
\colhead{($\mu$Jy)} &
\colhead{($\mu$Jy)} &
\colhead{($\mu$Jy)} &
\colhead{($\mu$Jy)}    &
\colhead{($\mu$Jy)}    &
\colhead{($\mu$Jy)}    &
\colhead{($\mu$Jy)}    &
\colhead{($\mu$Jy)}    &
\colhead{(mJy)}    &
\colhead{(mJy)}    &
\colhead{}
}
\startdata
 SWIRE &   LH\_01  & SWIRE J104148.93+592233.0                 & $<$0.441  & $<$0.326 & $<$0.486 & $<$0.821 &  \nodata &    5.4 &     12 &  $<$43 &   120 &    1351  &  $<$17.0  &  $<$90  &     1  \\
 SWIRE &   LH\_02  & SWIRE J104605.56+583742.4                 & $<$0.441  & $<$0.326 & $<$0.486 & $<$0.821 &  \nodata &     12 &     20 &     69 &   124 &    1407  &  $<$17.0  &  $<$90  &     1  \\
 SWIRE &   LH\_A4  & SWIRE J104409.95+585224.8                 &    0.360  &    1.038 &    1.186 &    1.312 &    1.823 &     64 &    150 &    391 &  1093 &    4134  &  $<$17.0  &  $<$90  &     2  \\
 SWIRE &   LH\_A5  & SWIRE J104453.07+585453.1                 &    0.236  &    0.333 &    0.409 &    0.611 & $<$1.000 &     69 &    141 &    292 &   467 &    1190  &  $<$17.0  &  $<$90  &     2  \\
 SWIRE &   LH\_A6  & SWIRE J104613.48+585941.4                 &    1.115  &    1.799 &    1.993 &    2.850 &    3.535 &     27 &     34 &     58 &   127 &    1147  &  $<$17.0  &  $<$90  &     2  \\
 SWIRE &   LH\_A8  & SWIRE J104528.29+591326.7                 & $<$0.441  &    0.517 &    1.042 &    2.223 &    3.654 &     32 &     45 &     91 &   207 &    2462  &      9.8  &  $<$90  &     2  \\
 SWIRE &   LH\_A11 & SWIRE J104314.93+585606.3                 & $<$0.411  & $<$2.473 & $<$2.324 & $<$3.269 &  \nodata &    9.2 &     22 &     60 &   117 &     968  &  $<$17.0  &  $<$90  &     2  \\
 SWIRE &   N1\_09  & SWIRE J160532.69+535226.4                 & $<$0.801  &    1.107 &    1.491 &    1.914 & $<$3.784 &     15 &     21 &  $<$43 &    73 &    2894  &      8.5  &  $<$90  &     1  \\
 SWIRE &   N2\_06  & SWIRE J163511.43+412256.8                 & $<$0.801  & $<$0.409 & $<$0.752 & $<$1.324 & $<$3.784 &     14 &     18 &     36 &   116 &    4231  &     10.3  &  $<$90  &     1  \\
 SWIRE &   N2\_08  & SWIRE J164216.93+410127.8\tablenotemark{b}& $<$0.801  & $<$0.409 & $<$0.752 & $<$1.324 & $<$3.784 &     72 &    176 &    475 &  1178 &    3984  &  $<$17.0  &   80.5  &     1  \\
 SWIRE &   N2\_09  & SWIRE J164401.40+405715.0                 & $<$0.801  & $<$0.409 & $<$0.752 & $<$1.324 & $<$3.784 &     39 &     58 &     86 &   143 &    3715  &     27.5  &    121  &     1  \\
 NDWFS &       B1  & SST24 J143001.91+334538.4\tablenotemark{c}&  \nodata  &    0.142 &    0.302 &    0.521 &  \nodata &      8 &     29 &    102 &    586 &   3800  &  $<$30.0  & \nodata &     1  \\
 NDWFS &       B2  & SST24 J143644.22+350627.4\tablenotemark{d}&  \nodata  &    0.325 &    0.525 &    0.753 &  \nodata &     43 &     97 &    380 &    732 &   2300  &  $<$30.0  & \nodata &     1  \\
 NDWFS &       B3  & SST24 J142827.19+354127.7                 &  \nodata  &    0.980 &    2.290 &    5.210 &  \nodata &    296 &    621 &   1601 &   3939 &  10550  &  \nodata  & \nodata &     1  \\
 NDWFS &       B4  & SST24 J142648.90+332927.2\tablenotemark{e}&  \nodata  &    0.514 &    0.912 &    1.194 &  \nodata &     53 &    178 &    493 &    911 &   2300  &  $<$30.0  & \nodata &     1  \\
 NDWFS &       B5  & SST24 J143508.49+334739.8                 &  \nodata  &    0.390 &    0.692 &    0.905 &  \nodata &     12 &     21 &     34 &    237 &   2600  &  $<$30.0  & \nodata &     1  \\
 FLS   & MIPS42    & SST24 J171758.44+592816.8                 &  \nodata  &  \nodata &    0.186 &  \nodata &  \nodata &   $<$9 &     30 &    103 &    680 &   4712  &     12.8  & \nodata &     3  \\
 FLS   & MIPS78    & SST24 J171538.18+592540.1                 &  \nodata  &  \nodata &    0.186 &  \nodata &  \nodata &   $<$9 &     39 &  $<$72 &    268 &   2973  &   $<$3.9  & \nodata &     3  \\
 FLS   & MIPS15840 & SST24 J171922.40+600500.4                 &  \nodata  &  \nodata &    0.462 &  \nodata &  \nodata &     18 &     25 &     62 &    197 &   1821  &   $<$4.8  & \nodata &     3  \\
 FLS   & MIPS22204 & SST24 J171844.38+592000.5                 &  \nodata  &  \nodata &    2.996 &  \nodata &  \nodata &  $<$27 &     39 &    132 &    478 &   4101  &     13.2  & \nodata &     3  \\
 FLS   & MIPS22303 & SST24 J171848.80+585115.1                 &  \nodata  &  \nodata &    0.395 &  \nodata &  \nodata &   $<$9 &  $<$12 &  $<$54 &    103 &   2030  &      7.2  & \nodata &     3  \\
\enddata
\tablecomments{Typical uncertainties to the optical and IR fluxes are $\sim$
  4\%, and 10\% of the measured fluxes, respectively. Upper
 limits correspond to 5$\sigma$.}
\tablenotetext{a}{The optical data correspond to U\gp \rp \ip\ z in SWIRE, to
 UBRI in NDWFS, and R in FLS.}
\tablenotetext{b}{Also known as ELAISC15 J164216.9+410128.}
\tablenotetext{c}{N. 9 in~\citet{houck05}.}
\tablenotetext{d}{Also known as NDWFS J143644.3+350627 and N. 13 in~\citet{houck05}.}
\tablenotetext{e}{Also known as FIRST J142648.9+332927 and ELAISR142650+332940B.}
\tablerefs{
 (1)~this work;
 (2)~\citet{weedman06a};
 (3)~IRAC data from~\citet{sajina07}, and R-band and MIPS[24] data from~\citet{yan07}.}
\end{deluxetable}

\begin{deluxetable}{l r ccc}
\tabletypesize{\scriptsize}
\tablecaption{IRS Low resolution Spectroscopy Observation Log\label{irs_log}}
\tablewidth{0pt}
\tablehead{
\colhead{Source}  & 
\colhead{Program} &
\colhead{SL1} &
\colhead{LL2} &
\colhead{LL1} \\
\colhead{Name}      & 
\colhead{ID}      & 
\colhead{(seconds)} &
\colhead{(seconds)} &
\colhead{(seconds)} 
} 
\startdata
 LH\_01                   &  136 &  3$\times$60  &  6$\times$120  &  6$\times$120  \\
 LH\_02                   &  136 &  3$\times$60  &  6$\times$120  &  6$\times$120  \\
 LH\_A4\tablenotemark{a}  &   15 &  2$\times$60  &  3$\times$120  &  3$\times$120  \\
 LH\_A5\tablenotemark{a}  &   15 &  3$\times$60  &  6$\times$120  &  6$\times$120  \\
 LH\_A6\tablenotemark{a}  &   15 &  3$\times$60  &  6$\times$120  &  6$\times$120  \\
 LH\_A8\tablenotemark{a}  &   15 &  2$\times$60  &  3$\times$120  &  3$\times$120  \\
 LH\_A11\tablenotemark{a} &   15 &  5$\times$60  &  6$\times$120  &  6$\times$120  \\
 N1\_09                   &   15 &  3$\times$60  &  3$\times$120  &  3$\times$120  \\
 N2\_06                   &   15 &  2$\times$60  &  2$\times$120  &  2$\times$120  \\
 N2\_08                   &   15 &  2$\times$60  &  2$\times$120  &  2$\times$120  \\
 N2\_09                   &   15 &  2$\times$60  &  2$\times$120  &  2$\times$120  \\
 B1                       &   12 &  1$\times$240 &  2$\times$120  &  2$\times$120  \\
 B2                       &   15 &  3$\times$60  &  3$\times$120  &  3$\times$120  \\
 B3                       &   15 &  2$\times$60  &  2$\times$120  &  2$\times$120  \\
 B4                       &   15 &  2$\times$60  &  3$\times$120  &  3$\times$120  \\
 B5                       &   15 &  3$\times$60  &  3$\times$120  &  3$\times$120  \\
 MIPS42                   & 3748 &  2$\times$240 &  2$\times$120  &  3$\times$120  \\
 MIPS78                   & 3748 &  2$\times$240 &  2$\times$120  &  3$\times$120  \\
 MIPS15840                & 3748 &  2$\times$240 &  6$\times$120  &  7$\times$120  \\
 MIPS22204                & 3748 &  2$\times$240 &  2$\times$120  &  3$\times$120  \\
 MIPS22303                & 3748 &       \nodata &  2$\times$120  &  3$\times$120  \\
\enddata
\tablecomments{IRS data processed with version 13.0 of the SSC pipeline.}
\tablenotetext{a}{IRS data processed with version 11.0 of the SSC pipeline.}
\end{deluxetable}

\begin{deluxetable}{l@{}r@{}c@{}c@{}c@{}c@{}c@{}c@{}c@{}c@{}c@{}c@{}c@{}c@{}c@{}c}
\tabletypesize{\scriptsize}
\rotate
\tablecaption{Infrared luminosities and optical depths~\label{lum_tab}}
\tablewidth{0pt}
\tablehead{
\colhead{Source ID}  & 
\colhead{$z$} & 
\colhead{$L_{5.5\mu m}$} & 
\colhead{$L_{5.8\mu m}$} & 
\colhead{$L_{6.0\mu m}$} & 
\colhead{$L_{IR}^{AGN}$\tablenotemark{a}} & 
\colhead{$L_{bol}^{AGN}$\tablenotemark{b}} &
\colhead{$L_{FIR}^{SB}$\tablenotemark{c}} &
\colhead{$L_{FIR}^{AGN+SB}$\tablenotemark{d}} &
\colhead{$L_{IR}^{AGN+SB}$\tablenotemark{e}} &
\colhead{$L_{bol}^{AGN+SB}$\tablenotemark{f}} &
\colhead{$f_{IR}^{AGN}$\tablenotemark{g}} &
\colhead{$f_{IR}^{SB}$\tablenotemark{h}} &
\colhead{$\tau_{Si}^{PL}$\tablenotemark{i}}  &
\colhead{$\tau_{Si}^{Temp}$\tablenotemark{l}}  &
\colhead{$SFR$\tablenotemark{m}} 
} 
\startdata
   LH\_01  &   2.84                   &  46.07  &   46.09  &   46.09  &   46.39  &   46.45  &   \nodata  &   \nodata  &   \nodata  &   \nodata  &   \nodata  &  \nodata  &  $>$0.79 & $>$0.37 &   \nodata \\
   LH\_02  &   2.12                   &  45.63  &   45.66  &   45.67  &   45.95  &   46.00  &   \nodata  &   \nodata  &   \nodata  &   \nodata  &   \nodata  &  \nodata  &     1.98 &    0.91 &   \nodata \\
   LH\_A4  &   2.54\tablenotemark{n}  &  46.45  &   46.45  &   46.45  &   46.69  &   46.83  &   \nodata  &   \nodata  &   \nodata  &   \nodata  &   \nodata  &  \nodata  &     0.87 &    0.57 &   \nodata \\
   LH\_A5  &   1.94\tablenotemark{o}  &  45.65  &   45.65  &   45.64  &   45.89  &   46.08  &   \nodata  &   \nodata  &   \nodata  &   \nodata  &   \nodata  &  \nodata  &     0.63 &    0.41 &   \nodata \\
   LH\_A6  &   2.20\tablenotemark{o}  &  45.67  &   45.68  &   45.68  &   45.92  &   46.02  &   \nodata  &   \nodata  &   \nodata  &   \nodata  &   \nodata  &  \nodata  &     0.43 &    0.18 &   \nodata \\
   LH\_A8  &   2.42\tablenotemark{o}  &  46.12  &   46.15  &   46.16  &   46.42  &   46.49  &     46.34  &     46.36  &     46.83  &     46.86  &      0.39  &     0.61  &     0.86 &    0.33 &       994 \\
   LH\_A11 &   2.30\tablenotemark{o}  &  45.70  &   45.72  &   45.73  &   46.01  &   46.08  &   \nodata  &   \nodata  &   \nodata  &   \nodata  &   \nodata  &  \nodata  &     0.85 &    0.38 &   \nodata \\
   N1\_09  &   2.75                   &  46.35  &   46.38  &   46.40  &   46.69  &   46.72  &     46.10  &     46.19  &     46.86  &     46.88  &      0.67  &     0.33  &  $>$0.86 & $>$0.42 &       572 \\
   N2\_06  &   2.96                   &  46.60  &   46.61  &   46.61  &   46.89  &   46.92  &     46.16  &     46.20  &     47.02  &     47.04  &      0.74  &     0.26  &  $>$0.67 & $>$0.20 &       657 \\
   N2\_08  &   2.40                   &  46.42  &   46.42  &   46.42  &   46.59  &   46.69  &     46.80  &     46.81  &     47.20  &     47.23  &      0.25  &     0.75  &     1.78 &    1.14 &      2868 \\
   N2\_09  &   1.98                   &  46.05  &   46.08  &   46.09  &   46.41  &   46.46  &     46.75  &     46.76  &     47.12  &     47.14  &      0.20  &     0.80  &     1.01 &    0.15 &      2556 \\
       B1  &   2.46                   &  46.41  &   46.41  &   46.41  &   46.56  &   46.62  &   \nodata  &   \nodata  &   \nodata  &   \nodata  &   \nodata  &  \nodata  &     2.36 &    1.49 &   \nodata \\
       B2  &   1.77                   &  45.91  &   45.91  &   45.90  &   46.09  &   46.20  &   \nodata  &   \nodata  &   \nodata  &   \nodata  &   \nodata  &  \nodata  &     0.88 &    0.71 &   \nodata \\
       B3  &  1.293\tablenotemark{n}  &  46.29  &   46.30  &   46.30  &   46.53  &   46.60  &   \nodata  &   \nodata  &   \nodata  &   \nodata  &   \nodata  &  \nodata  &     0.43 &    0.21 &   \nodata \\
       B4  &   1.82                   &  45.78  &   45.78  &   45.78  &   45.98  &   46.11  &   \nodata  &   \nodata  &   \nodata  &   \nodata  &   \nodata  &  \nodata  &     1.72 &    1.21 &   \nodata \\
       B5  &   2.00                   &  46.01  &   46.03  &   46.04  &   46.26  &   46.30  &   \nodata  &   \nodata  &   \nodata  &   \nodata  &   \nodata  &  \nodata  &     2.43 &    1.39 &   \nodata \\
 MIPS42    &   1.90\tablenotemark{p}  &  46.20  &   46.21  &   46.21  &   46.44  &   46.48  &     46.13  &     46.16  &     46.72  &     46.75  &      0.52  &     0.48  &     0.42 &    0.13 &       613 \\
 MIPS78    &   2.43\tablenotemark{p}  &  46.21  &   46.22  &   46.23  &   46.52  &   46.58  &   \nodata  &   \nodata  &   \nodata  &   \nodata  &   \nodata  &  \nodata  &     1.04 &    0.58 &   \nodata \\
 MIPS15840 &   2.23\tablenotemark{p}  &  45.96  &   45.97  &   45.97  &   46.21  &   46.28  &   \nodata  &   \nodata  &   \nodata  &   \nodata  &   \nodata  &  \nodata  &     0.37 &    0.10 &   \nodata \\
 MIPS22204 &   2.08                   &  46.17  &   46.19  &   46.20  &   46.48  &   46.55  &     46.18  &     46.22  &     46.77  &     46.81  &      0.52  &     0.48  &     1.04 &    0.58 &       688 \\
 MIPS22303 &   2.50\tablenotemark{p}  &  46.11  &   46.13  &   46.14  &   46.28  &   46.31  &     46.34  &     46.36  &     46.78  &     46.80  &      0.31  &     0.69  &     1.10 &    0.80 &       994 \\
\enddata
\tablecomments{ All luminosities are logarithm in unit of \ergs. Luminosities
for a starburst component are estimated only for sources detected at 70 or
160$\mu$m.}
\tablenotetext{a}{ \,Logarithm of the AGN IR luminosity obtained by integrating the torus model
                  between 3 and 1000$\mu$m.}
\tablenotetext{b}{ \,Logarithm of the AGN bolometric luminosity obtained by integrating the torus model
                  between 0.1 and 1000$\mu$m.}
\tablenotetext{c}{ \,Logarithm of the starburst component FIR luminosity between 42.5 and 122.5$\mu$m.}
\tablenotetext{d}{ \,Logarithm of the AGN and starburst FIR luminosities between 42.5 and 122.5$\mu$m.}
\tablenotetext{e}{ \,Logarithm of the AGN and starburst IR luminosities between 3 and 1000$\mu$m.}
\tablenotetext{f}{ \,Logarithm of the AGN and starburst bolometric luminosities.}
\tablenotetext{g}{ Fraction of AGN luminosity to the total IR luminosity.}
\tablenotetext{h}{ Fraction of starburst luminosity to the total IR luminosity.}
\tablenotetext{i}{ Silicate optical depth derived from $ln(F_{9.7\mu m}^{int}/F_{9.7\mu m}^{obs}$)
                  with $F_{9.7\mu m}^{int}$ estimated extrapolating a power-law model fit to the data at $\lambda<$7$\mu$m in the rest-frame.}
\tablenotetext{l}{ Silicate optical depth derived from $ln(F_{9.7\mu m}^{int}/F_{9.7\mu m}^{obs}$)
                  with $F_{9.7\mu m}^{int}$ estimated from a type 1 QSO template normalized at the observed 24$\mu$m flux
                  and redshifted at the redshift of the source.}
\tablenotetext{m}{ Star formation rate derived from $L^{SB}_{FIR}$ using the~\citet{kennicutt98} relationship.}
\tablenotetext{n}{ Optical spectroscopic redshift for LH\_A4 from~\citet{polletta06} and for B3 from~\citet{desai06}.}
\tablenotetext{o}{ LH\_A5: $z$=1.89, LH\_A6: 2.10, LH\_A8: 2.31, LH\_A11: 2.25 in~\citet{weedman06a}.}
\tablenotetext{p}{ MIPS42: $z$=1.95$\pm$0.07, MIPS78: $z$=2.65$\pm$0.1, MIPS15840: $z$=2.3$\pm$0.1, MIPS22303: $z$=2.34$\pm$0.14 in~\citet{yan07}.}
\end{deluxetable}

\begin{deluxetable}{l@{}c@{}c@{}c@{}c@{}c@{}c@{}c@{}c@{}c@{}c}
\tabletypesize{\scriptsize}
\tablecaption{Torus model parameters~\label{model_params}}
\tablewidth{0pt}
\tablehead{
\colhead{Source ID}  & 
\colhead{Model\tablenotemark{a}} & 
\colhead{$a$\tablenotemark{b}} & 
\colhead{$b$\tablenotemark{c}} & 
\colhead{$\theta$\tablenotemark{d}} &
\colhead{$N_o$\tablenotemark{e}}  &
\colhead{$N_o^{LOS}$\tablenotemark{f}}  &
\colhead{$\tau^{CA}_\mathrm{V}$\tablenotemark{g}} &
\colhead{$\tau_\mathrm{Si}^{PL}$\tablenotemark{h}} &
\colhead{$\tau_\mathrm{Si}^{Temp}$\tablenotemark{h}} &
\colhead{$\chi^2_{\nu}$\tablenotemark{i}} 
} 
\startdata
  LH\_01          &     T    &    3.0   &   1.5    &    60    &   12    &       6.0    &      0   &   0.29  &  0.01   &    0.016   \\
  LH\_02          &     T    &    3.0   &   1.0    &    90    &   26    &      26.0    &      0   &   1.00  &  0.32   &    5.085   \\
  LH\_A4          &     T    &    3.0   &   1.0    &    60    &   12    &       9.0    &      0   &   1.00  &  0.32   &    4.638   \\
                  &   T+C    &    2.0   &   1.0    &    15    &   26    &       0.0    &      6   &   0.62  &  0.35   &    0.575   \\
  LH\_A5          &     T    &    1.5   &   1.5    &    60    &   18    &      17.5    &      0   &   0.83  &  0.27   &    6.665   \\
                  &   T+C    &    2.0   &   1.0    &    30    &   26    &       0.2    &      7   &   0.62  &  0.36   &    0.703   \\
  LH\_A6          &     T    &    2.0   &   1.0    &    45    &   26    &       5.6    &      0   &   0.32  &  0.07   &    0.050   \\
  LH\_A8          &     T    &    3.0   &   1.0    &    60    &    9    &       6.5    &      0   &   0.81  &  0.26   &    0.078   \\
  LH\_A11         &     T    &    3.0   &   1.0    &    45    &   12    &       4.6    &      0   &   0.75  &  0.25   &    0.533   \\
  N1\_09          &     T    &    3.0   &   1.0    &    60    &   12    &       9.0    &      0   &   1.04  &  0.29   &    0.021   \\
  N2\_06          &     T    &    1.1   &   1.5    &    30    &   13    &       1.0    &      0   &   0.34  &  0.06   &    0.004   \\
  N2\_08          &     T    &    2.0   &   1.0    &    90    &   26    &      26.0    &      0   &   1.05  &  0.40   &   10.120   \\
                  &   T+C    &    2.0   &   1.0    &    30    &   26    &       0.2    &     21   &   2.33  &  1.35   &    2.515   \\
  N2\_09          &     T    &    3.0   &   1.5    &    45    &   8     &       5.2    &      0   &   0.49  &  0.04   &    2.307   \\
  B1              &     T    &    2.0   &   1.0    &    45    &   26    &       5.6    &      0   &   0.37  &  0.11   &    1.435   \\
                  &   T+C    &    3.0   &   1.0    &    15    &   26    &       0.0    &     25   &   2.94  &  1.64   &    0.042   \\
  B2              &     T    &    2.0   &   1.0    &    45    &   26    &       5.6    &      0   &   0.37  &  0.12   &    0.391   \\
                  &   T+C    &    2.0   &   1.0    &    30    &   26    &       0.2    &     18   &   1.85  &  1.07   &    0.540   \\
  B3              &     T    &    2.0   &   1.0    &    45    &   26    &       5.6    &      0   &   0.37  &  0.11   &    0.530   \\
  B4              &     T    &    2.0   &   1.0    &    30    &   26    &       0.2    &      0   &   0.01  &  0.00   &   10.000   \\
                  &   T+C    &    2.0   &   1.0    &    30    &   26    &       0.2    &     16   &   1.60  &  0.92   &    4.898   \\
  B5              &     T    &    2.0   &   1.0    &    45    &   26    &       5.6    &      0   &   0.56  &  0.14   &    0.006   \\
                  &   T+C    &    1.5   &   1.5    &    90    &    6    &       6.0    &     15   &   2.02  &  0.93   &    0.295   \\
  MIPS42          &     T    &    2.0   &   1.0    &    45    &   26    &       5.6    &      0   &   0.43  &  0.14   &    0.390   \\
  MIPS78          &     T    &    3.0   &   1.0    &    60    &   12    &       4.6    &      0   &   1.00  &  0.31   &    2.153   \\
                  &   T+C    &    3.0   &   1.0    &    30    &    9    &       0.3    &     10   &   0.93  &  0.40   &    0.003   \\
  MIPS15840       &     T    &    2.0   &   1.0    &    45    &   26    &       5.6    &      0   &   0.32  &  0.07   &    0.155   \\
  MIPS22204       &     T    &    3.0   &   1.0    &    45    &   12    &       4.6    &      0   &   0.75  &  0.25   &    2.957   \\
  MIPS22303       &     T    &    2.0   &   1.0    &    90    &   26    &      26.0    &      0   &   1.06  &  0.40   &    2.438   \\
                  &   T+C    &    2.0   &   1.0    &    90    &   26    &      26.0    &      4   &   1.54  &  0.68   &    0.775   \\
\enddata
\tablenotetext{a}{T: torus model, T+C: torus+cold absorber model.  The parameters
for the T+C model are given only when the torus only (T) model gives a poor
fit ($\chi^2_{\nu}>$0.1) to the observed SED and spectrum, or the T+C model
significantly improves the $\chi^2$.}
\tablenotetext{b}{Index of the torus density radial distribution, $n_r(r) \propto r^{-a}$,
where $r$ is the torus radius.}
\tablenotetext{c}{Index of the clouds vertical distribution, $H\propto r^b$, where $H$ is
the torus scale height.}
\tablenotetext{d}{Inclination of the torus axis with respect to the line of
sight in degrees.}
\tablenotetext{e}{Average number of clouds along an equatorial LOS.}
\tablenotetext{f}{Average number of clouds along the LOS, which
approximately corresponds to the optical depth of the torus in the
visual~\citep{natta84}.}
\tablenotetext{g}{Cold absorber (CA) optical depth in the visible.}
\tablenotetext{h}{Absorber (torus for model T and torus+cold absorber for
model T+C) optical depth associated with the $Si$ feature measured by
extrapolating a power-law model ($\tau_{Si}^{PL}$) or by fitting a QSO
template ($\tau_{Si}^{Temp}$) (see text).}
\tablenotetext{i}{Reduced $\chi^2$ obtained from the best-fit model and the
observed data at $\lambda>$1$\mu$m in the rest-frame.}

\end{deluxetable}

\begin{deluxetable}{lc ccc}
\tabletypesize{\scriptsize}
\tablecaption{X-ray properties\label{xray_data}}
\tablewidth{0pt}
\tablehead{
\colhead{Source ID}  & 
\colhead{F$_{0.3-8 keV}$\tablenotemark{a}} &
\colhead{\nh$^{eff}$\tablenotemark{b}} & 
\colhead{Log(L$_{0.3-8 keV}$)} &
\colhead{Log(L$_{0.3-8 keV}^{corr}$)\tablenotemark{c}}\\
\colhead{} &
\colhead{(10$^{-15}$ \ergcm2s)} &
\colhead{(10$^{22}$\cm2)} &
\colhead{(\ergs)} &
\colhead{(\ergs)} 
} 
\startdata
   LH\_02  &     4.51  &  0.12$^{+4.10}_{-0.00}$ &     44.07   &   44.08    \\
   LH\_A4  &     1.95  &  105$^{+192}_{-29}$     &     43.83   &   45.08    \\
   LH\_A5  &    19.31  &  9.4$^{+4.3}_{-3.1}$    &     44.61   &   45.06    \\
   LH\_A6  &    24.94  &  6.3$^{+3.5}_{-2.4}$    &     44.78   &   45.10    \\
   LH\_A8  &     4.45  &  40$^{+29}_{-17}$       &     44.19   &   45.00    \\
   LH\_A11 &  $<$1.0   &  \nodata                &  $<$43.46   &   \nodata  \\
       B1  &  $<$10.0  &  \nodata                &  $<$44.53   &   \nodata  \\
       B2  &  $<$10.0  &  \nodata                &  $<$44.21   &   \nodata  \\
       B3  &  $<$10.0  &  \nodata                &  $<$43.89   &   \nodata  \\
       B4  &  $<$10.0  &  \nodata                &  $<$44.23   &   \nodata  \\
       B5  &  $<$10.0  &  \nodata                &  $<$44.33   &   \nodata  \\
\enddata
\tablenotetext{a}{Broad-band (0.3--8 keV) X-ray flux.}
\tablenotetext{b}{Effective column density derived from the
hardness ratios assuming that the intrinsic spectrum is a power-law model
with photon index, $\Gamma$ equal to 1.7 and Galactic \nh\ 
(6$\times$10$^{19}$ \cm2). Uncertainties reflect only the statistical errors
from the observed counts and do not include uncertainties in the spectral
model.}
\tablenotetext{c}{Absorption-corrected X-ray luminosities derived assuming
the effective column density, \nh$^{eff}$, and an intrinsic spectrum modeled
with a power-law model with photon index, $\Gamma$=1.7.}
\tablecomments{For more details on the X-ray data see~\citet{polletta06}.
Upper limits to the X-ray fluxes correspond to 90\% completeness.}
\end{deluxetable}

\end{document}